\documentclass[aps,PRB,twocolumn,superscriptaddress]{revtex4}
\usepackage[T1]{fontenc}
\usepackage{times}
\usepackage{graphicx,color}
\usepackage{amsfonts,amsmath,amssymb,amsbsy}
\usepackage[colorlinks=true, linkcolor=blue, citecolor=blue, urlcolor=blue]{hyperref}
\usepackage{float}

\begin{document}

\title{Quantum geometry and $X$-wave magnets with $X=p,d,f,g,i$}
\author{Motohiko Ezawa}
\affiliation{Department of Applied Physics, The University of Tokyo, 7-3-1 Hongo, Tokyo
113-8656, Japan}

\begin{abstract}
Quantum geometry is a differential geometry based on quantum mechanics. It
is related to various transport and optical properties in condensed matter
physics. The Zeeman quantum geometry is a generalization of quantum geometry
including the spin degrees of freedom. It is related to electromagnetic
cross responses. Quantum geometry is generalized to non-Hermitian systems
and density matrices. Especially, the latter is quantum information
geometry, where the quantum Fisher information naturally arises as quantum
metric. We apply these results to the $X$-wave magnets, which include $d$%
-wave, $g$-wave and $i$-wave altermagnets as well as $p$-wave and $f$-wave
magnets. They have universal physics for anomalous Hall conductivity,
tunneling magneto-resistance and planar Hall effect. We also study
magneto-optical conductivity, magnetic circular dichroism and Friedel
oscillations in the $X$-wave magnets. Various analytic formulas are derived
in the case of two-band Hamiltonians. This paper presents a review of recent
progress together with some original results.
\end{abstract}

\date{\today }
\maketitle
\tableofcontents

\section{Introduction}

Quantum geometry is a differential geometry based on wave functions of
quantum mechanics\cite{Provost,Berry,Ma}. The key component is the quantum
geometric tensor derived from quantum distance under infinitesimal momentum
translation. Its real part gives the quantum metric, while its imaginary
part gives the Berry curvature. It is well established that the Hall
conductivity is related to the integral of the Berry curvature, whose
relation is known as the Thouless-Kohmoto-Nightingale-Nijs (TKNN) formula%
\cite{TKNN,NTW}. On the other hand, quantum metric is related to various
observables such as optical absorption\cite%
{AhnX,Holder,Bhalla,AhnNP,Souza,WChen2022,OnishiX,Sousa,Ghosh,WChen2024,WChen2025,EzawaQG,Oh,EllipticP}%
, nonlinear conductivity\cite%
{Sodeman,MaNature,CWang,Michishita,KamalDas,AGao,NWang,Kaplan,Sala,YFang,ZGong,EzawaMetricC,EzawaPNeel}
and bulk photovoltaic effects\cite{AhnX,AhnNP,WatanabeInject,PJerk}. Quantum
geometric tensor is observed experimentally\cite{Kang,KimS}. Recently,
quantum geometry is generalized to the Zeeman quantum geometry by
introducing spin translation in quantum distance. It is related to
electromagnetic responses.

Ferromagnets are useful for magnetic memory and spintronics devices.
However, there is a limitation of integration and fast dynamics due to the
stray field. On the other hand, there is no such limitation in
antiferromagnets. However, it is hard to readout the spin direction of
antiferromagnet due to net zero magnetization. $d$-wave magnets are
antiferromagnets whose electronic band structure has the $d$-wave splitting%
\cite{Naka,Gonza,NakaB,Bose,Hayami}. It is prominent that spin current can
be generated by applying electric field or thermal gradient without using
the spin-orbit interaction\cite{Naka,Gonza,NakaB,Bose,Hayami}. The notion of
the $d$-wave magnets is generalized to altermagnets\cite{SmejX,SmejX2}
including $g$-wave and $i$-wave altermagnets in two and three dimensions.
They break time-reversal symmetry. In addition, $p$-wave\cite%
{Hayami2020B,pwave} and $f$-wave magnets are also recognized, where
time-reversal symmetry is preserved. They are summarized in the $X$-wave
magnets\cite{Planar} with $X=p,d,f,g,i$ because they share universal features%
\cite{MTJ} irrespective of the presence or the absence of time-reversal
symmetry.

In this paper, we review recent progress of quantum geometry and the $X$%
-wave magnets. We derive compact analytical formulas of the Zeeman quantum
geometry based only on the two-band Hamiltonian without using
eigenfunctions. We also generalize the Zeeman quantum geometry to multiband
systems. Electromagnetic cross responses of the $X$-wave magnet with the
Rashba interaction is also studied. We also review quantum geometry defined
for non-Hermitian systems and density matrices. We note that there are some
recent review articles on quantum geometry\cite%
{TormaRev,LiuRev,JiangRev,YuRev} and altermagnets\cite%
{SmejX2,BaiRev,JungRev2024,NakaRev,Fukaya,JungRev2025,OhnoRev,BhoRev}.

This paper is composed as follows. In Sec.\ref{SecQG}, we construct quantum
geometry starting from the fidelity of the wave function, which measures the
similarity of the two wave functions. The quantum distance is defined based
on the fidelity. By expanding the quantum distance with an infinitesimal
momentum translation, we obtain the quantum geometric tensor. Its real part
gives the quantum metric, while its imaginary part gives the Berry
curvature. The integration of the Berry curvature leads to the Chern number,
which is an integer characterizing a topological insulator. By making a
Fourier transformation of the wave function from the momentum space to the
real space, we obtain the Wannier function. It is shown that the quantum
metric is related to the fluctuation of the position. We show an equality
for the quantum geometry, which gives a lower bound determined by the Chern
number. We summarize simple formulas to obtain the quantum metric and the
Berry curvature directly from the two-band Hamiltonian without using
information about the eigenfunctions. The quantum geometry is generalized to 
$N$-fold degenerate multiband systems.

In Sec.\ref{SecQGAp}, the TKNN formula is derived, which relates the Hall
conductivity to the Chern number. It explains the quantization of the Hall
conductivity for a topological insulator. We apply these results to the
simplest Dirac system in two dimensions. Especially, optical dichroism is
shown, where the optical absorption is selectively occurred depending on the
chirality of the Dirac system. The sum rule which relates the optical
conductivity and the quantum metric is derived. Bulk photovoltaic effects
including the injection current and the shift current are related to the
quantum metric. In addition, the second-order nonlinear conductivity is
shown to be related to the quantum metric.

In Sec.\ref{SecZeeman}, quantum geometry is generalized to include the spin
degrees of freedom, which is called Zeemann quantum geometry. The relation
between the Zeemann quantum geometry and electromagnetic cross responses are
clarified. Simple formulas obtaining the Zeeman quantum geometric tensor for
the two-band systems are derived based only on the Hamiltonian without using
knowledge of the eigenfunctions. Their results are applied to the Rashba
system. The Zeeman quantum geometry is generalized to $N$-fold degenerate
systems.

In Sec.\ref{SecNH}, quantum geometry is generalized to non-Hermitian
systems. Especially, the quantum metric and the Berry curvature are obtained
for the two-band system. They are studied in the Dirac system with a complex
mass term.

In Sec.\ref{SecQIG}, quantum geometry is generalized to density matrices,
which is known as the quantum information geometry. Especially, quantum
distance is related to quantum Fisher information. The Clam\'{e}r-Rao
inequality is explained, which gives the lower bound of the covariance of
the physical observable by the inverse of the quantum Fisher information. It
is shown that the quantum Fisher information is reduced to the classical
Fisher information for a pure state. Quantum geometry at thermal equilibrium
is derived, and it is shown that the Uhlmann curvature is reduced to the
Berry curvature at low temperature.

In Sec.\ref{SecX}, the notion of the $X$-wave magnets is introduced, which
includes the $d$-wave, $g$-wave and $i$-wave altermagnets as well as the $p$%
-wave and $f$-wave magnets. Their symmetry properties are summarized.\ The
quantum geometric properties are also calculated.

In Sec.\ref{SecXTra}, transport properties of the $X$-wave magnets are
summarized. The spin current generation by electric field and temperature
gradient is analytically studied. It is shown that only the $d$-wave
altermagnet has a linear response for the spin current. The analytic formula
for tunneling magnetoresistance is also studied. Finally, we study a coupled
system to the $X$-wave magnet and the Rashba system. It is shown that it is
impossible to detect the Neel vector of the $X$-wave magnet by anomalous
Hall conductivity based on the two-band model. Universal formula for the
planar Hall conductivity for the $X$-wave magnets is presented.
Electromagnetic responses of the $X$-wave magnets are also studied.

In Sec.\ref{SecLL}, we analytically derive Landau levels for the $p$-wave
magnet by using an analogy of the coherent state and for the $d$-wave
altermagnet by using an analogy of the squeezed state.

In Sec.\ref{SecFriedel}, we analytically derive the Friedel oscillation of
the spatial profile of the local density states.

Sec.\ref{SecSummary} is devoted to summaries, discusions and outlooks.

This paper contains both reviews and original results. Most of the parts are
review. The original ones are as follows. Two-band formulas for the Zeeman
quantum geometry in Sec.\ref{SecZeeman}.B. Non-Abelian Zeeman quantum
geometry in Sec.\ref{SecZeeman}.D. Zeeman quantum geometry induced cross
responses in Sec.\ref{SecZeeman}.E and F. Landau levels for the $p$-wave
magnet in Sec.\ref{SecLL}.A. The parts where the results are not original
but give some new interpretations are as follows: Spin current generation
and spin Nernst effect are derived analytically in Sec.\ref{SecXTra}.A. The
Landau levels for the $d$-wave altermagnet are derived by using the analogy
of the coherent state in Sec.\ref{SecLL}.A. The Friedel oscillation is
discussed for the $X$-wave magnet in Sec.\ref{SecFriedel}.

\section{Quantum geometry}

\label{SecQG}

Quantum geometry is a differential geometry based on the wave functions. The
basic notion is the fidelity, which measures the similarity of two wave
functions. It acts as the distance in the context of the differential
geometry. Starting from the fidelity with an infinitesimal translation in
the momentum space, we derive the quantum geometric tensor. Its real part
gives the quantum metric, while its imaginary part gives the Berry
curvature. The latter is related to the anomalous Hall effect and tological
insulators via the Chern number.

\subsection{Quantum distance and quantum geometric tensor}

The fidelity of the wave function $\left\vert \psi _{n}\left( \mathbf{k}%
\right) \right\rangle $ is defined by%
\begin{equation}
F_{n}\left( \mathbf{k},\mathbf{k}^{\prime }\right) \equiv \left\vert
\left\langle \psi _{n}\left( \mathbf{k}\right) \left\vert \psi _{n}\left( 
\mathbf{k}^{\prime }\right) \right\rangle \right. \right\vert ,
\label{Fidelity}
\end{equation}%
where the wave function is orthonormalized, $\left\langle \psi _{n}\left( 
\mathbf{k}\right) \left\vert \psi _{n}\left( \mathbf{k}^{\prime }\right)
\right\rangle \right. =\delta (\mathbf{k},\mathbf{k}^{\prime })$. It
satisfies $0\leq F_{n}\left( \mathbf{k},\mathbf{k}^{\prime }\right) \leq 1$.

The Hilbert-Schmidt distance is defined by%
\begin{equation}
s_{\text{HS}}\left( \mathbf{k},\mathbf{k}^{\prime }\right) \equiv \sqrt{%
1-F_{n}\left( \mathbf{k},\mathbf{k}^{\prime }\right) ^{2}},
\end{equation}%
where $s_{\text{HS}}\left( \mathbf{k},\mathbf{k}^{\prime }\right) =0$ when
the two wave functions are identical $\left\vert \psi _{n}\left( \mathbf{k}%
\right) \right\rangle =\left\vert \psi _{n}\left( \mathbf{k}^{\prime
}\right) \right\rangle $, while $s_{\text{HS}}\left( \mathbf{k},\mathbf{k}%
^{\prime }\right) =1$ when they are orthogonal $\left\langle \psi _{n}\left( 
\mathbf{k}\right) \left\vert \psi _{n}\left( \mathbf{k}^{\prime }\right)
\right\rangle \right. =0$.

The quantum geometric tensor $\mathcal{F}_{n}^{\mu \nu }$ is defined as
follows. We start with the quantum distance $ds_{\text{HS}}$ for the
infinitesimal translation $d\mathbf{k}$ of the momentum as\cite%
{Provost,Berry,Resta}%
\begin{equation}
ds_{\text{HS}}\left( \mathbf{k}\right) \equiv \sqrt{1-F_{n}\left( \mathbf{k},%
\mathbf{k}+d\mathbf{k}\right) ^{2}},
\end{equation}%
with%
\begin{equation}
F_{n}\left( \mathbf{k},\mathbf{k}+d\mathbf{k}\right) =|\langle \psi
_{n}\left( \mathbf{k}\right) |U_{d\mathbf{k}}|\psi _{n}(\mathbf{k})\rangle |,
\end{equation}%
where%
\begin{equation}
U_{d\mathbf{k}}\equiv e^{-id\mathbf{k}\cdot \mathbf{r}}  \label{Udk}
\end{equation}%
is the generator of the infinitesimal momentum translation $d\mathbf{k}$ and 
$r_{\mu }\mathbf{\equiv }i\partial /\partial _{k_{\mu }}$ is the position
operator. The quantum distance for the infinitesimal momentum is expanded in
terms of the quantum geometric tensor $\mathcal{F}_{n}^{\mu \nu }$ as%
\begin{equation}
\left( ds_{\text{HS}}\right) ^{2}=\sum_{\mu \nu }\mathcal{F}_{n}^{\mu \nu
}dk_{\mu }dk_{\nu },  \label{dsF}
\end{equation}%
where $\mu ,\nu $ stand for $x,y$, and 
\begin{equation}
\mathcal{F}_{n}^{\mu \nu }\left( \mathbf{k}\right) =\left\langle \partial
_{k_{\mu }}\psi _{n}\left( \mathbf{k}\right) \right\vert \left( 1-\left\vert
\psi _{n}\left( \mathbf{k}\right) \right\rangle \left\langle \psi _{n}\left( 
\mathbf{k}\right) \right\vert \right) \left\vert \partial _{k_{\nu }}\psi
_{n}\left( \mathbf{k}\right) \right\rangle ,  \label{F12}
\end{equation}%
with $\partial _{k_{\mu }}\equiv \frac{\partial }{\partial k_{\mu }}$. It is
called the Fubini-Study metric\cite{Fubini,Study}. It is rewritten as%
\begin{equation}
\mathcal{F}_{n}^{\mu \nu }\left( \mathbf{k}\right) =\left\langle \partial
_{k_{\mu }}\psi _{n}\left( \mathbf{k}\right) \right\vert \left(
1-P_{n}\left( \mathbf{k}\right) \right) \left\vert \partial _{k_{\nu }}\psi
_{n}\left( \mathbf{k}\right) \right\rangle ,
\end{equation}%
where we have defined the projection operator,%
\begin{equation}
P_{n}\left( \mathbf{k}\right) \equiv \left\vert \psi _{n}\left( \mathbf{k}%
\right) \right\rangle \left\langle \psi _{n}\left( \mathbf{k}\right)
\right\vert ,
\end{equation}%
satisfying the idenpotency condition $P\left( \mathbf{k}\right) ^{2}=P\left( 
\mathbf{k}\right) $. The presence of the projection operator is understood
as follows. $\left\vert \partial _{k_{\nu }}\psi _{n}\left( \mathbf{k}%
\right) \right\rangle $ is generally mapped out of the band $n$, where the
projection operator restricts $\left\vert \partial _{k_{\nu }}\psi
_{n}\left( \mathbf{k}\right) \right\rangle $ to the band $n$. $\mathcal{F}%
_{n}^{\mu \nu }$ is Hermitian, $\left( \mathcal{F}_{n}^{\mu \nu }\right)
^{\ast }=\mathcal{F}_{n}^{\nu \mu }$. The derivation of Eq.(\ref{F12}) is
shown in Appendix \ref{ApQ}.

The quantum metric is the real part of the quantum geometric tensor,%
\begin{equation}
g_{n}^{\mu \nu }\equiv \text{Re}\mathcal{F}_{n}^{\mu \nu }=\frac{\mathcal{F}%
_{n}^{\mu \nu }\left( \mathbf{k}\right) +\left( \mathcal{F}_{n}^{\mu \nu
}\left( \mathbf{k}\right) \right) ^{\ast }}{2},  \label{g12}
\end{equation}%
while the Berry curvature is the imaginary part of the quantum geometric
tensor,%
\begin{equation}
\Omega _{n}^{\mu \nu }\equiv 2\text{Im}\mathcal{F}_{n}^{\mu \nu }=i\left( 
\mathcal{F}_{n}^{\mu \nu }\left( \mathbf{k}\right) -\left( \mathcal{F}%
_{n}^{\mu \nu }\left( \mathbf{k}\right) \right) ^{\ast }\right) .
\label{BerryF}
\end{equation}%
There are relations%
\begin{equation}
g_{n}^{\mu \nu }=g_{n}^{\nu \mu },\qquad \Omega _{n}^{\mu \nu }=-\Omega
_{n}^{\nu \mu }.  \label{gOmega}
\end{equation}%
With the use of them, the quantum geometric tensor is decomposed into the
real and imaginary parts,%
\begin{equation}
\mathcal{F}_{n}^{\mu \nu }=g_{n}^{\mu \nu }-\frac{i}{2}\Omega _{n}^{\mu \nu
}.
\end{equation}

The Berry curvature does not contribute to the quantum distance%
\begin{equation}
\left( ds_{\text{HS}}\right) ^{2}=\sum_{\mu \nu }g_{n}^{\mu \nu }dk_{\mu
}dk_{\nu },
\end{equation}%
because of the symmetry $dk_{\mu }dk_{\nu }=dk_{\nu }dk_{\mu }$.

By introducing the Wilczek-Zee connection\cite{WZ},%
\begin{equation}
a_{nm}^{\mu }\left( \mathbf{k}\right) \equiv i\left\langle \psi _{n}\left( 
\mathbf{k}\right) \right\vert \partial _{k_{\mu }}\left\vert \psi _{m}\left( 
\mathbf{k}\right) \right\rangle ,  \label{WZa}
\end{equation}%
the quantum geometric tensor is rewritten as%
\begin{align}
& \mathcal{F}_{n}^{\mu \nu }\left( \mathbf{k}\right)  \notag \\
=& \sum_{m\neq n}\left\langle \partial _{k_{\mu }}\psi _{n}\left( \mathbf{k}%
\right) \left\vert \psi _{m}\left( \mathbf{k}\right) \right\rangle
\left\langle \psi _{m}\left( \mathbf{k}\right) \left\vert \partial _{k_{\nu
}}\psi _{n}\left( \mathbf{k}\right) \right\rangle \right. \right.  \notag \\
=& \sum_{m\neq n}\left( i\left\langle \psi _{m}\left( \mathbf{k}\right)
\left\vert \partial _{k_{\mu }}\psi _{n}\left( \mathbf{k}\right)
\right\rangle \right. \right) ^{\ast }i\left\langle \psi _{m}\left( \mathbf{k%
}\right) \left\vert \partial _{k_{\nu }}\psi _{n}\left( \mathbf{k}\right)
\right\rangle \right.  \notag \\
=& \sum_{m\neq n}a_{mn}^{\mu \ast }\left( \mathbf{k}\right) a_{mn}^{\nu
}\left( \mathbf{k}\right) .
\end{align}%
The Wilczek-Zee connection is Hermitian, 
\begin{equation}
a_{nm}^{\mu \ast }\left( \mathbf{k}\right) =a_{mn}^{\mu }\left( \mathbf{k}%
\right) ,  \label{AHermitian}
\end{equation}%
because%
\begin{align}
& i\frac{\partial }{\partial k_{\mu }}\left\langle \psi _{m}\left( \mathbf{k}%
\right) \left\vert \psi _{n}\left( \mathbf{k}\right) \right\rangle \right. 
\notag \\
=& i\left\langle \partial _{k_{\mu }}\psi _{m}\left( \mathbf{k}\right)
\left\vert \psi _{n}\left( \mathbf{k}\right) \right\rangle \right.
+i\left\langle \psi _{m}\left( \mathbf{k}\right) \left\vert \partial
_{k_{\mu }}\psi _{n}\left( \mathbf{k}\right) \right\rangle \right.  \notag \\
=& \left( -i\left\langle \psi _{n}\left( \mathbf{k}\right) \left\vert
\partial _{k_{\mu }}\psi _{m}\left( \mathbf{k}\right) \right\rangle \right.
\right) ^{\ast }+i\left\langle \psi _{m}\left( \mathbf{k}\right) \left\vert
\partial _{k_{\mu }}\psi _{n}\left( \mathbf{k}\right) \right\rangle \right. 
\notag \\
=& -a_{nm}^{\mu \ast }\left( \mathbf{k}\right) +a_{mn}^{\mu }\left( \mathbf{k%
}\right) =0.
\end{align}%
Then, the quantum geometric tensor is rewritten as%
\begin{equation}
\mathcal{F}_{n}^{\mu \nu }\left( \mathbf{k}\right) =\sum_{m\neq
n}a_{nm}^{\mu }\left( \mathbf{k}\right) a_{mn}^{\nu }\left( \mathbf{k}\right)
\label{Faa}
\end{equation}%
in terms of the Wilczek-Zee connection.

\subsection{Berry connection, Berry curvature and Chern number}

It follows that $\Omega _{n}^{xx}\left( \mathbf{k}\right) =\Omega
_{n}^{yy}\left( \mathbf{k}\right) =0$ from the antisymmetric property Eq.(%
\ref{gOmega}). Then, the nontrivial contributions are $\Omega
_{n}^{xy}\left( \mathbf{k}\right) $ and $\Omega _{n}^{yx}\left( \mathbf{k}%
\right) $. The Berry curvature (\ref{BerryF}) reads%
\begin{align}
& \Omega _{n}^{xy}\left( \mathbf{k}\right)  \notag \\
=& i\left\langle \partial _{k_{x}}\psi _{n}\left( \mathbf{k}\right)
\right\vert 1-P\left( \mathbf{k}\right) \left\vert \partial _{k_{y}}\psi
_{n}\left( \mathbf{k}\right) \right\rangle  \notag \\
& -i\left\langle \partial _{k_{y}}\psi _{n}\left( \mathbf{k}\right)
\right\vert 1-P\left( \mathbf{k}\right) \left\vert \partial _{k_{x}}\psi
_{n}\left( \mathbf{k}\right) \right\rangle  \notag \\
=& i\left\langle \partial _{k_{x}}\psi _{n}\left( \mathbf{k}\right)
\left\vert \partial _{k_{y}}\psi _{n}\left( \mathbf{k}\right) \right\rangle
\right. -i\left\langle \partial _{k_{y}}\psi _{n}\left( \mathbf{k}\right)
\left\vert \partial _{k_{x}}\psi _{n}\left( \mathbf{k}\right) \right\rangle
\right.  \notag \\
& -i\left\langle \partial _{k_{x}}\psi _{n}\left( \mathbf{k}\right)
\left\vert \psi _{n}\left( \mathbf{k}\right) \right\rangle \right.
\left\langle \psi _{n}\left( \mathbf{k}\right) \left\vert \partial
_{k_{y}}\psi _{n}\left( \mathbf{k}\right) \right\rangle \right.  \notag \\
& +i\left\langle \partial _{k_{y}}\psi _{n}\left( \mathbf{k}\right)
\left\vert \psi _{n}\left( \mathbf{k}\right) \right\rangle \right.
\left\langle \psi _{n}\left( \mathbf{k}\right) \left\vert \partial
_{k_{x}}\psi _{n}\left( \mathbf{k}\right) \right\rangle \right.  \notag \\
=& i\left\langle \partial _{k_{x}}\psi _{n}\left( \mathbf{k}\right)
\left\vert \partial _{k_{y}}\psi _{n}\left( \mathbf{k}\right) \right\rangle
\right. -i\left\langle \partial _{k_{y}}\psi _{n}\left( \mathbf{k}\right)
\left\vert \partial _{k_{x}}\psi _{n}\left( \mathbf{k}\right) \right\rangle
\right.  \notag \\
& +ia_{n}^{x\ast }\left( \mathbf{k}\right) a_{n}^{y}\left( \mathbf{k}\right)
-ia_{n}^{y\ast }\left( \mathbf{k}\right) a_{n}^{x}\left( \mathbf{k}\right) ,
\label{Berry1}
\end{align}%
where we have defined the Berry connection%
\begin{equation}
a_{n}^{\mu }\left( \mathbf{k}\right) \equiv a_{nn}^{\mu }\left( \mathbf{k}%
\right) ,
\end{equation}%
which is real, $a_{n}^{\mu \ast }\left( \mathbf{k}\right) =a_{n}^{\mu
}\left( \mathbf{k}\right) $.

Then, the Berry curvature (\ref{Berry1}) is rewritten as 
\begin{align}
& \Omega _{n}^{xy}\left( \mathbf{k}\right)  \notag \\
=& i\left\langle \partial _{k_{x}}\psi _{n}\left( \mathbf{k}\right)
\left\vert \partial _{k_{y}}\psi _{n}\left( \mathbf{k}\right) \right\rangle
\right. -i\left\langle \partial _{k_{y}}\psi _{n}\left( \mathbf{k}\right)
\left\vert \partial _{k_{x}}\psi _{n}\left( \mathbf{k}\right) \right\rangle
\right.  \notag \\
=& -2\text{Im}\left\langle \frac{\partial \psi _{n}\left( \mathbf{k}\right) 
}{\partial k_{x}}\left\vert \frac{\partial \psi _{n}\left( \mathbf{k}\right) 
}{\partial k_{y}}\right\rangle \right.  \notag \\
=& \frac{\partial a_{n}^{y}\left( \mathbf{k}\right) }{\partial k_{x}}-\frac{%
\partial a_{n}^{x}\left( \mathbf{k}\right) }{\partial k_{y}}=\left[ \nabla
\times \mathbf{a}_{n}\left( \mathbf{k}\right) \right] _{z}.
\end{align}%
Its integral over the whole Brillouin zone is quantized%
\begin{align}
\frac{1}{2\pi }\int d\mathbf{k}\Omega _{n}^{xy}\left( \mathbf{k}\right) =& 
\frac{1}{2\pi }\oint dk_{\mu }\nabla \times \Omega _{n}^{xy}\left( \mathbf{k}%
\right)  \notag \\
=& \frac{1}{2\pi }\oint dk_{\mu }a_{n}^{\mu }\left( \mathbf{k}\right) =%
\mathcal{C}_{n},  \label{ChernTKNN2}
\end{align}%
where $\mathcal{C}_{n}$ is the Chern number taking an integer.

We go on to prove that $\mathcal{C}_{n}$ is an integer\cite{Kohmoto}. The
formula (\ref{ChernTKNN2}) is interpreted as the integral of the Berry
curvature $\Omega _{n}^{xy}\left( \mathbf{k}\right) $ over the first
Brillouin zone, which is the Chern number. Using the Stokes theorem, this
can be rewritten as a contour integration along the boundary of the
Brillouin zone, 
\begin{equation}
\mathcal{C}_{n}=\frac{1}{2\pi }\int d\mathbf{k}\,\Omega _{n}^{xy}\left( 
\mathbf{k}\right) =\frac{1}{2\pi }\oint dk_{\mu }a_{n}^{\mu }\left( \mathbf{k%
}\right) .
\end{equation}%
Since it is a periodic system, it follows that $\mathcal{C}_{n}=0$ if $%
a_{n}^{\mu }\left( \mathbf{k}\right) $ is a regular function. However, the
gauge potential $a_{n}^{\mu }\left( \mathbf{k}\right) $ can be singular
though the magnetic field $\Omega _{n}^{xy}\left( \mathbf{k}\right) $ is
regular. In this case it is necessary to choose the contour integration to
avoid these singular points.\ Then, $\mathcal{C}_{n}$ counts the number of
singularities, following the argument familiar in the theory of Dirac
monopoles.

\subsection{Wannier function and polarization}

The Wannier function is defined by%
\begin{equation}
\left\vert w_{n}\left( \mathbf{r}\right) \right\rangle \equiv \frac{1}{2\pi }%
\int_{\text{BZ}}d\mathbf{k}e^{-i\mathbf{k}\cdot \mathbf{r}}\left\vert \psi
_{n}\left( \mathbf{k}\right) \right\rangle ,
\end{equation}%
or%
\begin{equation}
\left\vert \psi _{n}\left( \mathbf{k}\right) \right\rangle =\sum_{\mathbf{r}%
}e^{i\mathbf{k}\cdot \mathbf{r}}\left\vert w_{n}\left( \mathbf{r}\right)
\right\rangle .
\end{equation}%
Then, the Berry connection is rewritten as%
\begin{align}
a_{n}^{\mu }\left( \mathbf{k}\right) & \equiv i\left\langle \psi _{n}\left( 
\mathbf{k}\right) \right\vert \frac{d}{dk_{\mu }}\left\vert \psi _{n}\left( 
\mathbf{k}\right) \right\rangle  \notag \\
=& \sum_{\mathbf{r,r}^{\prime }}\left\langle w_{n}\left( \mathbf{r}^{\prime
}\right) \right\vert e^{-i\mathbf{k}\cdot \mathbf{r}^{\prime }}\frac{d}{%
dk_{\mu }}e^{i\mathbf{k}\cdot \mathbf{r}}\left\vert w_{n}\left( \mathbf{r}%
\right) \right\rangle  \notag \\
=& \sum_{\mathbf{r,r}^{\prime }}\left\langle w_{n}\left( \mathbf{r}^{\prime
}\right) \right\vert e^{i\mathbf{k}\cdot \left( \mathbf{r-r}^{\prime
}\right) }r_{\mu }\left\vert w_{n}\left( \mathbf{r}\right) \right\rangle .
\end{align}%
We integrate it over the Brillouin zone and obtain%
\begin{align}
& \frac{1}{2\pi }\int_{\text{BZ}}a_{n}^{\mu }\left( \mathbf{k}\right) d%
\mathbf{k}  \notag \\
=& \sum_{\mathbf{r,r}^{\prime }}\left\langle w_{n}\left( \mathbf{r}^{\prime
}\right) \right\vert \int_{\text{BZ}}e^{i\mathbf{k}\cdot \left( \mathbf{r-r}%
^{\prime }\right) }d\mathbf{k}r_{\mu }\left\vert w_{n}\left( \mathbf{r}%
\right) \right\rangle  \notag \\
=& \sum_{\mathbf{r,r}^{\prime }}\left\langle w_{n}\left( \mathbf{r}^{\prime
}\right) \right\vert \delta \left( \mathbf{r-r}^{\prime }\right) r_{\mu
}\left\vert w_{n}\left( \mathbf{r}\right) \right\rangle  \notag \\
=& \sum_{\mathbf{r}}\left\langle w_{n}\left( \mathbf{r}\right) \right\vert
r_{\mu }\left\vert w_{n}\left( \mathbf{r}\right) \right\rangle =\left\langle
r_{\mu }\right\rangle .
\end{align}%
It is the expectation value of the position $r_{\mu }$, which represents the
polarization.

The quantum metric is related to the fluctuation of the polarization\cite%
{Marzari},%
\begin{align}
&\frac{1}{2\pi }\int_{\text{BZ}}g_{n}^{\mu \mu }\left( \mathbf{k}\right) d%
\mathbf{k}  \notag \\
=&\sum_{\mathbf{r}}\left\langle w_{n}\left( \mathbf{r}\right) \right\vert
r_{\mu }^{2}\left\vert w_{n}\left( \mathbf{r}\right) \right\rangle
-\left\langle w_{n}\left( \mathbf{r}\right) \right\vert r_{\mu }\left\vert
w_{n}\left( \mathbf{r}\right) \right\rangle ^{2}  \notag \\
=&\left\langle r_{\mu }^{2}\right\rangle -\left\langle r_{\mu }\right\rangle
^{2}.
\end{align}

\subsection{Inequality}

There is an inequality for the quantum metric and the Berry curvature in two
dimensions\cite{Roy,Peotta},%
\begin{equation}
\frac{1}{2}\text{Tr}g_{n}^{\mu \nu }\left( \mathbf{k}\right) \geq \sqrt{\det
g_{n}^{\mu \nu }\left( \mathbf{k}\right) }\geq \frac{\left\vert \Omega
_{n}^{xy}\left( \mathbf{k}\right) \right\vert }{2},  \label{Ineq}
\end{equation}%
where the trace and the determinant are taken over $\mu $ and $\nu $. The
first inequality is proved by using the arithmetic geometric mean,%
\begin{equation}
\frac{g_{+}+g_{-}}{2}\geq \sqrt{g_{+}g_{-}},
\end{equation}%
where $g_{\pm }$ is the diagonal component of the diagonalized $g_{n}^{\mu
\nu }\left( \mathbf{k}\right) $,%
\begin{equation}
g_{n}^{\mu \nu }\left( \mathbf{k}\right) =\left( 
\begin{array}{cc}
g_{+} & 0 \\ 
0 & g_{-}%
\end{array}%
\right)
\end{equation}%
with%
\begin{equation}
g_{\pm }\equiv \frac{g_{n}^{xx}+g_{n}^{yy}\pm \sqrt{\left(
g_{n}^{xx}-g_{n}^{yy}\right) ^{2}+4\left( g_{n}^{xy}\right) ^{2}}}{2}.
\end{equation}%
The second inequality is proved by using the non-negativity of $\mathcal{F}%
_{n}^{\mu \nu }\left( \mathbf{k}\right) $,%
\begin{equation}
\det \mathcal{F}_{n}^{\mu \nu }\left( \mathbf{k}\right) \geq 0,
\end{equation}%
which is derived from the non-negativity of the quadratic form in Eq.(\ref%
{dsF}). It is equivalent to%
\begin{equation}
\det g_{n}^{\mu \nu }\left( \mathbf{k}\right) \geq \left( \frac{\Omega
_{n}^{xy}}{2}\right) ^{2},
\end{equation}%
where we have used%
\begin{equation}
\mathcal{F}_{n}^{\mu \nu }=\left( 
\begin{array}{cc}
g_{n}^{xx} & g_{n}^{xy}-i\Omega _{n}^{xy}/2 \\ 
g_{n}^{xy}+i\Omega _{n}^{xy}/2 & g_{n}^{yy}%
\end{array}%
\right) ,
\end{equation}%
which is derived from Eqs.(\ref{g12}), (\ref{BerryF}) and (\ref{gOmega}).

By integrating the inequality (\ref{Ineq}) over the whole Brillouin zone, we
obtain\cite{Roy,Peotta,Ozawa21,Mera}%
\begin{align}
& \frac{1}{2\pi }\int d\mathbf{k}\text{Tr}g_{n}^{\mu \nu }\left( \mathbf{k}%
\right)  \notag \\
& \geq 2\frac{1}{2\pi }\int d\mathbf{k}\sqrt{\det g_{n}^{\mu \nu }\left( 
\mathbf{k}\right) }  \notag \\
& \geq \frac{1}{2\pi }\int d\mathbf{k}\left\vert \Omega _{n}^{xy}\left( 
\mathbf{k}\right) \right\vert \geq \frac{1}{2\pi }\int d\mathbf{k}\Omega
_{n}^{xy}\left( \mathbf{k}\right) =\mathcal{C}_{n}.
\end{align}%
This gives a lower bound for $\int d\mathbf{k}$Tr$g_{n}^{\mu \nu }\left( 
\mathbf{k}\right) $, when the system is topological, $\mathcal{C}_{n}\geq 1$%
. There are some applications to this inequality\cite{Belli,OnishiX,Comb}.

The quantum volume is defined by the Riemann volume as\cite{Ozawa21}

\begin{equation}
\text{vol}_{g}\equiv \frac{1}{2\pi }\int d\mathbf{k}\sqrt{\det g_{n}^{\mu
\nu }\left( \mathbf{k}\right) }.
\end{equation}%
The quantum weight is defined by\cite{OnishiR}%
\begin{equation}
K^{\mu \nu }\equiv \frac{1}{2\pi }\int d\mathbf{k}g_{n}^{\mu \nu }\left( 
\mathbf{k}\right) ,
\end{equation}%
which measures quantum fluctuation in the center of mass. It is
experimentally determined from X-ray scattering.

\subsection{Quantum Geometry for two-band systems}

We consider a two-band system with the index $n=\pm $. The Hamiltonian is
generally given by 
\begin{equation}
H\left( \mathbf{k}\right) =h_{0}\left( \mathbf{k}\right) +\mathbf{\sigma }%
\cdot \mathbf{h}\left( \mathbf{k}\right) ,  \label{Hamil2}
\end{equation}%
where $\mathbf{h}\left( \mathbf{k}\right) $ is parametrized as the
normalized vector $\mathbf{n}\left( \mathbf{k}\right) $,%
\begin{equation}
\mathbf{n}\left( \mathbf{k}\right) \equiv \mathbf{h}\left( \mathbf{k}\right)
/|\mathbf{h}\left( \mathbf{k}\right) |=\left( \sin \theta \cos \phi ,\sin
\theta \sin \phi ,\cos \theta \right) ,  \label{dVector}
\end{equation}%
and $\sigma _{j}$ is the Pauli matrix with $j=x,y,z$. In this case, there
are simple forms in terms of $\mathbf{n}\left( \mathbf{k}\right) $.

The energy is given by%
\begin{equation}
\varepsilon _{\pm }=h_{0}\left( \mathbf{k}\right) \pm |\mathbf{h}\left( 
\mathbf{k}\right) |\text{.}
\end{equation}

The eigenfunctions $\psi _{\pm }\left( \mathbf{k}\right) $ of the
Hamiltonian (\ref{Hamil2}) corresponding to the energy $\varepsilon _{\pm }$
are given by%
\begin{align}
\psi _{+}\left( \mathbf{k}\right) =& e^{i\alpha \left( \mathbf{k}\right)
}\left( 
\begin{array}{c}
e^{-i\phi \left( \mathbf{k}\right) }\cos \frac{\theta \left( \mathbf{k}%
\right) }{2} \\ 
\sin \frac{\theta \left( \mathbf{k}\right) }{2}%
\end{array}%
\right) , \\
\psi _{-}\left( \mathbf{k}\right) =& e^{i\alpha \left( \mathbf{k}\right)
}\left( 
\begin{array}{c}
-e^{-i\phi \left( \mathbf{k}\right) }\sin \frac{\theta \left( \mathbf{k}%
\right) }{2} \\ 
\cos \frac{\theta \left( \mathbf{k}\right) }{2}%
\end{array}%
\right) ,
\end{align}%
where $\alpha \left( \mathbf{k}\right) $ {is a real function representing a
gauge degrees of freedom. }Then, the Wilczek-Zee connection (\ref{WZa}) is
calculated as%
\begin{align}
a_{+-}^{\mu }\left( \mathbf{k}\right) =& -i\left\langle \psi _{+}\left( 
\mathbf{k}\right) \right\vert \partial _{k_{\mu }}\psi _{-}\left( \mathbf{k}%
\right) \rangle  \notag \\
=& i\frac{\partial _{k_{\mu }}\theta }{2}+\frac{\partial _{k_{\mu }}\phi }{2}%
\sin \theta ,
\end{align}%
and%
\begin{align}
a_{-+}^{\mu }\left( \mathbf{k}\right) =& -i\left\langle \psi _{-}\left( 
\mathbf{k}\right) \right\vert \partial _{k_{\mu }}\psi _{+}\left( \mathbf{k}%
\right) \rangle  \notag \\
=& -i\frac{\partial _{k_{\mu }}\theta }{2}+\frac{\partial _{k_{\mu }}\phi }{2%
}\sin \theta ,
\end{align}%
irrespective of $\alpha $.

The quantum metric is explicitly given by\cite%
{Matsuura,Bleu,Gers,Robin,OnishiX,WChen2024,EzawaQG,Oh}%
\begin{align}
g_{\pm }^{\mu \nu }\left( \mathbf{k}\right) =&\pm \frac{1}{2}\left( \partial
_{k_{\mu }}\mathbf{n}\right) \cdot \left( \partial _{k_{\nu }}\mathbf{n}%
\right)  \notag \\
=&\pm \frac{1}{4}\left( \partial _{k_{\mu }}\theta \partial _{k_{\nu
}}\theta +\sin ^{2}\theta \partial _{k_{\mu }}\phi \partial _{k_{\nu }}\phi
\right)  \label{Gmn}
\end{align}%
with the normalized vector (\ref{dVector}). Two or three components of $%
\mathbf{n}$\ should be nonzero for nonzero $g^{\mu \nu }$. It is understood
as follows: If only one component $n_{z}\left( \mathbf{k}\right) $ is
nonzero, it is given by $n_{z}\left( \mathbf{k}\right) =h_{z}\left( \mathbf{k%
}\right) /\left\vert h_{z}\left( \mathbf{k}\right) \right\vert $, whose
derivative $\partial _{k_{\mu }}\mathbf{n}$ is singular.

The Berry curvature is explicitly given by\cite{Hsiang,Stic,Jiang,Bleu,Robin}%
\begin{align}
\Omega _{\pm }^{xy}\left( \mathbf{k}\right) =& \mp \frac{1}{2}\mathbf{n}%
\cdot \left( \partial _{k_{x}}\mathbf{n}\times \partial _{k_{y}}\mathbf{n}%
\right)  \notag \\
=& \pm \frac{1}{2}\sin \theta \left( \partial _{k_{x}}\phi \partial
_{k_{y}}\theta -\partial _{k_{x}}\theta \partial _{k_{y}}\phi \right) .
\label{Omegaxy}
\end{align}%
It is a solid angle of the vector $\mathbf{n}$. Hence, three components of $%
\mathbf{n}$ should be nonzero for nonzero $\Omega _{\pm }^{xy}$.

\subsection{Analogy of the theory of general relativity}

\label{SecQGR}

The quantum metric and the Berry curvature are mainly studied in the current
stage. In addition to them, there are following quantities that can be
defined in the analogy of the theory of general relativity\cite%
{Hete,WChen2024}. They are the Christoffel symbol%
\begin{equation}
\Gamma _{\mu \nu }^{\lambda }\equiv \frac{1}{2}g_{\lambda \sigma }\left(
\partial _{\mu }g_{\nu \sigma }+\partial _{\nu }g_{\sigma \mu }-\partial
_{\sigma }g_{\mu \nu }\right) ,
\end{equation}%
the Riemann tensor%
\begin{equation}
R_{\sigma \mu \nu }^{\rho }\equiv \partial _{\mu }\Gamma _{\nu \sigma
}^{\rho }-\partial _{\nu }\Gamma _{\mu \sigma }^{\rho }+\Gamma _{\mu \lambda
}^{\rho }\Gamma _{\nu \sigma }^{\lambda }-\Gamma _{\nu \lambda }^{\rho
}\Gamma _{\mu \sigma }^{\lambda },
\end{equation}%
the Ricci tensoer%
\begin{equation}
R_{\mu \nu }\equiv R_{\mu \lambda \nu }^{\lambda },
\end{equation}%
the Ricci scalar 
\begin{equation}
R\equiv g^{\mu \nu }R_{\nu \mu },
\end{equation}%
and the Einstein tensor%
\begin{equation}
G_{\mu \nu }\equiv R_{\mu \nu }-\frac{1}{2}Rg_{\mu \nu }.
\end{equation}

Especially, the Christoffel symbol is given by%
\begin{equation}
\Gamma _{\mu \nu }^{\lambda }=\frac{1}{4}\partial _{\lambda }\mathbf{n}\cdot
\partial _{\mu }\partial _{\nu }\mathbf{n}
\end{equation}%
for the two-band system\cite{WChen2025G}. Furthermore, the Riemann tensor
reads%
\begin{align}
R_{\sigma \mu \nu }^{\rho }=& \frac{1}{4}[\partial _{\mu }\partial _{\rho }%
\mathbf{n}\cdot \partial _{\nu }\partial _{\sigma }\mathbf{n+}\partial
_{\rho }\mathbf{n}\cdot \partial _{\mu }\partial _{\nu }\partial _{\sigma }%
\mathbf{n}  \notag \\
& -\partial _{\nu }\partial _{\rho }\mathbf{n}\cdot \partial _{\mu }\partial
_{\sigma }\mathbf{n}-\partial _{\rho }\mathbf{n}\cdot \partial _{\mu
}\partial _{\nu }\partial _{\sigma }\mathbf{n}]  \notag \\
& +\frac{1}{16}\left( \partial _{\rho }\mathbf{n}\cdot \partial _{\mu
}\partial _{\lambda }\mathbf{n}\right) \left( \partial _{\lambda }\mathbf{n}%
\cdot \partial _{\nu }\partial _{\sigma }\mathbf{n}\right)  \notag \\
& -\frac{1}{16}\left( \partial _{\rho }\mathbf{n}\cdot \partial _{\nu
}\partial _{\lambda }\mathbf{n}\right) \left( \partial _{\lambda }\mathbf{n}%
\cdot \partial _{\mu }\partial _{\sigma }\mathbf{n}\right) ,
\end{align}%
and the Ricci tensoer reads 
\begin{align}
R_{\mu \nu }=& \frac{1}{4}[\partial _{\lambda }\partial _{\lambda }\mathbf{n}%
\cdot \partial _{\nu }\partial _{\mu }\mathbf{n}-\partial _{\nu }\partial
_{\lambda }\mathbf{n}\cdot \partial _{\lambda }\partial _{\mu }\mathbf{n}] 
\notag \\
& +\frac{1}{16}\left( \partial _{\lambda }\mathbf{n}\cdot \partial _{\lambda
}\partial _{\lambda }\mathbf{n}\right) \left( \partial _{\lambda }\mathbf{n}%
\cdot \partial _{\nu }\partial _{\mu }\mathbf{n}\right) .  \notag \\
& -\frac{1}{16}\left( \partial _{\lambda }\mathbf{n}\cdot \partial _{\nu
}\partial _{\lambda }\mathbf{n}\right) \left( \partial _{\lambda }\mathbf{n}%
\cdot \partial _{\lambda }\partial _{\mu }\mathbf{n}\right) .
\end{align}

\subsection{Non-Abelian quantum geometry}

\label{SecQGMulti}

So far, we have studied the case where the target band is a single band. If
the target bands are $N$-fold degenerate, we consider the $N$-fold
degenerate wave function,%
\begin{equation}
\left\vert \psi \left( \mathbf{k}\right) \right\rangle
=\sum_{n=1}^{N}c_{n}\left\vert \psi _{n}\left( \mathbf{k}\right)
\right\rangle ,  \label{psi}
\end{equation}%
satisfying the normalization condition, $\sum_{n=1}^{N}\left\vert
c_{n}\right\vert ^{2}=1$. It is necessary to generalize quantum geometry to
the non-Abelian quantum geometry\cite{Ma}. The fidelity of the wave
functions $\left\vert \psi \left( \mathbf{k}\right) \right\rangle $ and $%
\left\vert \psi \left( \mathbf{k}+d\mathbf{k}\right) \right\rangle $ is
defined by%
\begin{equation}
F\left( \mathbf{k},\mathbf{k}+d\mathbf{k}\right) \equiv \left\vert
\left\langle \psi \left( \mathbf{k}\right) \right\vert U_{d\mathbf{k}%
}\left\vert \psi \left( \mathbf{k}\right) \right\rangle \right\vert
\end{equation}%
with $U_{d\mathbf{k}}$ given by Eq.(\ref{Udk}). The quantum distance is 
\begin{equation}
ds_{\text{HS}}\equiv \sqrt{1-F\left( \mathbf{k},\mathbf{k}+d\mathbf{k}%
\right) ^{2}}
\end{equation}%
The non-Abelian quantum geometric tensor is given by%
\begin{equation}
\left( ds_{\text{HS}}\right) ^{2}\equiv \sum_{\mu \nu }\sum_{nm}\mathcal{F}%
_{nm}^{\mu \nu }dk_{\mu }dk_{\nu }
\end{equation}%
with%
\begin{equation}
\mathcal{F}_{nm}^{\mu \nu }=\left\langle \partial _{k_{\mu }}\psi _{n}\left( 
\mathbf{k}\right) \right\vert \left( 1-P\left( \mathbf{k}\right) \right)
\left\vert \partial _{k_{\mu }}\psi _{m}\left( \mathbf{k}\right)
\right\rangle  \label{F12Mu}
\end{equation}%
where $\mathcal{P}\left( \mathbf{k}\right) $ is the projection operator onto
the target bands,%
\begin{equation}
\mathcal{P}\left( \mathbf{k}\right) \equiv \sum_{n=1}^{N}\left\vert \psi
_{n}\left( \mathbf{k}\right) \right\rangle \left\langle \psi _{n}\left( 
\mathbf{k}\right) \right\vert ,
\end{equation}%
which satisfies the idempotence relation $\mathcal{P}\left( \mathbf{k}%
\right) ^{2}=\mathcal{P}\left( \mathbf{k}\right) $. The derivation is shown
in Appendix \ref{ApQ2}.

The quantum metric is the real part of the quantum geometric tensor,%
\begin{equation}
g_{nm}^{\mu \nu }\equiv \text{Re}\mathcal{F}_{nm}^{\mu \nu }=\frac{\mathcal{F%
}_{mn}^{\mu \nu }\left( \mathbf{k}\right) +\left( \mathcal{F}_{mn}^{\mu \nu
}\left( \mathbf{k}\right) \right) ^{\ast }}{2},
\end{equation}%
while the Berry curvature is the imaginary part of the quantum geometric
tensor,%
\begin{equation}
\Omega _{nm}^{\mu \nu }\equiv 2\text{Im}\mathcal{F}_{nm}^{\mu \nu }=i\left( 
\mathcal{F}_{mn}^{\mu \nu }\left( \mathbf{k}\right) -\left( \mathcal{F}%
_{mn}^{\mu \nu }\left( \mathbf{k}\right) \right) ^{\ast }\right) .
\end{equation}%
It is further calculated as%
\begin{align}
& \Omega _{nm}^{\mu \nu }  \notag \\
=& i\left\langle \partial _{k_{\mu }}\psi _{n}\left( \mathbf{k}\right)
\right\vert 1-P\left( \mathbf{k}\right) \left\vert \partial _{k_{\nu }}\psi
_{m}\left( \mathbf{k}\right) \right\rangle  \notag \\
& -i\left\langle \partial _{k_{\nu }}\psi _{m}\left( \mathbf{k}\right)
\right\vert 1-P\left( \mathbf{k}\right) \left\vert \partial _{k_{\mu }}\psi
_{n}\left( \mathbf{k}\right) \right\rangle  \notag \\
=& i\left\langle \partial _{k_{\mu }}\psi _{n}\left( \mathbf{k}\right)
\left\vert \partial _{k_{\nu }}\psi _{m}\left( \mathbf{k}\right)
\right\rangle \right. -i\left\langle \partial _{k_{\nu }}\psi _{m}\left( 
\mathbf{k}\right) \left\vert \partial _{k_{\mu }}\psi _{n}\left( \mathbf{k}%
\right) \right\rangle \right.  \notag \\
& -i\sum_{n^{\prime }\neq n,m}\left\langle \partial _{k_{\mu }}\psi
_{n}\left( \mathbf{k}\right) \left\vert \psi _{n^{\prime }}\left( \mathbf{k}%
\right) \right\rangle \right. \left\langle \psi _{n^{\prime }}\left( \mathbf{%
k}\right) \left\vert \partial _{k_{\nu }}\psi _{m}\left( \mathbf{k}\right)
\right\rangle \right.  \notag \\
& +i\sum_{n^{\prime }\neq n,m}\left\langle \partial _{k_{\nu }}\psi
_{m}\left( \mathbf{k}\right) \left\vert \psi _{n^{\prime }}\left( \mathbf{k}%
\right) \right\rangle \right. \left\langle \psi _{n^{\prime }}\left( \mathbf{%
k}\right) \left\vert \partial _{k_{\mu }}\psi _{n}\left( \mathbf{k}\right)
\right\rangle \right.  \notag \\
=& i\left\langle \partial _{k_{\mu }}\psi _{n}\left( \mathbf{k}\right)
\left\vert \partial _{k_{\nu }}\psi _{m}\left( \mathbf{k}\right)
\right\rangle \right. -i\left\langle \partial _{k_{\nu }}\psi _{m}\left( 
\mathbf{k}\right) \left\vert \partial _{k_{\mu }}\psi _{n}\left( \mathbf{k}%
\right) \right\rangle \right.  \notag \\
& +\sum_{n^{\prime }\neq n,m}ia_{n^{\prime }n}^{\mu \ast }a_{n^{\prime
}m}^{\nu }-ia_{n^{\prime }m}^{\nu \ast }a_{n^{\prime }n}^{\mu }.
\end{align}%
With the use of the Hermiticity condition of the Berry connection (\ref%
{AHermitian}), we have%
\begin{align}
\Omega _{nm}^{\mu \nu }=& i\left\langle \partial _{k_{\mu }}\psi _{n}\left( 
\mathbf{k}\right) \left\vert \partial _{k_{\nu }}\psi _{m}\left( \mathbf{k}%
\right) \right\rangle \right.  \notag \\
& -i\left\langle \partial _{k_{\nu }}\psi _{m}\left( \mathbf{k}\right)
\left\vert \partial _{k_{\mu }}\psi _{n}\left( \mathbf{k}\right)
\right\rangle \right.  \notag \\
& -\sum_{n^{\prime }\neq n,m}ia_{nn^{\prime }}^{\mu }a_{n^{\prime }m}^{\nu
}+\sum_{n^{\prime }\neq n,m}ia_{mn^{\prime }}^{\nu }a_{n^{\prime }n}^{\mu } 
\notag \\
=& \nabla \times a_{nm}^{\mu }-\sum_{n^{\prime }\neq n,m}i\left[
a_{nn^{\prime }}^{\mu },a_{n^{\prime }m}^{\nu }\right] .
\end{align}%
There is an additional term $-i\left[ a_{nn^{\prime }}^{\mu },a_{n^{\prime
}m}^{\nu }\right] $ in the Wilczek-Zee connection comparing with the Berry
connection.

\section{Quantum geometry in condensed matter physics}

\label{SecQGAp}

Quantum geometric properties are observable in condensed matter physics.
First, we review that the Berry curvature is observable by the Hall
conductance. The Hall conductance is quantized for an insulator, which is
well described by the Chern number. On the other hand, the quantum metric
appears in the optical absorption, the bulk photovoltaic effect and
nonlinear conductivities.

\subsection{Thouless-Kohmoto-Nightingale-Nijs formula}

The Berry curvature is related to the Hall conductivity. Especially, the
Hall conductivity is quantized in an insulator and is proportional to the
Chern number, which is known as the TKNN formula\cite{TKNN}.

According to the Kubo formula, the Hall conductance is given by%
\begin{align}
\sigma _{xy}=-i\hbar e^{2}& \int d\mathbf{k}\sum_{n\neq m}\frac{f\left(
\varepsilon _{n}\left( \mathbf{k}\right) \right) }{\left( \varepsilon
_{n}\left( \mathbf{k}\right) -\varepsilon _{m}\left( \mathbf{k}\right)
\right) ^{2}}  \notag \\
& \times \lbrack \left\langle \psi _{n}\left( \mathbf{k}\right) \right\vert
v_{x}\left\vert \psi _{m}\left( \mathbf{k}\right) \right\rangle \left\langle
\psi _{m}\left( \mathbf{k}\right) \right\vert v_{y}\left\vert \psi
_{n}\left( \mathbf{k}\right) \right\rangle  \notag \\
& -\left\langle \psi _{n}\left( \mathbf{k}\right) \right\vert
v_{y}\left\vert \psi _{m}\left( \mathbf{k}\right) \right\rangle \left\langle
\psi _{m}\left( \mathbf{k}\right) \right\vert v_{x}\left\vert \psi
_{n}\left( \mathbf{k}\right) \right\rangle ],  \label{KTKubo}
\end{align}%
where $f\left( \varepsilon _{n}\left( \mathbf{k}\right) \right) =1/\left(
\exp \left[ \left( \varepsilon _{n}(\mathbf{k})-\mu \right) /\left( k_{\text{%
B}}T\right) +1\right] \right) $ is the Fermi distribution function and $\mu $
is the chemical potential. By using the Hellmann-Feynman theorem%
\begin{align}
& \left\langle \psi _{m}\left( \mathbf{k}\right) \right\vert v_{\mu
}\left\vert \psi _{n}\left( \mathbf{k}\right) \right\rangle  \notag \\
=& \frac{1}{\hbar }\partial _{k_{\mu }}\varepsilon _{n}\left( \mathbf{k}%
\right) \langle \psi _{m}\left( \mathbf{k}\right) |\psi _{n}\left( \mathbf{k}%
\right) \rangle  \notag \\
& +\frac{1}{\hbar }\left( \varepsilon _{n}\left( \mathbf{k}\right)
-\varepsilon _{m}\left( \mathbf{k}\right) \right) \left\langle \psi
_{m}\left( \mathbf{k}\right) \right\vert \partial _{k_{\mu }}\left\vert \psi
_{n}\left( \mathbf{k}\right) \right\rangle  \label{HF}
\end{align}
for the states subject to the orthonormalization condition, $\langle \psi
_{m}\left( \mathbf{k}\right) |\psi _{n}\left( \mathbf{k}\right) \rangle
=\delta _{mn}$, the Hall conductance is rewritten as%
\begin{align}
\sigma _{xy}=& -\frac{ie^{2}}{\hbar }\int d\mathbf{k}\sum_{n,m}f\left(
\varepsilon _{n}\left( \mathbf{k}\right) \right)  \notag \\
& \hspace{-15mm}\times \lbrack \langle \psi _{n}\left( \mathbf{k}\right)
|\partial _{k_{x}}\psi _{m}\left( \mathbf{k}\right) \rangle \left\langle
\psi _{m}\left( \mathbf{k}\right) \right\vert \partial _{k_{y}}\psi
_{n}\left( \mathbf{k}\right) \rangle  \notag \\
& -\left\langle \psi _{n}\left( \mathbf{k}\right) \right\vert \partial
_{k_{y}}\psi _{m}\left( \mathbf{k}\right) \rangle \left\langle \psi
_{m}\left( \mathbf{k}\right) \right\vert \partial _{k_{x}}\psi _{n}\left( 
\mathbf{k}\right) \rangle ].
\end{align}%
It should be noticed that the sum can be extended to include the states with 
$n=m$ in this formula, since 
\begin{align}
& \langle \psi _{n}\left( \mathbf{k}\right) |\partial _{k_{x}}\psi
_{n}\left( \mathbf{k}\right) \rangle \left\langle \psi _{n}\left( \mathbf{k}%
\right) \right\vert \partial _{k_{y}}\psi _{n}\left( \mathbf{k}\right)
\rangle  \notag \\
=& \left\langle \psi _{n}\left( \mathbf{k}\right) \right\vert \partial
_{k_{y}}\psi _{n}\left( \mathbf{k}\right) \rangle \left\langle \psi
_{n}\left( \mathbf{k}\right) \right\vert \partial _{k_{x}}\psi _{n}\left( 
\mathbf{k}\right) \rangle .
\end{align}%
The proof of the Hellmann-Feynman theorem is shown in Appendix \ref{ApHF}.

Making the use of the relation 
\begin{align}
& \langle \psi _{n}\left( \mathbf{k}\right) |\partial _{k_{x}}\psi
_{m}\left( \mathbf{k}\right) \rangle +\langle \partial _{k_{x}}\psi
_{n}\left( \mathbf{k}\right) |\psi _{m}\left( \mathbf{k}\right) \rangle 
\notag \\
=& \partial _{k_{x}}\langle \psi _{m}\left( \mathbf{k}\right) |\psi
_{n}\left( \mathbf{k}\right) \rangle =0,
\end{align}%
the Hall conductance is rewritten as%
\begin{align}
\sigma _{xy}=& -\frac{ie^{2}}{\hbar }\int d\mathbf{k}\sum_{n,m}f\left(
\varepsilon \left( \mathbf{k}\right) \right)  \notag \\
& \hspace{-15mm}\times \lbrack \langle \partial _{k_{x}}\psi _{n}\left( 
\mathbf{k}\right) |\psi _{m}\left( \mathbf{k}\right) \rangle \left\langle
\psi _{m}\left( \mathbf{k}\right) \right\vert \partial _{k_{y}}\psi
_{n}\left( \mathbf{k}\right) \rangle  \notag \\
& -\left\langle \partial _{k_{y}}\psi _{n}\left( \mathbf{k}\right)
\right\vert \psi _{m}\left( \mathbf{k}\right) \rangle \left\langle \psi
_{m}\left( \mathbf{k}\right) \right\vert \partial _{k_{x}}\psi _{n}\left( 
\mathbf{k}\right) \rangle ].
\end{align}%
Using the completeness condition $\sum_{m}|\psi _{m}\left( \mathbf{k}\right)
\rangle \left\langle \psi _{m}\left( \mathbf{k}\right) \right\vert =1$, we
obtain%
\begin{align}
\sigma _{xy}=& \frac{ie^{2}}{\hbar }\int d\mathbf{k}\sum_{n}f\left(
\varepsilon _{n}\left( \mathbf{k}\right) \right) [\langle \partial
_{k_{x}}\psi _{n}\left( \mathbf{k}\right) |\partial _{k_{y}}\psi _{n}\left( 
\mathbf{k}\right) \rangle  \notag \\
& -\left\langle \partial _{k_{y}}\psi _{n}\left( \mathbf{k}\right)
\right\vert \partial _{k_{x}}\psi _{n}\left( \mathbf{k}\right) \rangle ].
\end{align}%
It is rewritten as%
\begin{equation}
\sigma _{xy}=\frac{e^{2}}{2\pi h}\sum_{n}\int d\mathbf{k}\,f\left(
\varepsilon _{n}\left( \mathbf{k}\right) \right) \Omega _{n}^{xy}\left( 
\mathbf{k}\right) ,  \label{TKNNf}
\end{equation}%
where%
\begin{equation}
\Omega _{n}^{xy}\left( \mathbf{k}\right) =i\left[ \langle \partial
_{k_{x}}\psi _{n}\left( \mathbf{k}\right) |\partial _{k_{y}}\psi _{n}\left( 
\mathbf{k}\right) \rangle -\langle \partial _{k_{y}}\psi _{n}\left( \mathbf{k%
}\right) |\partial _{k_{x}}\psi _{n}\left( \mathbf{k}\right) \rangle \right]
.  \label{BerryC}
\end{equation}%
The conductance is the sum of the contributions from various bands indexed
by $n$. It is convenient to define a "gauge potential" in the momentum space
for each band index $n$ by%
\begin{equation}
a_{n}^{\mu }\left( \mathbf{k}\right) =-i\left\langle \psi _{n}\left( \mathbf{%
k}\right) \right\vert \partial _{k_{\mu }}\psi _{n}\left( \mathbf{k}\right)
\rangle ,  \label{BerryConne}
\end{equation}%
which is the Berry connection. We may rewrite (\ref{BerryC}) as 
\begin{equation}
\Omega _{n}^{xy}\left( \mathbf{k}\right) =\partial _{k_{x}}a_{n}^{y}\left( 
\mathbf{k}\right) -\partial _{k_{y}}a_{n}^{x}\left( \mathbf{k}\right) .
\label{BerryCurva}
\end{equation}%
This is the "magnetic field", which is the Berry curvature.

We consider the zero-temperature limit, where the Fermi distribution
function becomes a step function, $f\left( \varepsilon _{n}\left( \mathbf{k}%
\right) \right) =\theta \left( \mu -\varepsilon _{n}\right) $. The
conductance (\ref{TKNNf}) is simplified as 
\begin{equation}
\sigma _{xy}=\frac{e^{2}}{h}\sum_{n=1}^{N}\mathcal{C}_{n},  \label{TKNN}
\end{equation}%
becuase the integration is taken below the Fermi energy.

\subsection{Dirac system}

We apply the above formula to the Dirac Hamiltonian\cite{SiliceneReview},%
\begin{equation}
H^{\xi }=\hbar v_{\mathrm{F}}\left( \xi k_{x}\sigma _{x}+k_{y}\sigma
_{y}\right) +m\sigma _{z},  \label{DiracFree}
\end{equation}%
where $\xi =\pm 1$ is the valley degrees of freedom. It describes the
low-energy theory of graphene. It is expressed as $H^{\xi }=\mathbf{h}\cdot 
\mathbf{\sigma }$ with 
\begin{equation}
\mathbf{n}\left( \mathbf{k}\right) =\frac{1}{\sqrt{\left( \hbar v_{\text{F}%
}k\right) ^{2}+m^{2}}}\left( \xi \hbar v_{\mathrm{F}}k_{x},\hbar v_{\mathrm{F%
}}k_{y},m\right)
\end{equation}%
in the polar coordinate $\mathbf{k}=(k,\phi )$.

The quantum metrices are calculated as%
\begin{align}
g_{\pm }^{xx}\left( \mathbf{k}\right) =\mp & \frac{\left( \hbar v_{\text{F}%
}\right) ^{2}\left( \left( \hbar v_{\text{F}}k_{y}\right) ^{2}+m^{2}\right) 
}{\left( \left( \hbar v_{\text{F}}k\right) ^{2}+m^{2}\right) ^{2}}, \\
g_{\pm }^{yy}\left( \mathbf{k}\right) =\mp & \frac{\left( \hbar v_{\text{F}%
}\right) ^{2}\left( \left( \hbar v_{\text{F}}k_{x}\right) ^{2}+m^{2}\right) 
}{\left( \left( \hbar v_{\text{F}}k\right) ^{2}+m^{2}\right) ^{2}}, \\
g_{\pm }^{xy}\left( \mathbf{k}\right) =& \pm \frac{\left( \hbar v_{\text{F}%
}\right) ^{2}\left( \hbar v_{\text{F}}k_{x}\right) \left( \hbar v_{\text{F}%
}k_{y}\right) }{\left( \left( \hbar v_{\text{F}}k\right) ^{2}+m^{2}\right)
^{2}},
\end{align}%
while the Berry curvature is calculated as%
\begin{equation}
\Omega _{\pm }^{xy}\left( \mathbf{k}\right) =\pm \xi \frac{m\left( \hbar v_{%
\text{F}}\right) ^{2}}{\left( \left( \hbar v_{\text{F}}k\right)
^{2}+m^{2}\right) ^{3/2}}.
\end{equation}

When the chemical potential is within the gap $\mu <\left\vert m\right\vert $%
, we may calculate the Chern number explicitly as%
\begin{align}
\mathcal{C}^{\xi }=& \frac{1}{2\pi }\int_{0}^{\infty }2\pi kdk\Omega _{\pm
}^{xy}\left( \mathbf{k}\right)  \notag \\
=& -\frac{\xi }{2}\frac{m}{|m|^{3}}\int_{0}^{\infty }d\mathbf{k}\frac{(\hbar
v_{\text{F}})^{2}}{2\left( \left( \hbar v_{\text{F}}/m\right)
^{2}k^{2}+1\right) ^{3/2}}  \notag \\
=& -\frac{\xi }{4}\frac{m}{|m|}\int_{0}^{\infty }dx\frac{1}{\left(
x+1\right) ^{3/2}}  \notag \\
=& -\frac{\xi }{4}\frac{m}{|m|}\int_{1}^{\infty }dy\,y^{-\frac{3}{2}}  \notag
\\
=& \frac{\xi }{2}\frac{m}{|m|}\left. y^{-\frac{1}{2}}\right\vert
_{1}^{\infty }=-\frac{\xi }{2}\text{sgn}(m).  \label{ChernValley}
\end{align}

\begin{figure}[t]
\centerline{\includegraphics[width=0.48\textwidth]{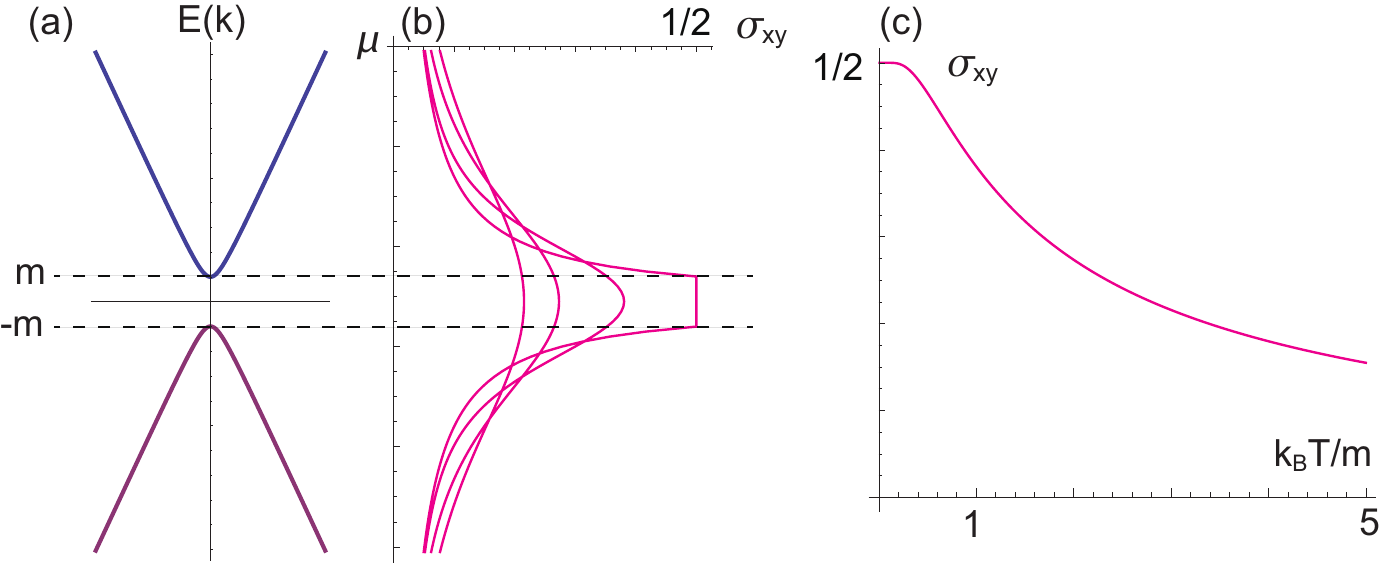}}
\caption{(a) The energy spectrum of the massive Dirac fermion with the mass $%
m$. (b) Hall conductance as a function of the chemical potential for the
massive Dirac Hamiltonian. (c) The temperature dependence of the Hall
conductance at the Fermi energy $\protect\mu =0$.}
\label{FigHall}
\end{figure}

When the chemical potential is among the gap, $\left\vert \mu \right\vert
<\left\vert m\right\vert $, the Chern number is already calculated as in (%
\ref{ChernValley}). When the chemical potential is below or above the gap, $%
\left\vert \mu \right\vert >\left\vert m\right\vert $, the integration is
taken in the region for $k>k_{\text{F}}$ with 
\begin{equation}
k_{\text{F}}=\frac{1}{\hbar v_{\text{F}}}\sqrt{\mu ^{2}-m^{2}},
\end{equation}%
which is the solution of $\varepsilon \left( \mathbf{k}\right) =\mu $. The
Hall conductivity is calculated as%
\begin{align}
\frac{\mathcal{\sigma }_{xy}^{\xi }}{e^{2}/h}=& \frac{1}{2\pi }\int_{k_{%
\text{F}}}^{\infty }2\pi kdk\Omega ^{\xi }\left( \mathbf{k}\right)  \notag \\
=& -\frac{\xi }{2}\frac{m}{|m|^{3}}\int_{k_{\text{F}}^{2}}^{\infty }d^{2}k%
\frac{(\hbar v_{\text{F}})^{2}}{2\left( \left( \hbar v_{\text{F}}/m\right)
^{2}k^{2}+1\right) ^{3/2}}  \notag \\
=& -\frac{\xi }{2}\frac{m}{|m|}\int_{\mu ^{2}/m^{2}-1}^{\infty }dx\frac{1}{%
2\left( x+1\right) ^{3/2}}  \notag \\
=& -\frac{\xi }{4}\frac{m}{|m|}\int_{\mu ^{2}/m^{2}}^{\infty }dy\,y^{-\frac{3%
}{2}}  \notag \\
=& \frac{\xi }{2}\frac{m}{|m|}\left. y^{-\frac{1}{2}}\right\vert _{\mu
^{2}/m^{2}}^{\infty }=-\frac{\xi }{2}\frac{|m|}{\left\vert \mu \right\vert }%
\text{sgn}(m).
\end{align}%
The TKNN formula dictates that the Hall conductance is proportional to the
Chern number. The Hall conductance is given by $\sigma _{xy}=\sum_{\xi
}\sigma _{xy}^{\xi }$ with%
\begin{equation}
\sigma _{xy}^{\xi }=\left\{ 
\begin{array}{ccc}
-\frac{\xi }{2}\frac{e^{2}}{h}\text{sgn}(m) & \text{for} & \left\vert \mu
\right\vert <\left\vert m\right\vert \\ 
&  &  \\ 
-\frac{\xi }{2}\frac{|m|}{\left\vert \mu \right\vert }\frac{e^{2}}{h}\text{%
sgn}(m) & \text{for} & \left\vert \mu \right\vert >\left\vert m\right\vert%
\end{array}%
\right.  \label{HallCon}
\end{equation}%
as a function of the chemical potential. It is quantized between the bulk
band gap, $|\mu |<|m|$, which is known as the quantum anomalous Hall effect.
On the other hand, the Hall conductance is anti-proportional to the chemical
potential outside the band gap, $|\mu |>|m|$. Eq.(\ref{HallCon}) is shown in
Fig.\ref{FigHall}.

We have%
\begin{equation}
\text{Tr}g_{n}^{\mu \nu }\left( \mathbf{k}\right) =\left( \hbar v_{\text{F}%
}\right) ^{2}\frac{2m^{2}+\left( \hbar v_{\text{F}}k\right) ^{2}}{2\left(
\left( \hbar v_{\text{F}}k\right) ^{2}+m^{2}\right) ^{2}},
\end{equation}%
and%
\begin{equation}
\sqrt{\det g_{n}^{\mu \nu }\left( \mathbf{k}\right) }=\frac{\left\vert
m\right\vert \left( \hbar v_{\text{F}}\right) ^{2}}{2\left( \left( \hbar v_{%
\text{F}}k\right) ^{2}+m^{2}\right) ^{3/2}}.
\end{equation}%
Especially, there is a relation%
\begin{equation}
2\sqrt{\det g_{n}^{\mu \nu }\left( \mathbf{k}\right) }=\left\vert \Omega
_{\pm }^{xy}\left( \mathbf{k}\right) \right\vert .
\end{equation}%
We can check the inequality (\ref{Ineq}) is satisfied. The quantum volume is
calculated as%
\begin{equation}
\int d\mathbf{k}\sqrt{\det g_{n}^{\mu \nu }\left( \mathbf{k}\right) }=\frac{1%
}{8}\left( \frac{\hbar v_{\text{F}}}{m}\right) ^{2}.
\end{equation}

A comment is in order. The Dirac system gives a half-integer of the Chern
number, which may contradict the fact that the Chern number is an integer.
This problem is solved by the Nielsen-Ninomiya theorem, which shows that the
number of Dirac cones must be even for the system defined for the Brillouin
zone. Then, the total Chern number is always an integer, which is relevant
in realistic materials.

\subsection{Optical absorption and elliptic dichroism}

We study optical inter-band transitions from the state $|\psi _{-}(\mathbf{k}%
)\rangle $ in the valence band to the state $|\psi _{+}(\mathbf{k})\rangle $
in the conduction band. We apply a beam of elliptical polarized light
perpendicular onto the sample, where the corresponding electromagnetic
potential is given by $\mathbf{A}(t)=(A_{x}\sin \omega t,A_{y}\cos \omega t)$%
. The electromagnetic potential is introduced into the Hamiltonian by way of
the minimal substitution, that is, by replacing the momentum $k_{j}$ with
the covariant momentum $P_{j}\equiv \hbar k_{j}+eA_{j}$.

Optical absorption is given by the optical conductivity%
\begin{align}
\sigma \left( \omega ;\vartheta \right) =& \sigma _{0}\int d\mathbf{k}\frac{%
f_{-}(\mathbf{k})-f_{+}(\mathbf{k})}{\varepsilon _{-}-\varepsilon _{+}}%
\left\vert P_{\vartheta }(\mathbf{k})\right\vert ^{2}  \notag \\
& \times \delta \left[ \varepsilon _{+}(\mathbf{k})-\varepsilon _{-}(\mathbf{%
k})-\hbar \omega \right] ,  \label{absorp}
\end{align}%
where we have set%
\begin{equation}
\sigma _{0}\equiv \frac{\pi e^{2}}{\left( 2\pi \right) ^{2}}.
\end{equation}

We start with the optical matrix element between the initial and final
states in the photo-emission process given by\cite%
{Yao08,Xiao,EzawaOpt,Li,TCIOpt,EllipticP} 
\begin{equation}
P_{i}(\mathbf{k})\equiv \hbar \left\langle \psi _{+}(\mathbf{k})\right\vert
v_{\mu }\left\vert \psi _{-}(\mathbf{k})\right\rangle ,  \label{EqP}
\end{equation}%
with the velocity%
\begin{equation}
v_{\mu }=\frac{1}{\hbar }\frac{\partial H\left( \mathbf{k}\right) }{\partial
k_{\mu }}.
\end{equation}%
The optical matrix element of the elliptic polarization is given by%
\begin{equation}
P_{\vartheta }(\mathbf{k})=P_{x}(\mathbf{k})\cos \vartheta +iP_{y}(\mathbf{k}%
)\sin \vartheta ,
\end{equation}%
where $\vartheta $ is the ellipticity of the injected beam, with $%
0<\vartheta <\pi $ for the right polarization and $-\pi <\vartheta <0$ for
the left polarization. $P_{\pi /4}(\mathbf{k})$ corresponds to the right
circularly polarized light, and $P_{-\pi /4}(\mathbf{k})$ corresponds to the
left circularly polarized light.

In the following, we assume that the temperature is absolutely zero. The
optical matrix element for the elliptic polarized light is expanded as%
\begin{align}
& \left\vert P_{\vartheta }(\mathbf{k})\right\vert ^{2}  \notag \\
=& \left\vert P_{x}(\mathbf{k})\right\vert ^{2}\cos ^{2}\vartheta
+\left\vert P_{y}(\mathbf{k})\right\vert ^{2}\sin ^{2}\vartheta  \notag \\
& +i\left[ P_{x}^{\ast }(\mathbf{k})P_{y}(\mathbf{k})-P_{y}^{\ast }(\mathbf{k%
})P_{x}(\mathbf{k})\right] \sin \vartheta \cos \vartheta .
\end{align}%
By using the Hellmann-Feynman theorem%
\begin{align}
& \left\langle \psi _{m}\left( \mathbf{k}\right) \right\vert v_{\mu
}\left\vert \psi _{n}\left( \mathbf{k}\right) \right\rangle  \notag \\
& =\frac{1}{\hbar }\left( \varepsilon _{n}\left( \mathbf{k}\right)
-\varepsilon _{m}\left( \mathbf{k}\right) \right) \left\langle \psi
_{m}\left( \mathbf{k}\right) \right\vert \partial _{k_{\mu }}\left\vert \psi
_{n}\left( \mathbf{k}\right) \right\rangle \quad
\end{align}%
for $m\neq n$, we have%
\begin{align}
& \frac{\left\vert P_{\mu }(\mathbf{k})\right\vert ^{2}}{\left( \varepsilon
_{n}\left( \mathbf{k}\right) -\varepsilon _{m}\left( \mathbf{k}\right)
\right) ^{2}}  \notag \\
=& \hbar ^{2}\left\langle \psi _{-}(\mathbf{k})\right\vert v_{\mu
}\left\vert \psi _{+}(\mathbf{k})\right\rangle \left\langle \psi _{+}(%
\mathbf{k})\right\vert v_{\mu }\left\vert \psi _{-}(\mathbf{k})\right\rangle
\notag \\
=& -\Delta ^{2}\left( \mathbf{k}\right) \left\langle \partial _{k_{\mu
}}\psi _{-}(\mathbf{k})\left\vert \psi _{+}(\mathbf{k})\right\rangle \right.
\left. \left\langle \psi _{+}(\mathbf{k})\right\vert \partial _{k_{\mu
}}\psi _{-}(\mathbf{k})\right\rangle  \notag \\
=& -\Delta ^{2}\left( \mathbf{k}\right) g_{-}^{\mu \mu }\left( \mathbf{k}%
\right) ,
\end{align}%
with $\mu =x,y$, and%
\begin{align}
& \frac{i\left[ P_{y}^{\ast }(\mathbf{k})P_{x}(\mathbf{k})-P_{x}^{\ast }(%
\mathbf{k})P_{y}(\mathbf{k})\right] }{\left( \varepsilon _{n}\left( \mathbf{k%
}\right) -\varepsilon _{m}\left( \mathbf{k}\right) \right) ^{2}}  \notag \\
& =i\Delta ^{2}\left( \mathbf{k}\right) [\left\langle \psi _{-}(\mathbf{k}%
)\right\vert v_{y}\left\vert \psi _{+}(\mathbf{k})\right\rangle \left\langle
\psi _{+}(\mathbf{k})\right\vert v_{x}\left\vert \psi _{-}(\mathbf{k}%
)\right\rangle  \notag \\
& \qquad \qquad -\left\langle \psi _{-}(\mathbf{k})\right\vert
v_{x}\left\vert \psi _{+}(\mathbf{k})\right\rangle \left\langle \psi _{+}(%
\mathbf{k})\right\vert v_{y}\left\vert \psi _{-}(\mathbf{k})\right\rangle ] 
\notag \\
& =i\Delta ^{2}\left( \mathbf{k}\right) \left[ \mathcal{F}_{-+}^{xy}\left( 
\mathbf{k}\right) -\mathcal{F}_{+-}^{yx}\left( \mathbf{k}\right) \right]
=\Delta ^{2}\left( \mathbf{k}\right) \Omega _{-}^{xy}\left( \mathbf{k}%
\right) .
\end{align}%
Then, the optical conductivity is rewritten by using the components of the
quantum geometric tensor\cite{EllipticP} as%
\begin{equation}
\sigma \left( \omega ;\vartheta \right) =\hbar \omega \sigma _{0}\int d%
\mathbf{k}f(\mathbf{k})G(\mathbf{k};\vartheta )\delta \left[ \varepsilon
_{+}(\mathbf{k})-\varepsilon _{-}(\mathbf{k})-\hbar \omega \right] ,
\label{OptG}
\end{equation}%
with%
\begin{equation}
G(\mathbf{k};\vartheta )\equiv g_{-}^{xx}(\mathbf{k})\cos ^{2}\vartheta
+g_{-}^{yy}(\mathbf{k})\sin ^{2}\vartheta +\Omega _{-}^{xy}(\mathbf{k})\sin
\vartheta \cos \vartheta .
\end{equation}

We study the optical absorption of the Dirac Hamiltonian (\ref{DiracFree})
at the band edge $\hbar \omega =\varepsilon _{+}(\mathbf{0})-\varepsilon
_{-}(\mathbf{0})$. It is simply given by%
\begin{align}
\frac{\sigma \left( \omega ;\vartheta \right) }{\hbar \omega \sigma _{0}}=&
G(\mathbf{0};\vartheta )  \notag \\
=& g_{-}^{xx}(\mathbf{0})\cos ^{2}\vartheta +g_{-}^{yy}(\mathbf{0})\sin
^{2}\vartheta +\Omega _{-}^{xy}(\mathbf{0})\sin \vartheta \cos \vartheta 
\notag \\
=& \frac{\left( \hbar v_{\text{F}}\right) ^{2}}{m^{2}}\left( 1-\frac{\xi }{2}%
\sin 2\vartheta \right) ,
\end{align}%
where we have used quantum metrices 
\begin{equation}
g_{-}^{xx}\left( \mathbf{0}\right) =g_{-}^{yy}\left( \mathbf{0}\right) =%
\frac{\left( \hbar v_{\text{F}}\right) ^{2}}{m^{2}},
\end{equation}%
and the Berry curvature%
\begin{equation}
\Omega _{-}^{xy}\left( \mathbf{0}\right) =-\xi \frac{\left( \hbar v_{\text{F}%
}\right) ^{2}}{m^{2}}.
\end{equation}%
It shows elliptic dichroism, where there is no absorption 
\begin{equation}
\sigma \left( \omega ;\vartheta \right) =0
\end{equation}%
for $\vartheta =\xi \pi /4$. Namely, the right or left circularly polarized
right is selectively absorbed depending on $\xi $.

\subsection{Sum rule}

The optical conductivity is given by%
\begin{align}
\sigma _{\text{D}}^{\mu \nu }\left( \omega ;\vartheta \right) =& \sigma
_{0}\int d\mathbf{k}\frac{f_{-}(\mathbf{k})-f_{+}(\mathbf{k})}{\varepsilon
_{-}-\varepsilon _{+}}  \notag \\
& \times \left\langle \psi _{-}(\mathbf{k})\right\vert v_{\mu }\left\vert
\psi _{+}(\mathbf{k})\right\rangle \left\langle \psi _{+}(\mathbf{k}%
)\right\vert v_{\nu }\left\vert \psi _{-}(\mathbf{k})\right\rangle  \notag \\
& \times \delta \left[ \varepsilon _{+}(\mathbf{k})-\varepsilon _{-}(\mathbf{%
k})-\hbar \omega \right] .
\end{align}%
By using the Hellmann-Feynman theorem, it is rewritten as%
\begin{align}
\sigma _{\text{D}}^{\mu \nu }\left( \omega ;\vartheta \right) =& \omega
\sigma _{0}\int d\mathbf{k}f_{-+}(\mathbf{k})a_{-+}^{\mu }\left( \mathbf{k}%
\right) a_{+-}^{\nu }\left( \mathbf{k}\right)  \notag \\
& \times \delta \left[ \varepsilon _{+}(\mathbf{k})-\varepsilon _{-}(\mathbf{%
k})-\hbar \omega \right] ,
\end{align}%
where $f_{-+}(\mathbf{k})\equiv f_{-}(\mathbf{k})-f_{+}(\mathbf{k})$. With
the use of Eq.(\ref{Faa}), the optical conductivity is related to the
quantum geometric tensor,%
\begin{align}
\sigma _{\text{D}}^{\mu \nu }\left( \omega ;\vartheta \right) =& \frac{%
\omega }{\hbar }\sigma _{0}\int d\mathbf{k}f_{-+}(\mathbf{k})\mathcal{F}%
_{-}^{\mu \nu }  \notag \\
& \times \delta \left[ \varepsilon _{+}(\mathbf{k})-\varepsilon _{-}(\mathbf{%
k})-\hbar \omega \right] .
\end{align}%
Then, we arrive at a relation between the quantum geometric tensor and the
conductivity\cite{Souza,OnishiX}%
\begin{equation}
\int_{0}^{\infty }\frac{d\omega }{\omega }\sigma _{\text{D}}^{\mu \nu
}\left( \omega \right) =\frac{\pi e^{2}}{\hbar \left( 2\pi \right) ^{2}}\int 
\mathcal{F}_{-}^{\mu \nu }d\mathbf{k},  \label{FlucD}
\end{equation}%
where $\sigma _{\text{D}}^{\mu \nu }\left( \omega \right) $ denotes the
dissipative (absorptive) component of the optical conductivity.

The real part reads\cite{Souza}%
\begin{equation}
\int_{0}^{\infty }\frac{d\omega }{\omega }\text{Re}\sigma _{\text{D}}^{\mu
\nu }\left( \omega \right) =\frac{\pi e^{2}}{\hbar \left( 2\pi \right) ^{2}}%
\int g_{-}^{\mu \nu }d\mathbf{k}.
\end{equation}%
The imaginary part is related to the Chern number,%
\begin{equation}
\int_{0}^{\infty }\frac{d\omega }{\omega }\text{Im}\sigma _{\text{D}}^{\mu
\nu }\left( \omega \right) =\frac{\pi e^{2}}{\hbar \left( 2\pi \right) ^{2}}%
\int \frac{\Omega _{-}^{xy}}{2}d\mathbf{k}=\frac{e^{2}}{2\hbar }\mathcal{C}%
_{-},
\end{equation}%
where we have used the TKNN formula%
\begin{equation}
\sigma _{\text{D}}^{xy}\left( 0\right) \equiv \sigma _{xy}=\frac{e^{2}}{h}%
\mathcal{C}_{-},
\end{equation}%
and the Kramers-Kronig relation. This relates the real and imaginary parts
of the optical Hall conductivity%
\begin{equation}
\int_{0}^{\infty }\frac{d\omega }{\omega }\text{Im}\sigma _{\text{D}}^{\mu
\nu }\left( \omega \right) =\frac{\pi }{2}\sigma _{xy}\left( 0\right) .
\end{equation}

\subsection{Bulk photovoltaic effects}

Photovoltaic currents are generated under photo-irradiation, which will be
useful for solar cell technologies. A $p$-$n$ junction presents a
conventional way to generate photocurrent. On the other hand, the bulk
photovoltaic photocurrent generation\ presents another way without using a
junction\cite{Beli,Kraut,Ave,Sipe,Frid}. The injection current\cite%
{Sipe,JuanNC,Juan,Ave,AhnX,AhnNP,WatanabeInject,Dai,EzawaVolta} and the
shift current\cite%
{Young,Young2,Ave,Kraut,Sipe,Juan,AhnX,MorimotoScAd,Kim,Barik,AhnNP,WatanabeInject,Dai,Yoshida,EzawaVolta}
are prominent, which are second order bulk photovoltaic currents.

The current density $J^{\mu }$ along the $\mu $ direction induced by the
applied electric field $E_{x}$ along the $x$ direction is expanded in a
power series of $E_{x}$ as%
\begin{equation}
J^{\mu }=\sum_{\ell =1}^{\infty }\sigma ^{\mu ;x^{\ell }}E_{x}^{\ell }\equiv
\sum_{\ell =1}^{\infty }J^{\mu ;x^{\ell }},
\end{equation}%
where $\mu =x,y,z$. We refer to $J^{\mu ;x^{\ell }}\equiv \sigma ^{\mu
;x^{\ell }}E_{x}^{\ell }$ as the $\ell $-th order current, where $\sigma
^{\mu ;x^{\ell }}$ is the $\ell $-th order conductivity. If there is
inversion symmetry in the system, the even order conductivities with even $%
\ell $ are prohibited, because%
\begin{equation}
J^{\mu }\mapsto -J^{\mu },\qquad E_{x}\mapsto -E_{x}  \label{Inversion}
\end{equation}%
under inversion symmetry.

The second order response has a form%
\begin{equation}
J^{\mu ;x^{2}}\left( \omega _{1}+\omega _{2}\right) =\sigma ^{\mu
;x^{2}}\left( \omega _{1}+\omega _{2};\omega _{1},\omega _{2}\right)
E_{x}\left( \omega _{1}\right) E_{x}\left( \omega _{2}\right) .
\end{equation}%
We investigate the direct current generation, 
\begin{equation}
J^{\mu ;x^{2}}\left( 0\right) =\sigma ^{\mu ;x^{2}}\left( 0;\omega ,-\omega
\right) E_{x}\left( \omega \right) E_{x}\left( -\omega \right) .
\end{equation}%
In the following, we use the abbreviation $J^{\mu ;x^{2}}\equiv J^{\mu
;x^{2}}\left( 0\right) $ and $\sigma ^{\mu ;x^{2}}\left( \omega \right)
\equiv \sigma ^{\mu ;x^{2}}\left( 0;\omega ,-\omega \right) $.

The conductivity of the injection current is in general given by the formula%
\cite{Sipe,JuanNC,Juan,Ave,AhnX,AhnNP,WatanabeInject,Dai,EzawaVolta}%
\begin{equation}
\sigma _{\text{inject}}^{\mu ;x^{2}}=-\tau \frac{2\pi e^{3}}{\hbar ^{2}}%
\sum_{n,m}\int d\mathbf{k}f_{nm}\Delta _{mn}^{\mu
}a_{nm}^{x}a_{mn}^{x}\delta \left( \omega _{mn}-\omega \right) ,
\label{Inject}
\end{equation}%
where $\tau $ is the relaxation time, $f_{nm}=f_{n}-f_{m}$ with the Fermi
distribution function $f_{n}$ for the band $n$, $a_{nm}^{x}$ is the
Wilczek-Zee connection, $\omega _{nm}\equiv \left( \varepsilon
_{n}-\varepsilon _{m}\right) /\hbar $, and%
\begin{equation}
\Delta _{mn}^{\mu }=v_{m}^{\mu }-v_{n}^{\mu }
\end{equation}%
is the difference of the velocities defined by 
\begin{equation}
v_{m}^{\mu }=\frac{1}{\hbar }\left\langle m\right\vert \partial _{k_{\mu
}}H\left\vert m\right\rangle
\end{equation}%
with the Hamiltonian $H$. The injection current is induced when the
velocities are different ($\Delta _{mn}^{\mu }\neq 0$) between the
conduction band $n$ and the valence band\textsl{\ }$m$ along the $\mu $
direction. The derivation is shown in Appendix \ref{ApInject}.

The shift current is in general given by the formula\cite%
{Young,Young2,Ave,Sipe,Juan,AhnX,MorimotoScAd,Kim,Barik,AhnNP,WatanabeInject,Dai,Yoshida,EzawaVolta}%
\begin{equation}
\sigma _{\text{shift}}^{\mu ;x^{2}}=-\frac{\pi e^{3}}{\hbar ^{2}}%
\sum_{n,m}\int d\mathbf{k}f_{nm}R_{mn}^{\mu ,x}a_{nm}^{x}a_{mn}^{x}\delta
\left( \omega _{mn}-\omega \right) ,  \label{Shift}
\end{equation}%
where 
\begin{equation}
R_{mn}^{\mu ,x}=a_{mm}^{x}-a_{nn}^{x}+i\partial _{k_{\mu }}\log a_{mn}^{x}
\label{shift}
\end{equation}%
is the shift vector\cite{Sipe}. The shift vector is gauge invariant although
the Wilczek-Zee connection is not gauge invariant. The shift vector
describes the difference of the mean position of the Wannier function
between two bands $m$ and $n$. The integrand in Eq.(\ref{shift}) is
rewritten as%
\begin{equation}
R_{mn}^{\mu ,x}a_{nm}^{x}a_{mn}^{x}=ia_{mn}^{x}a_{nm,\mu }^{x},
\end{equation}%
where we have defined the covariant derivative,%
\begin{equation}
\nabla _{k_{\mu }}a_{nm}^{x}\equiv a_{nm,\mu }^{x}\equiv \frac{\partial
a_{nm}^{x}}{\partial k_{\mu }}-ia_{nm}^{x}\left(
a_{nn}^{x}-a_{mm}^{x}\right) .
\end{equation}%
The shift current is induced when the mean positions are different ($%
R_{mn}^{\mu ,x}\neq 0$) between the conduction band $n$ and the valence band 
$m$. The derivation is shown in Appendix \ref{ApShift}.

In the following, we concentrate on the longitudinal conductivities by
setting $\mu =x$.

The injection current is rewritten in terms of the quantum metric\cite%
{AhnX,AhnNP,PJerk} as%
\begin{equation}
\sigma _{\text{inject}}^{x;x^{2}}=-\tau \frac{2\pi e^{3}}{\hbar ^{2}}\int d%
\mathbf{k}f_{-+}\Delta _{+-}^{x}g_{-}^{xx}\delta \left( \omega _{+-}-\omega
\right) .
\end{equation}%
while the shift current is rewritten in terms of the quantum metric\cite%
{WatanabeInject,PJerk} as%
\begin{equation}
\sigma _{\text{shift}}^{x;x^{2}}=-\frac{\pi e^{3}}{\hbar ^{2}V}\int d\mathbf{%
k}f_{-+}R_{+-}^{x,x}g_{-}^{xx}\delta \left( \omega _{+-}-\omega \right) .
\label{Rgxx}
\end{equation}

\subsection{Nonlinear conductivity}

The second-order nonlinear conductivity $\sigma ^{\mu \nu ;\rho }$ is
expanded in terms of the electron relaxation time $\tau $ as\cite{Kaplan}

\begin{equation}
\sigma ^{\mu \nu ;\rho }=\sigma _{\text{NLDrude}}^{\mu \nu ;\rho }+\sigma _{%
\text{Dipole}}^{\mu \nu ;\rho }+\sigma _{\text{Metric}}^{\mu \nu ;\rho },
\end{equation}%
where%
\begin{equation}
\sigma _{\text{Metric}}^{\mu \nu ;\rho }\propto \tau ^{0},\quad \sigma _{%
\text{Dipole}}^{\mu \nu ;\rho }\propto \tau ,\quad \sigma _{\text{NLDrude}%
}^{\mu \nu ;\rho }\propto \tau ^{2}.
\end{equation}

First, $\sigma _{\text{NLDrude}}^{\mu \nu ;\rho }$ is the nonlinear Drude
conductivity\cite{Ideue,NLDrude},%
\begin{equation}
\sigma _{\text{NLDrude}}^{\mu \nu ;\rho }=-\frac{e^{3}\tau ^{2}}{\hbar ^{3}}%
\sum_{n}\int d\mathbf{k}f_{n}\frac{\partial ^{3}\varepsilon _{n}}{\partial
k_{\mu }\partial k_{\nu }\partial k_{\rho }}.  \label{NLDrude}
\end{equation}%
It is also an extrinsic nonlinear conductivity.

Second, $\sigma _{\text{Dipole}}^{\mu \nu ;\rho }$ is the nonlinear
transverse conductivity induced by the Berry curvature dipole (BCD)\cite%
{Sodeman},%
\begin{equation}
\sigma _{\text{Dipole}}^{\mu \nu ;\rho }=-\frac{e^{3}\tau }{\hbar ^{2}}%
\sum_{n}\int d\mathbf{k}f_{n}\left( \frac{\partial \Omega _{n}^{\nu \rho }}{%
\partial k_{\mu }}+\frac{\partial \Omega _{n}^{\mu \rho }}{\partial k_{\nu }}%
\right) .
\end{equation}%
It is an extrinsic nonlinear conductivity, since it vanishes as $\tau
\rightarrow 0$.

Third, only the term $\sigma _{\text{Metric}}^{\mu \nu ;\rho }$ survives in
the dirty limit $\tau \rightarrow 0$, which is the intrinsic nonlinear
conductivity. It is the quantum-metric induced nonlinear conductivity. There
are still debates\cite{Qiang,ZGuo} on the coefficients of the quantum-metric
induced nonlinear conductivity.

(i) The Luttinger-Kohn approach\cite{Kaplan} gives 
\begin{align}
& \sigma _{\text{Metric}}^{\mu \nu ;\rho }  \notag \\
& =-\frac{e^{3}}{\hbar }\sum_{n}\int d\mathbf{k}f_{n}\left( 2\frac{\partial
G_{n}^{\mu \nu }}{\partial k_{\rho }}-\frac{1}{2}\left( \frac{\partial
G_{n}^{\nu \rho }}{\partial k_{\mu }}+\frac{\partial G_{n}^{\mu \rho }}{%
\partial k_{\nu }}\right) \right) ,  \label{Metric}
\end{align}%
where $G_{n}^{ab}$ is the band--energy normalized quantum metric or the
Berry connection polarizability. It is given as\cite%
{Gao,HLiu,CWang,KamalDas,Kaplan,Teresa}%
\begin{equation}
G_{n}^{\mu \nu }=2\text{Re}\sum_{m\neq n}\frac{a_{nm}^{\mu }\left( \mathbf{k}%
\right) a_{mn}^{\nu }\left( \mathbf{k}\right) }{\varepsilon _{n}\left( 
\mathbf{k}\right) -\varepsilon _{m}\left( \mathbf{k}\right) }.
\label{BMetric}
\end{equation}

(ii) The wave packet dynamics approach\cite{Gao} gives%
\begin{align}
& \sigma _{\text{Metric}}^{\mu \nu ;\rho }  \notag \\
& =-\frac{e^{3}}{\hbar }\sum_{n}\int d\mathbf{k}f_{n}\left( \frac{\partial
G_{n}^{\mu \nu }}{\partial k_{\rho }}-\frac{1}{2}\left( \frac{\partial
G_{n}^{\nu \rho }}{\partial k_{\mu }}+\frac{\partial G_{n}^{\mu \rho }}{%
\partial k_{\nu }}\right) \right) .
\end{align}

(iii) The quantum kinetics approach\cite{KamalDas} gives 
\begin{align}
& \sigma _{\text{Metric}}^{\mu \nu ;\rho }  \notag \\
& =-\frac{e^{3}}{\hbar }\sum_{n}\int d\mathbf{k}f_{n}\left( \frac{1}{2}\frac{%
\partial G_{n}^{\mu \nu }}{\partial k_{\rho }}-\left( \frac{\partial
G_{n}^{\nu \rho }}{\partial k_{\mu }}+\frac{\partial G_{n}^{\mu \rho }}{%
\partial k_{\nu }}\right) \right) .
\end{align}

(iv) The intrinsic Ohmic conductivity\cite{Ohmic} gives%
\begin{align}
& \sigma _{\text{Metric}}^{\mu \nu ;\rho }  \notag \\
& =\frac{e^{3}}{\hbar }\sum_{n}\int d\mathbf{k}f_{n}\left( \frac{\partial
G_{n}^{\mu \nu }}{\partial k_{\rho }}+\frac{\partial G_{n}^{\nu \rho }}{%
\partial k_{\mu }}+\frac{\partial G_{n}^{\mu \rho }}{\partial k_{\nu }}%
\right) .
\end{align}

\section{Zeeman Quantum geometry for momentum and spin}

\label{SecZeeman}

So far, we have considered quantum geometry for momentum translation.
However, there is also a spin degree of freedom for electrons. It is
important to study the effect of spin rotation in the context of
spintronics. This is achieved by generalizing quantum geometry to Zeeman
quantum geometry\cite{Xiang2,Xiang25,Chak,Xiang3}, where local spin rotation
between two adjacent wave functions is also taken into account.

The Zeeman quantum geometric tensor $g_{nm}^{\mu \nu }$ is defined by the
quantum distance $ds_{\text{HS}}$ for the infinitesimal translation $d%
\mathbf{k}$ of the momentum and infinitesimal spin rotation $d\mathbf{\theta 
}$ as\cite{Xiang2,Xiang25,Chak,Xiang3} 
\begin{equation}
ds_{\text{HS}}\left( \mathbf{k}\right) \equiv \sqrt{1-\left\vert
\left\langle \psi _{n}\left( \mathbf{k}\right) \right\vert U_{d\mathbf{%
\theta }}U_{d\mathbf{k}}\left\vert \psi _{n}\left( \mathbf{k}\right)
\right\rangle \right\vert ^{2}},
\end{equation}%
where 
\begin{equation}
U_{d\mathbf{\theta }}\equiv e^{-\frac{i}{2}d\mathbf{\theta }\cdot \mathbf{%
\sigma }},\quad U_{d\mathbf{k}}\equiv e^{-id\mathbf{k}\cdot \mathbf{r}}
\label{Udt}
\end{equation}%
are the generators of the spin angular momentum $d\mathbf{\theta }$ and the
momentum $d\mathbf{k}$. The Zeeman quantum geometric tensor $z_{nm}^{\mu \nu
}$ and the spin quantum geometric tensor $s_{nm}^{\mu \nu }$ are given by%
\begin{align}
\left( ds_{\text{HS}}\right) ^{2}=& \sum_{\mu \nu }\sum_{n\neq m}(\mathcal{F}%
_{nm}^{\mu \nu }dk_{\mu }dk_{\nu }+\frac{s_{nm}^{\mu \nu }}{4}d\theta _{\mu
}d\theta _{\nu }  \notag \\
& +\frac{z_{nm}^{\mu \nu }+z_{mn}^{\mu \nu }}{2}dk_{\mu }d\theta _{\nu }),
\label{dsZ}
\end{align}%
where 
\begin{equation}
\mathcal{F}_{nm}^{\mu \nu }\equiv a_{nm}^{\mu }a_{mn}^{\nu }
\end{equation}%
is the quantum geometric tensor,%
\begin{equation}
s_{nm}^{\mu \nu }\equiv s_{nm}^{\mu }s_{mn}^{\nu }  \label{SGT}
\end{equation}%
is the spin quantum geometric tensor, and%
\begin{equation}
z_{nm}^{\mu \nu }\equiv a_{nm}^{\mu }s_{mn}^{\nu }  \label{Zz12}
\end{equation}%
is the Zeeman quantum geometric tensor, where we have defined the
expectation value of the spin operator,%
\begin{equation}
s_{nm}^{\mu }\equiv \left\langle \psi _{n}\left( \mathbf{k}\right)
\right\vert \sigma _{\mu }\left\vert \psi _{n}\left( \mathbf{k}\right)
\right\rangle .
\end{equation}%
The derivation from Eq.(\ref{dsZ})\symbol{126}Eq.(\ref{Zz12}) is shown in
Appendix \ref{ApZQ}.

As in the case of the quantum metric and the Berry curvature, the Zeeman
quantum metric $\mathcal{Q}_{nm}^{\mu \nu }$ is defined by the real part of
the Zeeman quantum geometric tensor,%
\begin{equation}
\mathcal{Q}_{nm}^{\mu \nu }\equiv \text{Re}z_{nm}^{\mu \nu }=\left(
a_{nm}^{\mu }s_{mn}^{\nu }+a_{mn}^{\mu }s_{nm}^{\nu }\right) /2,  \label{Qab}
\end{equation}%
and the Zeeman Berry curvature $\mathcal{Z}_{nm}^{\mu \nu }$ by the
imaginary part of the Zeeman quantum geometric tensor, 
\begin{equation}
\mathcal{Z}_{nm}^{\mu \nu }\equiv 2\text{Im}z_{nm}^{\mu \nu }=i\left(
a_{nm}^{\mu }s_{mn}^{\nu }-a_{mn}^{\mu }s_{nm}^{\nu }\right) .  \label{Zab}
\end{equation}%
We also define the spin quantum metric $\mathcal{S}_{nm}^{\mu \nu }$ by the
real part of the spin quantum geometric tensor,%
\begin{equation}
\mathcal{S}_{nm}^{\mu \nu }\equiv \text{Re}s_{nm}^{\mu \nu }\equiv \left(
s_{nm}^{\mu }s_{mn}^{\nu }+s_{mn}^{\mu }s_{nm}^{\nu }\right) /2,
\label{SQGT}
\end{equation}%
and the spin Berry curvature $\mathcal{A}_{nm}^{\mu \nu }$ by%
\begin{equation}
\mathcal{A}_{nm}^{\mu \nu }\equiv 2\text{Im}s_{nm}^{\mu \nu }\equiv i\left(
s_{nm}^{\mu }s_{mn}^{\nu }-s_{mn}^{\mu }s_{nm}^{\nu }\right) .
\label{SBerry}
\end{equation}

\subsection{Responses originated from the Zeeman geometry}

In the linear response theory, we obtain the response of the current $J^{\mu
;\nu }$ or the spin polarization $S^{\mu ;\nu }$ by applying electric field $%
E^{\nu }$ or magnetic field $B^{\nu }$.

We apply alternating electric field $E^{\nu }\left( t\right) $\ and
alternating magnetic field $B^{\nu }\left( t\right) $, 
\begin{align}
E^{\nu }\left( t\right) =& \frac{1}{2}E^{\nu }\sum_{\omega _{1}=\pm \omega
}e^{-i\omega _{1}t}, \\
B^{\nu }\left( t\right) =& \frac{1}{2}B^{\nu }\sum_{\omega _{1}=\pm \omega
}e^{-i\omega _{1}t}.
\end{align}%
The Hamiltonian for the external fields is given by%
\begin{equation}
H_{1}=E^{\nu }\left( t\right) a^{\nu }-g\mu _{\text{B}}B^{\nu }\left(
t\right) \sigma ^{\nu }.
\end{equation}%
We solve the quantum Liouville equation%
\begin{equation}
i\frac{\partial \rho }{\partial t}=\left[ H^{\prime },\rho \right]
\end{equation}%
with $H^{\prime }=H+H_{1}$, where $H$ is the non-perturbated Hamiltonian.

The first-order solution of the density matrix is given by\cite{Xiang25}%
\begin{align}
\rho _{mn}^{\mu \nu }=& \frac{1}{2}\sum_{\omega _{1}=\pm \omega }\left[
i\delta _{mn}\partial _{k_{\mu }}f_{m}+f_{nm}a_{mn}^{\mu }\right] \frac{%
E^{\nu }e^{-i\omega _{1}t}}{\hbar \omega _{1}-\varepsilon _{mn}+i\eta } 
\notag \\
& -\frac{g\mu _{\text{B}}}{2}\sum_{\omega _{1}=\pm \omega }\frac{%
f_{nm}s_{mn}^{\mu }B^{\nu }e^{-i\omega _{1}t}}{\hbar \omega _{1}-\varepsilon
_{mn}+i\eta },  \label{Rho1}
\end{align}%
where $f_{nm}\equiv f_{n}-f_{m}$ and $\varepsilon _{mn}\equiv \varepsilon
_{m}-\varepsilon _{n}$.

In the linear response theory, the current $J^{\mu ;\nu }$ is given by%
\begin{equation}
J^{\mu ;\nu }=\sum_{nm}\int d\mathbf{k}v_{nm}^{\mu }\rho _{mn}^{\mu \nu },
\label{Jab}
\end{equation}%
where $v_{nm}^{\mu }$\ is the velocity operator. By using the
Hellmann--Feynman theorem, it is rewritten as%
\begin{equation}
v_{nm}^{\mu }=i\varepsilon _{nm}a_{nm}^{\mu }.
\end{equation}%
On the other hand, the spin $S^{\mu ;\nu }$ polarization is given by 
\begin{equation}
S^{\mu ;\nu }=\sum_{nm}\int d\mathbf{k}s_{nm}^{\mu }\rho _{mn}^{\mu \nu }.
\label{Sab}
\end{equation}

There are four types of responses, $J^{\mu ;\nu }/E^{\nu }$, $J^{\mu ;\nu
}/B^{\nu }$, $S^{\mu ;\nu }/E^{\nu }$ and $S^{\mu ;\nu }/B^{\nu }$, where $%
J^{\mu ;\nu }/E^{\nu }$ and $S^{\mu ;\nu }/B^{\nu }$ are direct responses,
while $J^{\mu ;\nu }/B^{\nu }$ and $S^{\mu ;\nu }/E^{\nu }$ are cross
responses.

(i) The current $J^{\mu ;\nu }$ is induced by electric field $E^{\nu }$ as%
\cite{Xiang2}%
\begin{align}
& \frac{J^{\mu ;\nu }}{E^{\nu }}  \notag \\
=& \int d\mathbf{k}\sum_{n>m}f_{nm}\left[ \Omega _{nm}^{\mu \nu }\cos \omega
t+2g_{nm}^{\mu \nu }\frac{\omega \sin \omega t}{\varepsilon _{mn}}\right] .
\label{JE}
\end{align}%
The first term represents the Hall current generated by the Berry curvature $%
\Omega _{nm}^{\mu \nu }$. The quantum metric $g_{nm}^{\mu \nu }$ in the
second term\ appears as an oscillating response\cite{XiangL}.

(ii) The current $J^{\mu ;\nu }$ is induced by magnetic field $B^{\nu }$ as%
\cite{Xiang25}

\begin{align}
& \frac{J^{\mu ;\nu }}{B^{\nu }}  \notag \\
=& g\mu _{\text{B}}\int d\mathbf{k}\sum_{n>m}f_{nm}\left[ \mathcal{Z}%
_{nm}^{\mu \nu }\cos \omega t+2\mathcal{Q}_{nm}^{\mu \nu }\frac{\omega \sin
\omega t}{\varepsilon _{mn}}\right] .  \label{JB}
\end{align}%
The Zeeman quantum geometric tensor $\mathcal{Z}_{nm}^{\mu \nu }$\
contributes to the cross response between the magnetic field $B^{\nu }$ and
the current $J^{\mu ;\nu }$, where $J^{\mu ;\nu }$ is called the intrinsic
gyrotropic magnetic current $J^{\mu ;\nu }$. The Zeeman quantum metric $%
\mathcal{Q}_{nm}^{\mu \nu }$ in the second term\ appears as an oscillating
response.

(iii) Spin polarization $S^{\mu ;\nu }$ is induced by electric field $E^{\nu
}$ as\cite{Xiang2}

\begin{align}
& \frac{S^{\mu ;\nu }}{E^{\nu }}  \notag \\
=& -\int d\mathbf{k}\sum_{n>m}f_{nm}\left[ \frac{2\mathcal{Q}_{nm}^{\mu \nu }%
}{\varepsilon _{mn}}\cos \omega t+\mathcal{Z}_{nm}^{\mu \nu }\frac{\omega
\sin \omega t}{\varepsilon _{mn}^{2}}\right] .  \label{SE}
\end{align}%
The Zeeman quantum metric $\mathcal{Q}_{nm}^{\mu \nu }$\ contributes to the
cross response between the electric field $E^{\nu }$ and the spin
polarization $S^{\mu ;\nu }$. The Zeeman quantum geometric tensor $\mathcal{Z%
}_{nm}^{\mu \nu }$ in the second term\ appears as an oscillating response.

(iv) Spin polarization $S^{\mu ;\nu }$ is also induced by magnetic field $%
B^{\nu }$ as

\begin{align}
& \frac{S^{\mu ;\nu }}{B^{\nu }}  \notag \\
=& g\mu _{\text{B}}\int d\mathbf{k}\sum_{n>m}f_{nm}\left[ \frac{\mathcal{S}%
_{nm}^{\mu \nu }}{\varepsilon _{mn}}\cos \omega t+2\mathcal{A}_{nm}^{\mu \nu
}\frac{\omega \sin \omega t}{\varepsilon _{mn}^{2}}\right] .  \label{SB}
\end{align}%
It is interesting that there may be off-diagonal response if $\mathcal{S}%
_{nm}^{\mu \nu }$ or $\mathcal{A}_{nm}^{\mu \nu }$ has off-diagonal
components. Detailed derivations of Eqs.(\ref{JE}), (\ref{JB}), (\ref{SE})
and (\ref{SB}) are given in Appendix \ref{ApCrossRes}.

We may call the first terms in Eqs.(\ref{JE}), (\ref{JB}), (\ref{SE}) and (%
\ref{SB}) the static terms and the second terms the dynamic terms, because
the second terms vanish for $\omega =0$.

\subsection{Zeeman Quantum Geometry for two-band systems}

We consider the two-band system (\ref{Hamil2}) with the index $n=\pm $. The
simple relations for the Zeeman quantum metric are derived as%
\begin{equation}
\mathcal{Z}_{+-}^{\mu \nu }=-\mathcal{Z}_{-+}^{\mu \nu }=\frac{\partial
n_{\nu }}{\partial k_{\mu }}.  \label{Z}
\end{equation}%
Two or three components of $\mathbf{n}$\ should be nonzero for nonzero $%
\mathcal{Z}_{+-}^{\mu \nu }$. In general, we have $\mathcal{Z}_{+-}^{\mu \nu
}\neq \mathcal{Z}_{+-}^{\nu \mu }$.

The Zeeman Berry curvature is derived as%
\begin{equation}
\mathcal{Q}_{+-}^{\mu \nu }=\frac{1}{2}\sum_{\rho \sigma }\varepsilon _{\nu
\rho \sigma }n_{\rho }\frac{\partial n_{\sigma }}{\partial k_{\mu }},
\label{Q}
\end{equation}%
Two or three components of $\mathbf{n}$\ should be nonzero for nonzero $%
\mathcal{Q}_{+-}^{\mu \nu }$.

The spin quantum metric is%
\begin{equation}
\mathcal{S}_{+-}^{\mu \nu }=\delta _{\mu \nu }-n_{\mu }n_{\nu },  \label{S}
\end{equation}%
while the spin Berry curvature is%
\begin{equation}
\mathcal{A}_{+-}^{\mu \nu }=-2\sum_{\rho }\varepsilon _{\mu \nu \rho
}n_{\rho }.  \label{A}
\end{equation}%
They are defined even when only a single component of $\mathbf{n}$ is
nonzero.

\subsection{Rashba system}

As an example of the two-band system, we consider the Rashba system
described by%
\begin{equation}
H\left( \mathbf{k}\right) =\frac{\hbar ^{2}\mathbf{k}^{2}}{2m}+\lambda
\left( \mathbf{k}\times \mathbf{\sigma }\right) _{z}+B\sigma _{z},
\end{equation}%
where $\lambda $ is the strength of the Rashba interaction and $B$ is static
magnetic field applied perpendicularly to the plane.

The energy is given by%
\begin{equation}
E_{\chi }=\frac{\hbar ^{2}k^{2}}{2m}+\chi \sqrt{\lambda ^{2}k^{2}+B^{2}}
\end{equation}%
for the lower band with $\chi =-1$ and for the upper band with $\chi =1$.
The Fermi surface is given by%
\begin{equation}
k_{\eta }^{\chi }=\sqrt{m}\sqrt{\mu +m\lambda ^{2}+\eta \sqrt{m^{2}\lambda
^{4}+2\mu m\lambda ^{2}+B^{2}}},
\end{equation}%
where $\eta =-1$ describes the inner Fermi surface and $\eta =1$ describes
the outer Fermi surface.

The Berry curvature is given by%
\begin{equation}
\Omega ^{xy}=-\frac{B\lambda ^{2}}{2\left( \lambda ^{2}k^{2}+B^{2}\right)
^{3/2}},
\end{equation}%
while the quantum metrices are given by%
\begin{align}
g_{+-}^{xx} =&\frac{\lambda ^{2}\left( \lambda ^{2}k_{y}^{2}+B^{2}\right) }{%
2\left( \lambda ^{2}k^{2}+B^{2}\right) ^{2}}, \\
g_{+-}^{xy} =&g_{+-}^{yx}=\frac{-\lambda ^{4}k_{x}k_{y}}{2\left( \lambda
^{2}k^{2}+B^{2}\right) ^{2}}, \\
g_{+-}^{yy} =&\frac{\lambda ^{2}\left( \lambda ^{2}k_{x}^{2}+B^{2}\right) }{%
2\left( \lambda ^{2}k^{2}+B^{2}\right) ^{2}}.
\end{align}%
There is no contribution from the quantum metric $g_{+-}^{xy}$ in the Hall
current (\ref{JE}) because $g_{+-}^{xy}\propto \sin 2\phi $ leads to $\int
g_{+-}^{xy}d\phi =0$.

The Zeeman Berry curvatures are calculated as%
\begin{align}
\mathcal{Z}_{+-}^{xx} =&\frac{\partial n_{x}}{\partial k_{x}}=\frac{\lambda
^{3}k_{x}k_{y}}{\left( \lambda ^{2}k^{2}+B^{2}\right) ^{3/2}}, \\
\mathcal{Z}_{+-}^{xy} =&\frac{\partial n_{y}}{\partial k_{x}}=\frac{\lambda
\left( \lambda ^{2}k_{y}^{2}+B^{2}\right) }{\left( \lambda
^{2}k^{2}+B^{2}\right) ^{3/2}}, \\
\mathcal{Z}_{+-}^{yx} =&\frac{\partial n_{x}}{\partial k_{y}}=-\frac{\lambda
\left( \lambda ^{2}k_{x}^{2}+B^{2}\right) }{\left( \lambda
^{2}k^{2}+B^{2}\right) ^{3/2}}, \\
\mathcal{Z}_{+-}^{xz} =&\frac{\partial n_{z}}{\partial k_{x}}=-\frac{\lambda
^{2}Bk_{x}}{\left( \lambda ^{2}k^{2}+B^{2}\right) ^{3/2}}, \\
\mathcal{Z}_{+-}^{yz} =&\frac{\partial n_{z}}{\partial k_{y}}=-\frac{\lambda
^{2}Bk_{y}}{\left( \lambda ^{2}k^{2}+B^{2}\right) ^{3/2}}.
\end{align}%
Only the Zeeman Berry curvature $\mathcal{Z}_{+-}^{xy}$ contributes to a
nonzero response because we have%
\begin{equation}
\int \mathcal{Z}_{+-}^{xx}d\phi =\int \mathcal{Z}_{+-}^{xz}d\phi =\int 
\mathcal{Z}_{+-}^{yz}d\phi =0.
\end{equation}%
We note that $\mathcal{Z}_{+-}^{xx}$ and $\mathcal{Z}_{+-}^{xy}$ are
singular at $k=0$, while $\mathcal{Z}_{+-}^{xz}$ and $\mathcal{Z}_{+-}^{yz}$
are zero for $B=0$.

The static intrinsic gyrotropic magnetic current at zero temperature is
calculated as%
\begin{align}
\frac{J^{x;y}}{B^{y}}=& g\mu _{\text{B}}\int d\mathbf{k}\sum_{n>m}f_{nm}%
\mathcal{Z}_{nm}^{xy}  \notag \\
=& g\mu _{\text{B}}\int_{k_{-}^{\chi }}^{k_{+}^{\chi }}kdkd\phi \frac{%
\lambda \left( \lambda ^{2}k_{y}^{2}+B^{2}\right) }{\left( \lambda
^{2}k^{2}+B^{2}\right) ^{3/2}}  \notag \\
=& g\mu _{\text{B}}\sum_{\eta =\pm 1}\pi \lambda \frac{\left( k_{\eta
}^{\chi }\right) ^{2}}{\sqrt{\lambda ^{2}\left( k_{\eta }^{\chi }\right)
^{2}+B^{2}}}.
\end{align}%
The electric-filed induced spin polarization at zero temperature is
calculated as%
\begin{align}
\frac{S^{\mu ;\nu }}{E^{\nu }}=& -\int d\mathbf{k}f_{+-}\mathcal{Z}%
_{+-}^{\nu \mu }\frac{\omega }{\varepsilon _{+-}^{2}}\sin \left( \omega
t\right)  \notag \\
=& \sum_{\eta =\pm 1}\eta \pi \frac{\lambda \left( k_{\eta }^{\chi }\right)
^{2}+4B^{2}}{\left( \lambda ^{2}\left( k_{\eta }^{\chi }\right)
^{2}+B^{2}\right) ^{3/2}}\omega \sin \left( \omega t\right) .
\end{align}

The Zeeman quantum metrices are calculated as%
\begin{align}
\mathcal{Q}_{+-}^{xx} =&\mathcal{Q}_{+-}^{yy}-\frac{B\lambda }{2\left(
\lambda ^{2}k^{2}+B^{2}\right) }, \\
\mathcal{Q}_{+-}^{xy} =&\mathcal{Q}_{+-}^{yx}=0, \\
\mathcal{Q}_{+-}^{xz} =&-\frac{\lambda ^{2}k_{y}}{2\left( \lambda
^{2}k^{2}+B^{2}\right) }.
\end{align}%
Only the Zeeman quantum metric $\mathcal{Q}_{+-}^{xx}$ contributes to a
nonzero response because%
\begin{equation}
\int \mathcal{Q}_{+-}^{xz}d\phi =0.
\end{equation}%
The static spin polarization is induced by electric field as

\begin{align}
\frac{S^{x;x}}{E^{x}}=& -\int d\mathbf{k}f_{+-}\frac{2\mathcal{Q}_{+-}^{xx}}{%
\varepsilon _{+-}}  \notag \\
=& -\sum_{\eta =\pm 1}\frac{\eta \pi B}{\lambda \sqrt{\lambda ^{2}\left(
k_{\eta }^{\chi }\right) ^{2}+B^{2}}}.
\end{align}%
The dynamic intrinsic gyrotropic magnetic current is obtained as

\begin{align}
\frac{J^{x;x}}{B^{x}}=& g\mu _{\text{B}}\int d\mathbf{k}\sum_{n>m}f_{nm}2%
\mathcal{Q}_{nm}^{xx}\frac{\omega }{\varepsilon _{mn}}\sin \omega t  \notag
\\
=& g\mu _{\text{B}}\sum_{\eta =\pm 1}\frac{\eta 2\pi B}{\lambda \sqrt{%
\lambda ^{2}\left( k_{\eta }^{\chi }\right) ^{2}+B^{2}}}\omega \sin \omega t.
\end{align}

The spin Berry curvatures are calculated as%
\begin{align}
\mathcal{A}_{+-}^{xy}=& -2n_{z}=-\frac{2B}{\sqrt{\lambda ^{2}k^{2}+B^{2}}},
\\
\mathcal{A}_{+-}^{yz}=& -2n_{x}=\frac{2\lambda k_{y}}{\sqrt{\lambda
^{2}k^{2}+B^{2}}}, \\
\mathcal{A}_{+-}^{zx}=& -2n_{y}=-\frac{2\lambda k_{x}}{\sqrt{\lambda
^{2}k^{2}+B^{2}}},
\end{align}%
among which there is only nonzero contribution from $\mathcal{A}_{+-}^{xy}$
because%
\begin{equation}
\int d\phi \mathcal{A}_{+-}^{yz}=\int d\phi \mathcal{A}_{+-}^{zx}=0.
\end{equation}

Spin polarization is also induced by magnetic field as

\begin{align}
\frac{S^{x;y}}{B^{y}}=& g\mu _{\text{B}}\int d\mathbf{k}f_{+-}2\mathcal{A}%
_{+-}^{xy}\frac{\omega }{\varepsilon _{+-}^{2}}\sin \omega t  \notag \\
=& g\mu _{\text{B}}\sum_{\eta =\pm 1}\frac{\eta 4\pi B}{\lambda ^{2}\sqrt{%
\lambda ^{2}\left( k_{\eta }^{\chi }\right) ^{2}+B^{2}}}.
\end{align}

The diagonal spin quantum metrices are calculated as%
\begin{align}
\mathcal{S}_{+-}^{xx}=& \frac{\lambda ^{2}k_{x}^{2}+B^{2}}{\lambda
^{2}k^{2}+B^{2}},  \label{RSxx} \\
\mathcal{S}_{+-}^{yy}=& \frac{\lambda ^{2}k_{y}^{2}+B^{2}}{\lambda
^{2}k^{2}+B^{2}},  \label{RSyy} \\
\mathcal{S}_{+-}^{zz}=& \frac{\lambda ^{2}k^{2}}{\lambda ^{2}k^{2}+B^{2}},
\label{RSzz}
\end{align}%
all of which contribute to nonzero responses.

The off-diagonal spin quantum metrices are calculated as%
\begin{align}
\mathcal{S}_{+-}^{xy}=& \frac{2\lambda ^{2}k_{x}k_{y}}{\lambda
^{2}k^{2}+B^{2}}, \\
\mathcal{S}_{+-}^{yz}=& -\frac{2B\lambda k_{x}}{\lambda ^{2}k^{2}+B^{2}}, \\
\mathcal{S}_{+-}^{zx}=& \frac{2B\lambda k_{y}}{\lambda ^{2}k^{2}+B^{2}},
\end{align}%
all of which do not contribute to the off-diagonal responses because%
\begin{equation}
\int \mathcal{S}_{+-}^{xy}d\phi =\int \mathcal{S}_{+-}^{yz}d\phi =\int 
\mathcal{S}_{+-}^{zx}d\phi =0.
\end{equation}

\subsection{Non-Abelian Zeeman quantum geometry}

\label{SecZeemanMulti}

We generalize the Zeeman quantum geometry to the multiband systems. The
Zeeman quantum geometric tensor $g_{nm}^{\mu \nu }$ is defined by the
quantum distance $ds_{\text{HS}}$ for the infinitesimal translation $d%
\mathbf{k}$ of the momentum as 
\begin{equation}
\left( ds_{\text{HS}}\right) ^{2}\equiv \sqrt{1-\left\vert \left\langle \psi
\left( \mathbf{k}\right) \right\vert U_{d\mathbf{\theta }}U_{d\mathbf{k}%
}\left\vert \psi \left( \mathbf{k}\right) \right\rangle \right\vert ^{2}},
\end{equation}%
where $\left\vert \psi \left( \mathbf{k}\right) \right\rangle $ is given by
Eq.(\ref{psi}), $U_{d\mathbf{\theta }}$ and $U_{d\mathbf{k}}$ are given by
Eq.(\ref{Udt}). The Zeeman quantum geometric tensor is given by%
\begin{align}
\left( ds_{\text{HS}}\right) ^{2}=& \sum_{\mu \nu }\sum_{n\neq m}g_{nm}^{\mu
\nu }dk_{\mu }dk_{\nu }+\frac{s_{nm}^{\mu \nu }}{4}d\theta _{\mu }d\theta
_{\nu }  \notag \\
& +\frac{z_{nm}^{\mu \nu }+z_{mn}^{\mu \nu }}{2}dk_{\mu }d\theta _{\nu },
\end{align}%
where 
\begin{equation}
g_{nm}^{\mu \nu }\equiv \left\langle \partial _{k_{\mu }}\psi _{n}\left( 
\mathbf{k}\right) \right\vert \left( 1-P\left( \mathbf{k}\right) \right)
\left\vert \partial _{k_{\mu }}\psi _{m}\left( \mathbf{k}\right)
\right\rangle
\end{equation}%
is the quantum metric,%
\begin{equation}
s_{nm}^{\mu \nu }\equiv \left\langle \psi _{n}\left( \mathbf{k}\right)
\right\vert \sigma _{\mu }\left( 1-P\left( \mathbf{k}\right) \right) \sigma
_{\nu }\left\vert \psi _{m}\left( \mathbf{k}\right) \right\rangle
\end{equation}%
is the spin geometric tensor, and 
\begin{equation}
z_{nm}^{\mu \nu }\equiv \left\langle \partial _{k_{\mu }}\psi _{n}\left( 
\mathbf{k}\right) \right\vert \left( 1-P\left( \mathbf{k}\right) \right)
\sigma _{\nu }\left\vert \psi _{m}\left( \mathbf{k}\right) \right\rangle
\end{equation}%
is the Zeeman geometric tensor. The derivations are shown in Appendix \ref%
{ApZQ2}.

\section{Quantum geometry for non-Hermitian systems}

\label{SecNH}

\subsection{Open quantum system and non-Hermitian Hamiltonian}

Non-Hermitian systems attract much attention. It is required that the
Hamiltonian is Hermitian in quantum mechanics. However, it becomes
non-Hermitian if we consider an open quantum system. An open quantum system
is described by the Lindblad equation for the density matrix $\rho $,%
\begin{equation}
\frac{d\rho }{dt}=-\frac{i}{\hbar }\left[ H,\rho \right] +\sum_{k}\gamma
_{k}\left( L_{k}\rho L_{k}^{\dagger }-\frac{1}{2}\left\{ L_{k}^{\dagger
}L_{k},\rho \right\} \right) ,
\end{equation}%
where $L_{k}$ is the Lindblad operator and $\gamma $ is the strength of the
dissipation. The Lindblad equation is rewritten in the form of%
\begin{equation}
\frac{d\rho }{dt}=-\frac{i}{\hbar }\left( H_{\text{eff}}\rho -\rho H_{\text{%
eff}}^{\dagger }\right) +\sum_{k}\gamma _{k}L_{k}\rho L_{k}^{\dagger },
\end{equation}%
where we have introduced a non-Hermitian Hamiltonian by%
\begin{equation}
H_{\text{eff}}\equiv H-\frac{i\hbar }{2}\sum_{k\gamma }\gamma
_{k}L_{k}^{\dagger }L_{k}.
\end{equation}

\subsection{Non-Hermitian quantum geometry}

Quantum geometry is generalized to non-Hermitian systems\cite%
{Silb,Robin,Ye,DongH,Beh}, where $H^{\dagger }\neq H$. There are right and
left eigenfunctions and eigenvalues,%
\begin{equation}
H=\varepsilon _{n}^{\text{R}}\left\vert \psi _{n}^{\text{R}}\left( \mathbf{k}%
\right) \right\rangle ,\quad \left\langle \psi _{n}^{\text{L}}\left( \mathbf{%
k}\right) \right\vert H=\left\langle \psi _{n}^{\text{L}}\left( \mathbf{k}%
\right) \right\vert \varepsilon _{n}^{\text{L}},
\end{equation}%
where%
\begin{equation*}
\left( \left\vert \psi _{n}^{\text{R}}\left( \mathbf{k}\right) \right\rangle
\right) ^{\ast }\neq \left\langle \psi _{n}^{\text{L}}\left( \mathbf{k}%
\right) \right\vert
\end{equation*}%
in general. However, it is straightforward to generalize quantum geometry to
non-Hermitian case. The eigen functions are orthonormalized,%
\begin{equation}
\left\langle \psi _{n}^{\text{L}}\left( \mathbf{k}\right) \left\vert \psi
_{n}^{\text{R}}\left( \mathbf{k}^{\prime }\right) \right\rangle \right.
=\delta \left( \mathbf{k},\mathbf{k}^{\prime }\right) .
\end{equation}

If the target bands are $N$-fold degenerate, the right and left
eigenfunctions are given by

\begin{align}
\left\vert \psi ^{\text{R}}\left( \mathbf{k}\right) \right\rangle =&
\sum_{n=1}^{N}c_{n}^{\text{R}}\left\vert \psi _{n}^{\text{R}}\left( \mathbf{k%
}\right) \right\rangle , \\
\left\langle \psi ^{\text{L}}\left( \mathbf{k}\right) \right\vert =&
\sum_{n=1}^{N}c_{n}^{\text{L}}\left\langle \psi _{n}^{\text{L}}\left( 
\mathbf{k}\right) \right\vert .
\end{align}

The fidelity is defined by%
\begin{equation}
F\left( \mathbf{k},\mathbf{k}^{\prime }\right) \equiv \sqrt{\left\langle
\psi ^{\text{L}}\left( \mathbf{k}\right) \left\vert \psi ^{\text{R}}\left( 
\mathbf{k}^{\prime }\right) \right\rangle \right. \left\langle \psi ^{\text{L%
}}\left( \mathbf{k}^{\prime }\right) \left\vert \psi ^{\text{R}}\left( 
\mathbf{k}\right) \right\rangle \right. }.
\end{equation}

The Hilbert-Schmidt distance is defined by%
\begin{equation}
s_{\text{HS}}\left( \mathbf{k},\mathbf{k}^{\prime }\right) \equiv \sqrt{%
1-F\left( \mathbf{k},\mathbf{k}^{\prime }\right) ^{2}}.
\end{equation}%
The quantum distance with the infinitesimal momentum translation $d\mathbf{k}
$ is given by%
\begin{equation}
ds_{\text{HS}}\left( \mathbf{k}\right) \equiv \sqrt{1-F\left( \mathbf{k,k}+d%
\mathbf{k}\right) ^{2}}
\end{equation}%
with%
\begin{equation}
F\left( \mathbf{k}\right) =\sqrt{\left\langle \psi ^{\text{L}}\left( \mathbf{%
k}\right) \right\vert U_{d\mathbf{k}}\left\vert \psi ^{\text{R}}\left( 
\mathbf{k}\right) \right\rangle \left\langle \psi ^{\text{L}}\left( \mathbf{k%
}\right) \right\vert U_{d\mathbf{k}}^{-1}\left\vert \psi ^{\text{R}}\left( 
\mathbf{k}\right) \right\rangle },
\end{equation}%
\newline
where $U_{d\mathbf{k}}$ is given by Eq.(\ref{Udk}). The quantum distance for
the infinitesimal momentum is expanded in terms of the quantum geometric
tensor $\mathcal{F}_{n}^{\mu \nu }$ as%
\begin{equation}
\left( ds_{\text{HS}}\right) ^{2}=\sum_{\mu \nu }\sum_{nm}\mathcal{F}%
_{nm}^{\mu \nu }dk_{\mu }dk_{\nu },
\end{equation}%
The quantum geometric tensor is 
\begin{equation}
\mathcal{F}_{nm}^{\mu \nu }\left( \mathbf{k}\right) =\left\langle \partial
_{k_{\mu }}\psi _{n}^{\text{L}}\left( \mathbf{k}\right) \right\vert \left(
1-P\left( \mathbf{k}\right) \right) \left\vert \partial _{k_{\nu }}\psi
_{m}^{\text{R}}\left( \mathbf{k}\right) \right\rangle
\end{equation}%
with the projection operator%
\begin{equation}
P\left( \mathbf{k}\right) \equiv \sum_{n=1}^{N}\left\vert \psi _{n}^{\text{R}%
}\left( \mathbf{k}\right) \right\rangle \left\langle \psi _{n}^{\text{L}%
}\left( \mathbf{k}\right) \right\vert .
\end{equation}

The quantum metric is the real part of the quantum geometric tensor,%
\begin{equation}
g_{nm}^{\mu \nu }\equiv \text{Re}\mathcal{F}_{nm}^{\mu \nu }=\frac{\mathcal{F%
}_{mn}^{\mu \nu }\left( \mathbf{k}\right) +\left( \mathcal{F}_{mn}^{\mu \nu
}\left( \mathbf{k}\right) \right) ^{\ast }}{2},
\end{equation}%
while the Berry curvature is the imaginary part of the quantum geometric
tensor,%
\begin{equation}
\Omega _{nm}^{\mu \nu }\equiv 2\text{Im}\mathcal{F}_{nm}^{\mu \nu }=i\left( 
\mathcal{F}_{mn}^{\mu \nu }\left( \mathbf{k}\right) -\left( \mathcal{F}%
_{mn}^{\mu \nu }\left( \mathbf{k}\right) \right) ^{\ast }\right) .
\end{equation}%
The Wilczek-Zee connection is given by%
\begin{equation}
a_{nm}^{\mu }\left( \mathbf{k}\right) \equiv -i\left\langle \psi _{n}^{\text{%
L}}\right\vert \partial _{k_{\mu }}\left\vert \psi _{nm}^{\text{R}%
}\right\rangle ,
\end{equation}%
while the Berry curvature is given by\cite{KohmotoNH,Zhu,Yin,Lieu,SDLiang}%
\begin{equation}
\Omega _{n}^{\mu \nu }\left( \mathbf{k}\right) =\nabla \times \mathbf{a}%
_{n}\left( \mathbf{k}\right) -i\sum_{n^{\prime }}\left[ a_{nn^{\prime
}}^{\mu },a_{n^{\prime }m}^{\nu }\right] .
\end{equation}%
We note that there are some other generalizations to non-Hermitian systems%
\cite{Silb,Robin,Ye,DongH,Beh,Pal}.

\subsection{Two-band systems}

We consider a two-band system with the index $n=\pm $. The Hamiltonian is
generally given by 
\begin{equation}
H\left( \mathbf{k}\right) =h_{0}\left( \mathbf{k}\right) +\mathbf{\sigma }%
\cdot \mathbf{h}\left( \mathbf{k}\right) ,
\end{equation}%
where 
\begin{align}
h_{0}\left( \mathbf{k}\right) =&h_{0\text{Re}}\left( \mathbf{k}\right) +ih_{0%
\text{Im}}\left( \mathbf{k}\right) , \\
\mathbf{h}\left( \mathbf{k}\right) =&\mathbf{h}_{\text{Re}}\left( \mathbf{k}%
\right) +i\mathbf{h}_{\text{Im}}\left( \mathbf{k}\right) ,
\end{align}%
where $h_{0}\left( \mathbf{k}\right) $ and $\mathbf{h}\left( \mathbf{k}%
\right) $ are complex functions with $h_{0\text{Re}}$, $h_{0\text{Im}}$, $%
\mathbf{h}_{\text{Re}}\left( \mathbf{k}\right) $ and $\mathbf{h}_{\text{Im}%
}\left( \mathbf{k}\right) $ being real functions. The energy is given by%
\begin{equation}
\varepsilon _{\pm }=h_{0}\left( \mathbf{k}\right) \pm \sqrt{\mathbf{h}\left( 
\mathbf{k}\right) ^{2}},
\end{equation}%
where%
\begin{equation}
\mathbf{h}\left( \mathbf{k}\right) ^{2}\equiv h_{x}^{2}\left( \mathbf{k}%
\right) +h_{y}^{2}\left( \mathbf{k}\right) +h_{z}^{2}\left( \mathbf{k}%
\right) .
\end{equation}%
Its right and left eigenvalues are given by\cite{JiangNH}%
\begin{align}
\left\vert \psi _{\pm }^{\text{R}}\right\rangle =&\frac{1}{\sqrt{%
2\varepsilon _{\pm }\left( \varepsilon _{\pm }-h_{z}\right) }}\left(
h_{x}-ih_{y},\varepsilon _{\pm }-h_{z}\right) ^{T}, \\
\left\langle \psi _{\pm }^{\text{L}}\right\vert =&\frac{1}{\sqrt{%
2\varepsilon _{\pm }\left( \varepsilon _{\pm }-h_{z}\right) }}\left(
h_{x}+ih_{y},\varepsilon _{\pm }-h_{z}\right) ,
\end{align}%
which satisfy the biorthogal condition, $\left\langle \psi _{\pm }^{\text{L}%
}\left\vert \psi _{\pm }^{\text{R}}\right\rangle \right. =1$. The
non-Hermitian Berry connection is calculated as\cite{JiangNH}%
\begin{equation}
a_{\pm }^{\mu }=\frac{h_{x}\partial _{k_{\mu }}h_{y}-h_{y}\partial _{k_{\mu
}}h_{x}}{\varepsilon _{\pm }\left( \varepsilon _{\pm }-h_{z}\right) }.
\end{equation}%
The non-Hermitian Berry curvature reads\cite{EzawaNHChern}%
\begin{equation}
\Omega _{\pm }\left( \mathbf{k}\right) =\nabla \times a_{\pm }\left( \mathbf{%
k}\right) =\frac{1}{2\varepsilon _{\pm }^{3/2}}\varepsilon _{\mu \upsilon
\rho }h_{\mu }\partial _{k_{x}}h_{\nu }\partial _{k_{y}}h_{\rho }.
\label{BerryNH}
\end{equation}%
It is further simplified as%
\begin{equation}
\Omega _{\pm }\left( \mathbf{k}\right) =\mp \frac{1}{2}\mathbf{n}\cdot
\left( \partial _{k_{x}}\mathbf{n}\times \partial _{k_{y}}\mathbf{n}\right)
\end{equation}%
with%
\begin{equation}
\mathbf{n\equiv }\frac{\mathbf{h}\left( \mathbf{k}\right) }{\sqrt{\mathbf{h}%
\left( \mathbf{k}\right) ^{2}}}.
\end{equation}%
It is identical to Eq.(\ref{Omegaxy}) for the Hermitian system. Note that 
\begin{equation}
\sqrt{\mathbf{h}\left( \mathbf{k}\right) ^{2}}\neq \left\vert \mathbf{h}%
\left( \mathbf{k}\right) \right\vert
\end{equation}%
for non-Hermitian systems. The non-Hermitian Chern number is defined by\cite%
{KohmotoNH,Fu} 
\begin{equation}
C_{\pm }=\frac{1}{2\pi }\int \Omega _{\pm }\left( \mathbf{k}\right) d^{2}k,
\end{equation}%
where the integration is over the Brillouin zone.

We can check%
\begin{equation}
g_{\pm }^{\mu \nu }\left( \mathbf{k}\right) =\pm \frac{1}{2}\left( \partial
_{k_{\mu }}\mathbf{n}\right) \cdot \left( \partial _{k_{\nu }}\mathbf{n}%
\right)
\end{equation}%
even for the non-Hermitian systems. It is identical to Eq.(\ref{Gmn}).

\subsection{Dirac system with a complex mass}

We consider a Dirac Hamiltonian with a complex mass $m=B+i\gamma $, whose
Hamiltonian is given by%
\begin{equation}
H\left( \mathbf{k}\right) =\lambda \left( \mathbf{k}\times \mathbf{\sigma }%
\right) _{z}+\left( B+i\gamma \right) \sigma _{z},
\end{equation}%
where $\gamma $ is real. The Berry curvature (\ref{BerryNH}) is calculated as%
\begin{equation}
\Omega _{\pm }\left( \mathbf{k}\right) =\pm \frac{\left( m+i\gamma \right)
\lambda ^{2}}{2\left( \lambda ^{2}k^{2}+\left( m+i\gamma \right) ^{2}\right)
^{3/2}},
\end{equation}%
which leads to the Chern number\cite{EzawaHighC}%
\begin{equation}
C_{\pm }=\pm \frac{m+i\gamma }{2\left\vert m+i\gamma \right\vert },
\end{equation}%
where $\left\vert C_{\pm }\right\vert =1/2$.

The quantum metrices are calculated as%
\begin{align}
g_{\pm }^{xx}\left( \mathbf{k}\right) =\mp & \frac{\lambda ^{2}\left( \left(
\lambda k_{y}\right) ^{2}+\left( m+i\gamma \right) ^{2}\right) }{\left(
\left( \lambda k\right) ^{2}+\left( m+i\gamma \right) ^{2}\right) ^{2}}, \\
g_{\pm }^{yy}\left( \mathbf{k}\right) =\mp & \frac{\lambda ^{2}\left( \left(
\lambda k_{x}\right) ^{2}+\left( m+i\gamma \right) ^{2}\right) }{\left(
\left( \lambda k\right) ^{2}+\left( m+i\gamma \right) ^{2}\right) ^{2}}, \\
g_{\pm }^{xy}\left( \mathbf{k}\right) =& \pm \frac{\left( \lambda \right)
^{2}\left( \lambda k_{x}\right) \left( \lambda k_{y}\right) }{\left( \left(
\lambda k\right) ^{2}+\left( m+i\gamma \right) ^{2}\right) ^{2}},
\end{align}%
The quantum volume is calculated as%
\begin{equation}
\int d\mathbf{k}\sqrt{\det g_{n}^{\mu \nu }\left( \mathbf{k}\right) }=\frac{1%
}{8}\left( \frac{\lambda }{m+i\gamma }\right) ^{2}.
\end{equation}%
It is negative for pure imaginary mass $m=0$ and $\gamma \neq 0$.

\section{Quantum information geometry}

\label{SecQIG}

\subsection{Uhlmann quantum geometry for density matrix}

So far, quantum geometry is constructed for the wave function. It means that
it is constructed for pure states. On the other hand, mixed states are
important for quantum information, finite temperature system and open
quantum systems. They are described by the density matrix%
\begin{equation}
\rho \left( \mathbf{k}\right) =\sum_{n=1}^{N}p_{n}\left\vert \psi _{n}\left( 
\mathbf{k}\right) \right\rangle \left\langle \psi _{n}\left( \mathbf{k}%
\right) \right\vert ,  \label{rhopphi}
\end{equation}%
where $p_{n}=\exp \left( -\varepsilon _{n}/k_{\text{B}}T\right) $. Quantum
geometry for it is constructed by Uhlmann\cite%
{Uhl1,Uhl2,Hubner,Caro18,Leon,XHou,Ji,WeiU}. Starting from the fidelity of
for the density matrix, we naturally obtain quantum Fisher information,
which gives the lower boundary of the quantum fluctuation. Hence, quantum
geometry for density matrices is called quantum information geometry.

We start with the Uhlmann fidelity\cite{Hubner2,Uhl3} for the density matrix
defined by%
\begin{equation}
F\left( \mathbf{k},\mathbf{k}^{\prime }\right) =\text{Tr}\sqrt{\sqrt{\rho
\left( \mathbf{k}\right) }\rho (\mathbf{k}^{\prime })\sqrt{\rho \left( 
\mathbf{k}\right) }}.  \label{UhlFidelity}
\end{equation}%
It is symmetric 
\begin{equation}
F\left( \mathbf{k},\mathbf{k}^{\prime }\right) =F\left( \mathbf{k}^{\prime },%
\mathbf{k}\right) ,
\end{equation}%
because $\sqrt{\rho \left( \mathbf{k}\right) }\rho (\mathbf{k}^{\prime })%
\sqrt{\rho \left( \mathbf{k}\right) }$ and $\sqrt{\rho (\mathbf{k}^{\prime })%
}\rho \left( \mathbf{k}\right) \sqrt{\rho (\mathbf{k}^{\prime })}$ have the
same eigenvalues. It is bounded as%
\begin{equation}
0\leq F\left( \mathbf{k},\mathbf{k}^{\prime }\right) \leq 1,
\end{equation}%
where%
\begin{equation}
F\left( \mathbf{k},\mathbf{k}\right) =1,
\end{equation}%
becuase%
\begin{equation}
\text{Tr}\rho \left( \mathbf{k}\right) =1.
\end{equation}%
It reduces to the fidelity (\ref{Fidelity}) for the pure states,%
\begin{align}
F\left( \mathbf{k},\mathbf{k}^{\prime }\right) & =\text{Tr}\sqrt{\rho \left( 
\mathbf{k}\right) \rho (\mathbf{k}^{\prime })}  \notag \\
& =\text{Tr}\sqrt{\left\vert \psi \left( \mathbf{k}\right) \right\rangle
\left\langle \psi \left( \mathbf{k}\right) \right\vert \psi (\mathbf{k}%
^{\prime })\rangle \langle \psi (\mathbf{k}^{\prime }|}  \notag \\
& =\sqrt{\langle \psi \left( \mathbf{k}\right) |\psi (\mathbf{k}^{\prime
})\rangle \langle \psi (\mathbf{k}^{\prime })|\psi \left( \mathbf{k}\right)
\rangle }  \notag \\
& =\left\vert \langle \psi \left( \mathbf{k}\right) |\psi (\mathbf{k}%
^{\prime })\rangle \right\vert .
\end{align}%
With the use of the Uhlmann fidelity (\ref{UhlFidelity}), the Bures distance%
\cite{Bures} is defined by%
\begin{equation}
s_{\text{B}}=\sqrt{1-F\left( \mathbf{k},\mathbf{k}^{\prime }\right) ^{2}}.
\label{Bures}
\end{equation}

We make a purification of the density matrix with the use of the amplitude $%
W $ satisfying 
\begin{equation}
\rho =WW^{\dagger },  \label{rhoWW}
\end{equation}%
where the amplitude $W$ is given by%
\begin{align}
W =&\left( \sqrt{p_{1}}\left\vert \psi _{1}\left( \mathbf{k}\right)
\right\rangle ,\sqrt{p_{2}}\left\vert \psi _{2}\left( \mathbf{k}\right)
\right\rangle ,\cdots ,\sqrt{p_{N}}\left\vert \psi _{N}\left( \mathbf{k}%
\right) \right\rangle \right) , \\
W^{\dagger } =&\left( 
\begin{array}{c}
\sqrt{p_{1}}\left\langle \psi _{1}\left( \mathbf{k}\right) \right\vert \\ 
\sqrt{p_{2}}\left\langle \psi _{2}\left( \mathbf{k}\right) \right\vert \\ 
\vdots \\ 
\sqrt{p_{N}}\left\langle \psi _{N}\left( \mathbf{k}\right) \right\vert%
\end{array}%
\right) .
\end{align}%
Hence, purification is always possible using the spectral decomposition. It
is identical to represent%
\begin{equation}
\rho =\text{Tr}_{\text{E}}\left\vert \Psi \right\rangle \left\langle \Psi
\right\vert  \label{rhoPsi}
\end{equation}%
for the extended Hamiltonian 
\begin{equation}
H_{\text{ext}}\equiv H\otimes H_{\text{E}},
\end{equation}%
by attaching an environment Hamiltonian $H_{\text{E}}$, where the trace is
taken over the environment. It means that the mixed states are represented
by a pure state $\left\vert \Psi \right\rangle $ if we prepare a large
system described by $H_{\text{ext}}$. Comparing Eq.(\ref{rhoWW}) and Eq.(\ref%
{rhoPsi}), $W$ and $W^{\dagger }$ correspond to $\left\vert \Psi
\right\rangle $ and $\left\langle \Psi \right\vert $, respectively.

There is a gauge degrees of freedom $U^{\prime }$ in $W$ as%
\begin{equation}
W=\sqrt{\rho }U^{\prime },
\end{equation}%
where $U$ is a unitary matrix. By using the purification, the Uhlmann
fidelity is rewritten as%
\begin{align}
F\left( \mathbf{k},\mathbf{k}^{\prime }\right) =&\text{Tr}\sqrt{W^{\dagger
}\left( \mathbf{k}\right) W\left( \mathbf{k}^{\prime }\right) W^{\dagger
}\left( \mathbf{k}^{\prime }\right) W\left( \mathbf{k}\right) }  \notag \\
=&\text{Tr}\left\vert W^{\dagger }\left( \mathbf{k}\right) W\left( \mathbf{k}%
^{\prime }\right) \right\vert .
\end{align}%
The Bures distance (\ref{Bures}) is rewritten as%
\begin{align}
s_{\text{B}}^{2}=& 2\left( 1-\text{Tr}\left\vert W^{\dagger }\left( \mathbf{k%
}\right) W\left( \mathbf{k}^{\prime }\right) \right\vert \right)  \notag \\
=& 2\text{Tr}\left( W\left( \mathbf{k}^{\prime }\right) -W\left( \mathbf{k}%
\right) \right) \left( W\left( \mathbf{k}^{\prime }\right) -W\left( \mathbf{k%
}\right) \right) ^{\dagger },
\end{align}%
where we have used the normalization condition for the density matrix,%
\begin{equation}
\text{Tr}\rho \left( \mathbf{k}\right) =\text{Tr}W\left( \mathbf{k}\right)
W^{\dagger }\left( \mathbf{k}\right) =1,
\end{equation}%
and the Uhlmann parallel transport condition\cite{Uhl1}%
\begin{equation}
W^{\dagger }\left( \mathbf{k}\right) W\left( \mathbf{k}^{\prime }\right)
=W^{\dagger }\left( \mathbf{k}^{\prime }\right) W\left( \mathbf{k}\right) >0.
\label{Paral}
\end{equation}%
The Bures distance for the infinitesimal distance $d\mathbf{k}$ is given by
setting $\mathbf{k}^{\prime }=\mathbf{k}+d\mathbf{k}$ as%
\begin{equation}
\left( ds_{\text{B}}\right) ^{2}=2\text{Tr}dWdW^{\dagger }=2\text{Tr}%
\left\vert dW\right\vert ^{2}.  \label{dBures}
\end{equation}

The differential form of the Uhlmann parallel transport condition (\ref%
{Paral}) is given by%
\begin{equation}
W^{\dagger }dW=\left( dW^{\dagger }\right) W.  \label{WdW}
\end{equation}%
It is fulfilled by the ansatz%
\begin{equation}
dW=\frac{1}{2}\mathcal{L}W,\qquad \mathcal{L}^{\dagger }=\mathcal{L}.
\label{dWLW}
\end{equation}%
Indeed, the left and right hand sides of Eq.(\ref{WdW}) are given by 
\begin{align}
W^{\dagger }dW=& \frac{1}{2}W^{\dagger }\mathcal{L}W, \\
\left( dW^{\dagger }\right) W=& \frac{1}{2}\left( \mathcal{L}W\right)
^{\dagger }W=\frac{1}{2}W^{\dagger }\mathcal{L}^{\dagger }W=\frac{1}{2}%
W^{\dagger }\mathcal{L}W,
\end{align}%
and identical with the use of Eq.(\ref{dWLW}). $\mathcal{L}$ is called the
symmetric logarithmic derivative (SLD)\cite{Braun,Braun2,Matteo}.

Then, the Bures distance (\ref{dBures}) reads%
\begin{align}
\left( ds_{\text{B}}\right) ^{2}=& 2\text{Tr}\left\vert dW\right\vert ^{2}=2%
\text{Tr}\left\vert \frac{1}{2}\mathcal{L}W\right\vert ^{2}  \notag \\
=& \frac{1}{2}\text{Tr}\mathcal{L}WW^{\dagger }\mathcal{L}=\frac{1}{2}\text{%
Tr}\mathcal{L}\rho \mathcal{L}  \notag \\
=& \frac{1}{2}\text{Tr}\rho \mathcal{L}^{2}=\frac{1}{2}\left\langle \mathcal{%
L}^{2}\right\rangle .
\end{align}%
By using one form of the SLD,%
\begin{equation}
\mathcal{L}=\mathcal{L}^{\mu }dk_{\mu },  \label{L1form}
\end{equation}%
we obtain the Uhlmann quantum geometric tensor $\mathcal{F}_{\text{U}}^{\mu
\nu }$ as%
\begin{equation}
\left( ds_{\text{B}}\right) ^{2}=\text{Tr}\left[ \rho \sum_{\mu \nu }%
\mathcal{L}^{\mu }\mathcal{L}^{\nu }dk_{\mu }dk_{\nu }\right] \equiv
\sum_{\mu \nu }^{\mu \nu }\mathcal{F}_{\text{U}}^{\mu \nu }dk_{\mu }dk_{\nu }
\end{equation}%
with%
\begin{equation}
\mathcal{F}_{\text{U}}^{\mu \nu }\equiv \text{Tr}\left[ \rho \mathcal{L}%
^{\mu }\mathcal{L}^{\nu }\right] =\left\langle \mathcal{L}^{\mu }\mathcal{L}%
^{\nu }\right\rangle .
\end{equation}%
The Uhlmann quantum geometric tensor $\mathcal{F}_{\text{U}}^{\mu \nu }$ is
decomposed as%
\begin{equation}
\mathcal{F}_{\text{U}}^{\mu \nu }=\mathcal{F}_{\text{QFisher}}^{\mu \nu }+i%
\mathcal{\bar{U}}^{\mu \nu }
\end{equation}%
with the SLD quantum Fisher information\cite{Hels,Matteo,Ji}%
\begin{equation}
\mathcal{F}_{\text{QFisher}}^{\mu \nu }\equiv \frac{1}{2}\text{Tr}\left[
\rho \left\{ \mathcal{L}^{\mu },\mathcal{L}^{\nu }\right\} \right] =\frac{1}{%
2}\left\langle \left\{ \mathcal{L}^{\mu },\mathcal{L}^{\nu }\right\}
\right\rangle ,
\end{equation}%
and the mean Uhlmann curvature\cite{Leon,Ji}%
\begin{equation}
\mathcal{\bar{U}}^{\mu \nu }\equiv -\frac{i}{2}\text{Tr}\left[ \rho \left[ 
\mathcal{L}^{\mu },\mathcal{L}^{\nu }\right] \right] =-\frac{i}{2}%
\left\langle \left[ \mathcal{L}^{\mu },\mathcal{L}^{\nu }\right]
\right\rangle .
\end{equation}%
We note that 
\begin{equation}
\mathcal{U}^{\mu \nu }\equiv -\frac{i}{2}\left[ \mathcal{L}^{\mu },\mathcal{L%
}^{\nu }\right]
\end{equation}
is the Uhlmann curvature, which satisfies%
\begin{equation}
\mathcal{\bar{U}}^{\mu \nu }=\text{Tr}\left[ \rho \mathcal{U}^{\mu \nu }%
\right] =\left\langle \mathcal{U}^{\mu \nu }\right\rangle .
\end{equation}

It follows from Eq.(\ref{rhoWW}) that 
\begin{align}
d\rho =& d\left( WW^{\dagger }\right) =\left( dW\right) W^{\dagger
}+WdW^{\dagger }  \notag \\
=& \frac{1}{2}\mathcal{L}WW^{\dagger }+W\frac{1}{2}\left( \mathcal{L}%
W\right) ^{\dagger }  \notag \\
=& \frac{1}{2}\left( \mathcal{L}\rho +\rho \mathcal{L}\right) =\frac{1}{2}%
\left\{ \mathcal{L},\rho \right\} .  \label{drhoSLD}
\end{align}%
By using Eq.(\ref{rhopphi}) and Eq.(\ref{drhoSLD}), the SLD $\mathcal{L}%
^{\mu }$ is explicitly given by\cite{YMZhang,JLiu,JLiu2,Matteo,Ji}%
\begin{equation}
\mathcal{L}^{\mu }=\sum_{n}\frac{\partial \ln p_{n}}{\partial k_{\mu }}%
\left\vert \psi _{n}\right\rangle \left\langle \psi _{n}\right\vert
+2i\sum_{n\neq m}\frac{p_{m}-p_{n}}{p_{m}+p_{n}}a_{nm}^{\mu }\left\vert \psi
_{n}\right\rangle \left\langle \psi _{m}\right\vert .  \label{SLD}
\end{equation}%
The derivation is shown in Appendix.\ref{ApU}.

\subsection{Classical Fisher information}

The symmetric logarithmic derivative is decomposed into the classical part $%
\mathcal{L}_{\text{C}}^{\mu }$ and the quantum part $\mathcal{L}_{\text{Q}%
}^{\mu }$\ as\cite{Ji}%
\begin{align}
\mathcal{L}^{\mu }=& \mathcal{L}_{\text{C}}^{\mu }+\mathcal{L}_{\text{Q}%
}^{\mu }, \\
\mathcal{L}_{\text{C}}^{\mu }\equiv & \sum_{n}\frac{\partial \ln p_{n}}{%
\partial k_{\mu }}\left\vert \psi _{n}\right\rangle \left\langle \psi
_{n}\right\vert , \\
\mathcal{L}_{\text{Q}}^{\mu }\equiv & 2i\sum_{n\neq m}\frac{p_{m}-p_{n}}{%
p_{m}+p_{n}}a_{nm}^{\mu }\left\vert \psi _{n}\right\rangle \left\langle \psi
_{m}\right\vert .
\end{align}%
When the SLD $\mathcal{L}^{\mu }$ commutes with the density matrix $\rho $,%
\begin{equation}
\left[ \mathcal{L}^{\mu },\rho \right] =0,
\end{equation}%
we have $\mathcal{L}_{\text{Q}}^{\mu }=0$ and $\mathcal{L}^{\mu }=\mathcal{L}%
_{\text{C}}^{\mu }$. Then, we obtain the classical Fisher information based
on the classical part of the SLD as 
\begin{align}
& \frac{1}{2}\text{Tr}\left[ \rho \left\{ \mathcal{L}_{\text{C}}^{\mu },%
\mathcal{L}_{\text{C}}^{\nu }\right\} \right]  \notag \\
=& \frac{1}{2}\sum_{n=1}^{N}p_{n}\left\vert \psi _{n}\left( \mathbf{k}%
\right) \right\rangle \left\langle \psi _{n}\left( \mathbf{k}\right)
\right\vert  \notag \\
& \times (\sum_{n^{\prime }}\frac{\partial \ln p_{n^{\prime }}}{\partial
k_{\mu }}\left\vert \psi _{n^{\prime }}\right\rangle \left\langle \psi
_{n^{\prime }}\right\vert \sum_{n^{\prime \prime }}\frac{\partial \ln
p_{n^{\prime \prime }}}{\partial k_{\nu }}\left\vert \psi _{n^{\prime \prime
}}\right\rangle \left\langle \psi _{n^{\prime \prime }}\right\vert  \notag \\
& +\sum_{n^{\prime \prime }}\frac{\partial \ln p_{n^{\prime \prime }}}{%
\partial k_{\nu }}\left\vert \psi _{n^{\prime \prime }}\right\rangle
\left\langle \psi _{n^{\prime \prime }}\right\vert \sum_{n^{\prime }}\frac{%
\partial \ln p_{n^{\prime }}}{\partial k_{\mu }}\left\vert \psi _{n^{\prime
}}\right\rangle \left\langle \psi _{n^{\prime }}\right\vert )  \notag \\
=& \sum_{n,n^{\prime },n^{\prime \prime }=1}^{N}p_{n}\frac{\partial \ln
p_{n^{\prime }}}{\partial k_{\mu }}\frac{\partial \ln p_{n^{\prime \prime }}%
}{\partial k_{\nu }}\delta _{nn^{\prime }}\delta _{n^{\prime }n^{\prime
\prime }}  \notag \\
=& \sum_{n=1}^{N}p_{n}\frac{\partial \ln p_{n}}{\partial k_{\mu }}\frac{%
\partial \ln p_{n}}{\partial k_{\nu }}\equiv \mathcal{F}_{\text{CFisher}%
}^{\mu \nu }.  \label{CFisher}
\end{align}

On the other hand, the classical part of the mean Uhlmann curvature is zero,%
\begin{equation}
\mathcal{\bar{U}}^{\mu \nu }\equiv -\frac{i}{2}\text{Tr}\left[ \rho \left[ 
\mathcal{L}_{\text{C}}^{\mu },\mathcal{L}_{\text{C}}^{\nu }\right] \right]
=0.
\end{equation}

\subsection{Quantum Cram\'{e}r-Rao inequality}

There is an inequality known as the Quantum Cram\'{e}r-Rao inequality\cite%
{Hels,Braun,Braun2},%
\begin{equation}
\mathcal{F}_{\text{CFisher}}\leq \mathcal{F}_{\text{QFisher}},  \label{CR}
\end{equation}%
where we have defined%
\begin{align}
\mathcal{F}_{\text{CFisher}} =&\sum_{\mu \nu }a_{\mu }\mathcal{F}_{\text{%
CFisher}}^{\mu \nu }a_{\nu }, \\
\mathcal{F}_{\text{QFisher}} =&\sum_{\mu \nu }a_{\mu }\mathcal{F}_{\text{%
QFisher}}^{\mu \nu }a_{\nu }
\end{align}%
for an arbitrary set of parameters $a_{\mu }$. The proof is shown in
Appendix.\ref{ApQCR}.

On the other hand, the classical Cram\'{e}r-Rao inequality states that the
covariance is bounded by the inverse of the classical Fisher information
matrix, 
\begin{align}
&\sum_{\mu \nu }a_{\mu }p_{n}\left( k_{\mu }-\left\langle k_{\mu
}\right\rangle \right) \left( k_{\nu }-\left\langle k_{\nu }\right\rangle
\right) a_{\nu }  \notag \\
\geq &\frac{1}{N}\sum_{\mu \nu }a_{\mu }\left( \mathcal{F}_{\text{CFisher}%
}^{-1}\right) ^{\mu \nu }a_{\nu }.
\end{align}%
Combining the quantum and classical Cram\'{e}r-Rao inequalities, the lower
bound of the covariance is determined by the inverse of the quantum Fisher
information.

\subsection{Quantum Fisher information for a pure state and quantum metric}

By inserting Eq.(\ref{L1form}) to Eq.(\ref{drhoSLD}), we obtain%
\begin{equation}
d\rho =\frac{1}{2}\left\{ \mathcal{L},\rho \right\} =\frac{1}{2}\left\{ 
\mathcal{L}^{\mu }dk_{\mu },\rho \right\} ,
\end{equation}%
which leads to%
\begin{equation}
\frac{\partial \rho }{\partial k_{\mu }}=\frac{1}{2}\left\{ \mathcal{L}^{\mu
},\rho \right\} .  \label{SLDrho}
\end{equation}

We show that the quantum Fisher information reproduces the quantum metric
when the density matrix $\rho $ describes a pure state $\left\vert \psi
_{n}\right\rangle $. By using the relation $\rho ^{2}=\rho $, we have%
\begin{equation}
\frac{\partial \rho }{\partial k_{\mu }}=\frac{\partial \rho ^{2}}{\partial
k_{\mu }}=\rho \frac{\partial \rho }{\partial k_{\mu }}+\frac{\partial \rho 
}{\partial k_{\mu }}\rho =\left\{ \frac{\partial \rho }{\partial k_{\mu }}%
,\rho \right\} .
\end{equation}%
Comparing it with Eq.(\ref{SLDrho}), we have%
\begin{equation}
\mathcal{L}^{\mu }=2\frac{\partial \rho }{\partial k_{\mu }}.
\end{equation}%
Then, the quantum Fisher information is rewritten as%
\begin{equation}
\mathcal{F}_{\text{QFisher}}^{\mu \nu }=2\text{Tr}\left[ \rho \left\{ \frac{%
\partial \rho }{\partial k_{\mu }},\frac{\partial \rho }{\partial k_{\nu }}%
\right\} \right] =2\left\langle \left\{ \frac{\partial \rho }{\partial
k_{\mu }},\frac{\partial \rho }{\partial k_{\nu }}\right\} \right\rangle .
\label{FisherRho}
\end{equation}%
By inserting 
\begin{equation}
\rho =\left\vert \psi _{n}\right\rangle \left\langle \psi _{n}\right\vert ,
\end{equation}%
the quantum Fisher information is reduced to the quantum metric%
\begin{align}
\mathcal{F}_{\text{QFisher}}^{\mu \nu } =&2\text{Tr}\frac{\partial
\left\langle \psi _{n}\right\vert }{\partial k_{\mu }}\frac{\partial
\left\vert \psi _{n}\right\rangle }{\partial k_{\nu }}-\frac{\partial
\left\langle \psi _{n}\right\vert }{\partial k_{\mu }}\left\vert \psi
_{n}\right\rangle \left\langle \psi _{n}\right\vert \frac{\partial
\left\vert \psi _{n}\right\rangle }{\partial k_{\nu }}  \notag \\
&+\frac{\partial \left\langle \psi _{n}\right\vert }{\partial k_{\nu }}\frac{%
\partial \left\vert \psi _{n}\right\rangle }{\partial k_{\mu }}-\frac{%
\partial \left\langle \psi _{n}\right\vert }{\partial k_{\nu }}\left\vert
\psi _{n}\right\rangle \left\langle \psi _{n}\right\vert \frac{\partial
\left\vert \psi _{n}\right\rangle }{\partial k_{\mu }}  \notag \\
=&2\left( \mathcal{F}_{n}^{\mu \nu }+\mathcal{F}_{n}^{\nu \mu }\right)
=4g_{n}^{\mu \nu }.  \label{FIpure}
\end{align}%
while the mean Uhlmann curvature is reduced to the Berry curvature%
\begin{align}
\mathcal{\bar{U}}^{\mu \nu } =&-2i\text{Tr}\frac{\partial \left\langle \psi
_{n}\right\vert }{\partial k_{\mu }}\frac{\partial \left\vert \psi
_{n}\right\rangle }{\partial k_{\nu }}-\frac{\partial \left\langle \psi
_{n}\right\vert }{\partial k_{\mu }}\left\vert \psi _{n}\right\rangle
\left\langle \psi _{n}\right\vert \frac{\partial \left\vert \psi
_{n}\right\rangle }{\partial k_{\nu }}  \notag \\
&-\frac{\partial \left\langle \psi _{n}\right\vert }{\partial k_{\nu }}\frac{%
\partial \left\vert \psi _{n}\right\rangle }{\partial k_{\mu }}+\frac{%
\partial \left\langle \psi _{n}\right\vert }{\partial k_{\nu }}\left\vert
\psi _{n}\right\rangle \left\langle \psi _{n}\right\vert \frac{\partial
\left\vert \psi _{n}\right\rangle }{\partial k_{\mu }}  \notag \\
=&-2i\left( \mathcal{F}_{n}^{\mu \nu }+\mathcal{F}_{n}^{\nu \mu }\right)
=i4\Omega _{n}^{\mu \nu }.  \label{UhlPure}
\end{align}%
Then, the quantum Fisher information is reduced to the quantum metric, while
the mean Uhlmann curvature is reduced to the Berry curvature. Thus, the
Uhlmann quantum geometry is a generalization of quantum geometry. The
derivation of Eq.(\ref{FIpure}) is shown in Appendix.\ref{ApQFpure}.

\subsection{Fluctuation-dissipation theorem}

There is a fluctuation-dissipation theorem\cite{Ji},%
\begin{equation}
-\frac{1}{2\hbar }g_{\text{Bures}}^{\mu \nu }=\frac{\tanh ^{2}\frac{\hbar
\omega }{2k_{\text{B}}T}}{1-e^{-\hbar \omega /k_{\text{B}}T}}\frac{\chi
^{\mu \nu }}{\left( \hbar \omega \right) ^{2}},
\end{equation}%
where $\chi ^{\mu \nu }$ is the susceptibility. It is reduced to Eq.(\ref%
{FlucD}) for the pure state.

\subsection{Quantum geometry at thermal equilibrium}

At a thermal equilibrium, the probability is given by%
\begin{equation}
p_{n}=\exp \left[ -\beta \varepsilon _{n}\right] ,
\end{equation}%
where $\beta \equiv 1/k_{\text{B}}T$ is the inverse temperature. We consider
a two-band system with $n=\pm $. The SLD operator (\ref{SLD}) is given by

\begin{align}
\mathcal{L}^{\mu } =&-\beta \sum_{n=\pm }\frac{\partial \varepsilon _{n}}{%
\partial k_{\mu }}\left\vert \psi _{n}\right\rangle \left\langle \psi
_{n}\right\vert  \notag \\
&\hspace{-0.4in}-2i\tanh \frac{\Delta \varepsilon }{2}(a_{-+}^{\mu
}\left\vert \psi _{-}\right\rangle \left\langle \psi _{+}\right\vert
-a_{+-}^{\mu }\left\vert \psi _{+}\right\rangle \left\langle \psi
_{-}\right\vert )
\end{align}%
with%
\begin{equation}
\Delta \varepsilon =\varepsilon _{+}-\varepsilon _{-},
\end{equation}%
where we have used%
\begin{equation}
\frac{e^{-\beta \varepsilon _{+}}-e^{-\beta \varepsilon _{-}}}{e^{-\beta
\varepsilon _{+}}+e^{-\beta \varepsilon _{-}}}=\tanh \frac{\Delta
\varepsilon }{2}.
\end{equation}%
It is rewritten as%
\begin{equation}
\mathcal{L}^{\mu }=\Psi ^{\dagger }\left( 
\begin{array}{cc}
-\beta \frac{\partial \varepsilon _{+}}{\partial k_{\mu }} & 2i\tanh \frac{%
\beta \Delta \varepsilon }{2}a_{+-}^{\mu } \\ 
-2i\tanh \frac{\beta \Delta \varepsilon }{2}a_{-+}^{\mu } & -\beta \frac{%
\partial \varepsilon _{-}}{\partial k_{\mu }}%
\end{array}%
\right) \Psi
\end{equation}%
with%
\begin{equation}
\Psi \equiv \left( 
\begin{array}{c}
\left\langle \psi _{+}\right\vert \\ 
\left\langle \psi _{-}\right\vert%
\end{array}%
\right) .
\end{equation}%
The Uhlmann curvature is calculated as%
\begin{equation}
\mathcal{U}^{\mu \nu }=-\frac{i}{2}\left[ \mathcal{L}^{\mu },\mathcal{L}%
^{\nu }\right] =2\tanh \frac{\beta \Delta \varepsilon }{2}\left( 
\begin{array}{cc}
\mathcal{U}^{++} & \mathcal{U}^{+-} \\ 
\mathcal{U}^{+-} & -\mathcal{U}^{++}%
\end{array}%
\right)
\end{equation}%
with%
\begin{align}
\mathcal{U}^{++} =&\Omega ^{\mu \nu }\tanh \frac{\beta \Delta \varepsilon }{2%
}, \\
\mathcal{U}^{+-} =&\frac{\partial \varepsilon _{+}}{\partial k_{\mu }}%
a_{+-}^{\nu }-\frac{\partial \varepsilon _{+}}{\partial k_{\nu }}a_{+-}^{\mu
}.
\end{align}%
$\mathcal{U}^{++}$ recovers the Berry curvature at the low temperature
limit, $\beta \rightarrow 0$,%
\begin{equation}
\mathcal{U}^{++}=-\lim_{\beta \rightarrow 0}2\Omega ^{\mu \nu }\tanh ^{2}%
\frac{\beta \Delta \varepsilon }{2}=-2\Omega ^{\mu \nu }.
\end{equation}

We define the quantum Fisher information density by%
\begin{equation}
F_{\text{QFisher}}^{\mu \nu }\equiv \frac{1}{2}\left\{ \mathcal{L}^{\mu },%
\mathcal{L}^{\nu }\right\} ,
\end{equation}%
which satisfies%
\begin{equation}
\mathcal{F}_{\text{QFisher}}^{\mu \nu }\equiv \text{Tr}\left[ \rho F_{\text{%
QFisher}}^{\mu \nu }\right] =\left\langle F_{\text{QFisher}}^{\mu \nu
}\right\rangle .
\end{equation}%
It is explicitly given by%
\begin{equation}
F_{\text{QFisher}}^{\mu \nu }=\left( \frac{\partial \varepsilon _{+}}{%
\partial k_{\mu }}\frac{\partial \varepsilon _{+}}{\partial k_{\nu }}%
+4g^{\mu \nu }\tanh ^{2}\frac{\beta \Delta \varepsilon }{2}\right) I_{2},
\end{equation}%
where $I_{2}$ is the 2 by 2 identity matrix.

\section{$X$-wave magnets}

\label{SecX}

\subsection{Fermi surface symmetry}

\begin{figure*}[t]
\centerline{\includegraphics[width=0.98\textwidth]{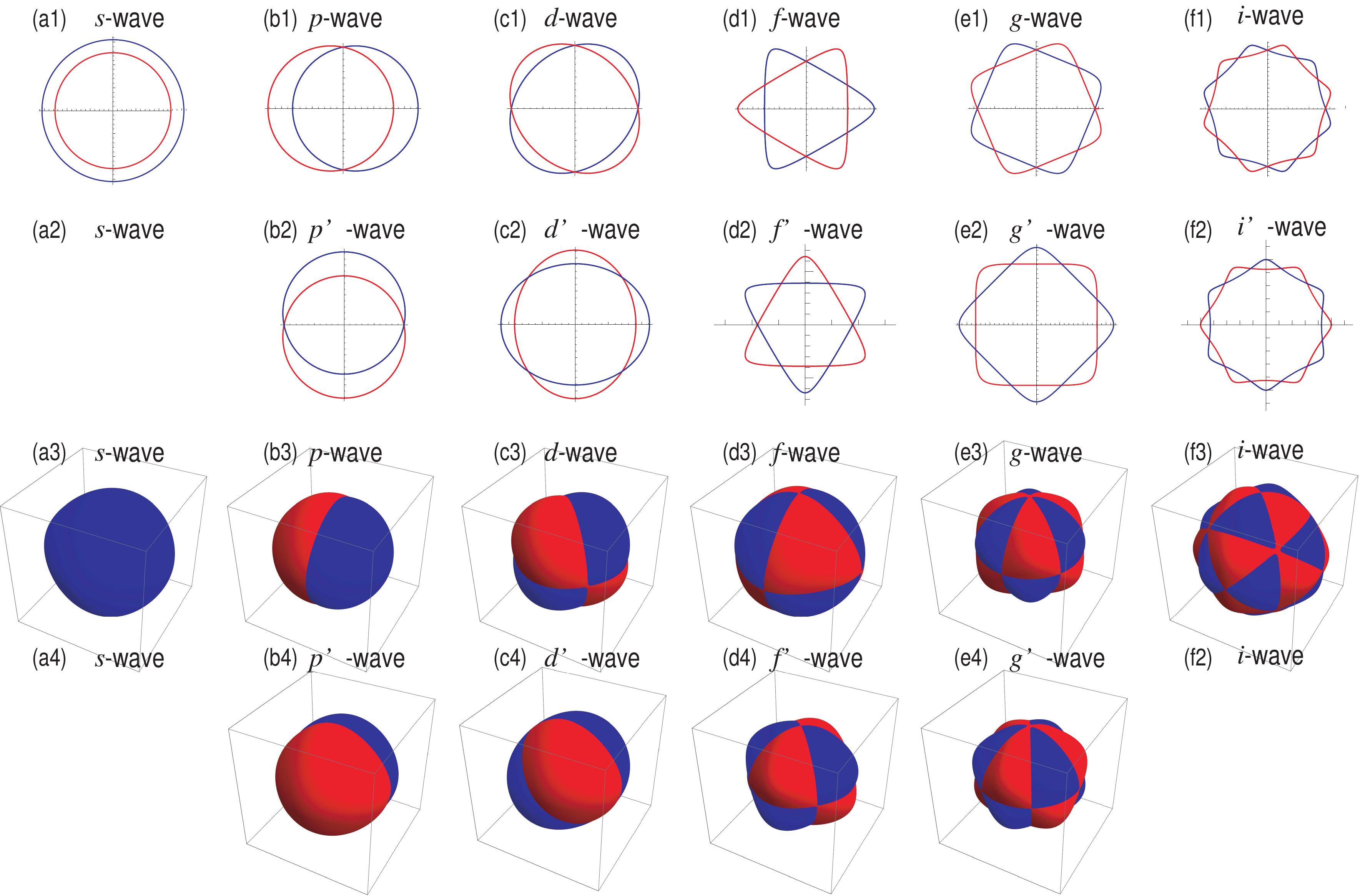}}
\caption{Fermi surfaces in two and three dimensions. (a1), (a3) $s$-wave
magnet; (b1), (b3) $p$-wave magnet; (c1), (c3) $d$-wave altermagnet; (c2) $%
d^{\prime }$-wave altermagnet; (d1), (d3) $f$-wave magnet; (d2) $f^{\prime }$%
-wave magnet; (e1), (e3) $g$-wave altermagnet; (e2) $g^{\prime }$-wave$\ $%
altermagnet; ((f1) and (f3) $i$-wave altermagnet. Red (blue) curves indicate
up (down)-spin Fermi surfaces.}
\label{FigSurface}
\end{figure*}

The Fermi surface of electrons coupled with a ferromagnet is known to have
the $s$-wave symmetry as shown in Fig.\ref{FigSurface}(a). Recently proposed
altermagnets generalize it to Fermi surfaces possessing higher-wave
symmetries\cite{SmejX,SmejX2}. The Fermi surface has $0,1,2,3,4$ and $6$\
nodes for the $s$-wave, $p$-wave, $d$-wave, $f$-wave, $g$-wave and $i$-wave
symmetry, respectively. The $d$-wave altermagnet has the Fermi surface with
the $d$-wave symmetry as shown in Fig.\ref{FigSurface}(c1). In the similar
way, the Fermi surface of the $g$-wave altermagnet is shown in Fig.\ref%
{FigSurface}(e1) and that of the $i$-wave alatermagnet is shown in Fig.\ref%
{FigSurface}(f1). Altermagnets break time-reversal symmetry. On the other
hand, the $p$-wave magnet preserves time-reversal symmetry. Its Fermi
surface has the $p$-wave symmetry as shown in Fig.\ref{FigSurface}(b1). In a
similar way, the $f$-wave magnet has the Fermi surface with the $f$-wave
symmetry as shown in Fig.\ref{FigSurface}(d1). We call them the $X$-wave
magnets with $X=p,d,f,g,i$. We note that there are no $h$-wave magnets
because of the incompatibility between the five-fold rotational symmetry and
the lattice symmetry. 

The simplest expressions on magnetic terms with higher symmetries in two
dimensions are summarized as follows. The $X$-wave magnet is characterized
by a function $f_{X}^{2\text{D}}\left( \mathbf{k}\right) $, which reads\cite%
{SmejX,SmejX2,GI,Planar,MTJ},%
\begin{align}
f_{s}^{2\text{D}}\left( \mathbf{k}\right) & =1, \\
f_{p}^{2\text{D}}\left( \mathbf{k}\right) & =ak_{x}=ak\cos \phi ,
\label{f2p} \\
f_{d}^{2\text{D}}\left( \mathbf{k}\right) & =2a^{2}k_{x}k_{y}=a^{2}k^{2}\sin
2\phi ,  \label{f2d} \\
f_{f}^{2\text{D}}\left( \mathbf{k}\right) & =a^{3}k_{x}\left(
k_{x}^{2}-3k_{y}^{2}\right) =a^{3}k^{3}\cos 3\phi , \\
f_{g}^{2\text{D}}\left( \mathbf{k}\right) & =4a^{4}k_{x}k_{y}\left(
k_{x}^{2}-k_{y}^{2}\right) =a^{4}k^{4}\sin 4\phi , \\
f_{i}^{2\text{D}}\left( \mathbf{k}\right) & =2a^{6}k_{x}k_{y}\left(
3k_{x}^{2}-k_{y}^{2}\right) \left( k_{x}^{2}-3k_{y}^{2}\right)
=a^{6}k^{6}\sin 6\phi ,
\end{align}%
where $k_{x}=k\cos \phi $, $k_{y}=k\sin \phi $. We note that the $d$-wave
altermagnet described by the function $f_{d}^{2\text{D}}\left( \mathbf{k}%
\right) $ is commonly called the $d_{xy}$-wave altermagnet. The $X$-wave
magnet has $N_{X}$ nodes in the band structure, where $N_{X}=1,2,3,4,6$ for $%
X=p,d,f,g,i$, respectively.

There is another type of the $X^{\prime }$-wave magnet in two dimensions
characterized by a function $f_{X^{\prime }}^{2\text{D}}\left( \mathbf{k}%
\right) $ such that 
\begin{align}
f_{p^{\prime }}^{2\text{D}}\left( \mathbf{k}\right) =& ak_{y}=ak\sin \phi ,
\\
f_{d^{\prime }}^{2\text{D}}\left( \mathbf{k}\right) =& a^{2}\left(
k_{x}^{2}-k_{y}^{2}\right) =a^{2}k^{2}\cos 2\phi , \\
f_{f^{\prime }}^{2\text{D}}\left( \mathbf{k}\right) =& a^{3}k_{y}\left(
3k_{x}^{2}-k_{y}^{2}\right) =a^{3}k^{3}\sin 3\phi , \\
f_{g^{\prime }}^{2\text{D}}\left( \mathbf{k}\right) =& a^{4}\left(
k_{x}^{2}-k_{y}^{2}-2k_{x}k_{y}\right) \left(
k_{x}^{2}-k_{y}^{2}+2k_{x}k_{y}\right)  \notag \\
=& a^{4}k^{4}\cos 4\phi , \\
f_{i^{\prime }}^{2\text{D}}\left( \mathbf{k}\right) =& 2a^{6}\left(
k_{x}^{2}-k_{y}^{2}\right) \left(
k_{x}^{4}+k_{y}^{4}-14k_{x}^{2}k_{y}^{2}\right)  \notag \\
=& a^{6}k^{6}\cos 6\phi ,
\end{align}%
where $X^{\prime }=p^{\prime },d^{\prime },f^{\prime },g^{\prime },i^{\prime
}$. We note that the $d^{\prime }$-wave altermagnet described by the
function $f_{d^{\prime }}^{2\text{D}}\left( \mathbf{k}\right) $ is commonly
called the $d_{x^{2}-y^{2}}$-wave altermagnet, which is obtained by rotating
the $d_{xy}$ altermagnet by 45 degrees.

The $X$-wave magnet and the $X^{\prime }$-wave magnet are summarized by the
function in the form of%
\begin{equation}
f_{X,X^{\prime }}^{2\text{D}}\left( \mathbf{k}\right) =\left( ak\right)
^{N_{X}}\sin N_{X}\phi ,
\end{equation}%
or%
\begin{equation}
f_{X,X^{\prime }}^{\text{2D}}\left( \mathbf{k}\right) =\left( ak\right)
^{N_{X}}\cos N_{X}\phi .
\end{equation}

The magnetic terms of the $X$-wave magnet in three dimensions are given by 
\begin{align}
f_{d}^{3\text{D}}\left( \mathbf{k}\right) & =a^{2}k_{z}\left(
k_{x}+k_{y}\right) , \\
f_{f}^{3\text{D}}\left( \mathbf{k}\right) &
=2a^{3}k_{x}k_{y}k_{z}=a^{2}k^{2}\cos \theta \sin 2\phi , \\
f_{g}^{3\text{D}}\left( \mathbf{k}\right) & =a^{4}k_{z}k_{x}\left(
k_{x}^{2}-3k_{y}^{2}\right) =a^{3}k^{3}\cos \theta \cos 3\phi , \\
f_{i}^{3\text{D}}\left( \mathbf{k}\right) & =a^{6}\left(
k_{x}^{2}-k_{y}^{2}\right) \left( k_{y}^{2}-k_{z}^{2}\right) \left(
k_{z}^{2}-k_{x}^{2}\right) .  \label{EqY}
\end{align}

There are relations between the $X$-wave magnet between two and three
dimensions,%
\begin{align}
f_{f}^{3\text{D}}\left( \mathbf{k}\right) =a& k_{z}f_{d}^{2\text{D}}\left( 
\mathbf{k}\right) , \\
f_{g}^{3\text{D}}\left( \mathbf{k}\right) =a& k_{z}f_{f}^{2\text{D}}\left( 
\mathbf{k}\right) .
\end{align}%
There are also $X^{\prime }$-wave magnets in three dimensions\cite{Roig},%
\begin{align}
f_{d^{\prime }}^{3\text{D}}\left( \mathbf{k}\right) =& a^{2}k_{y}\left(
k_{x}+k_{z}\right) , \\
f_{f^{\prime }}^{3\text{D}}\left( \mathbf{k}\right) =& a^{3}k_{z}\left(
k_{x}^{2}-k_{y}^{2}\right) =a^{2}k^{2}\cos \theta \cos 2\phi , \\
f_{g^{\prime }}^{3\text{D}}\left( \mathbf{k}\right) =& a^{4}k_{z}k_{y}\left(
3k_{x}^{2}-k_{y}^{2}\right) =a^{3}k^{3}\cos \theta \sin 3\phi .
\end{align}

We set $a=1$ in the following.

\subsection{Model Hamiltonian}

We consider the two-band Hamiltonian described\cite{Planar} by%
\begin{equation}
H\left( \mathbf{k}\right) =\frac{\hbar ^{2}k^{2}}{2m}+Jf_{X}\left( \mathbf{k}%
\right) \mathbf{n}\cdot \mathbf{\sigma }+\lambda \left( k_{x}\sigma
_{y}-k_{y}\sigma _{x}\right) +\mathbf{B}\cdot \mathbf{\sigma }.
\label{ModelHamil}
\end{equation}%
The first term represents the kinetic energy, making the system metallic.
The second term represents the $X$-wave term with the $X$-wave function $%
f_{X}\left( \mathbf{k}\right) $, where $\mathbf{n}$ is the direction of the
spin-splitting of the band structure, and $J$ is the coupling constant
induced by the $X$-wave magnet. The third term represents the Rashba
interaction introduced by making an interface between the $X$-wave magnet
and the substrate, where $\lambda $ is the magnitude of the Rashba
interaction. The fourth term is the magnetic field term. The Rashba
interaction is introduced by placing an altermagnet on the substrate\cite%
{SmejRev,SmejX,SmejX2,Zu2023,Gho,Li2023,EzawaAlter,EzawaMetricC}.

\subsection{Symmtery}

We summarize symmetry properties of the $X$-wave magnets\cite{BhoRev}.

\subsubsection{Spin diagonal case}

We consider the Hamiltonian (\ref{ModelHamil}) by setting $\lambda =0$, $%
\mathbf{n}=\left( 0,0,1\right) $ and $\mathbf{B}=0$, where the spin is a
good quantum number, $\sigma _{z}=s=\pm 1$. The Hamiltonian is diagonal with
respect to the spin $\mathbf{\sigma }$, where $s=\pm 1$. Let us use $%
s=\uparrow \downarrow $\ within indices and $s=\pm 1$\ in equations.

The energy is given by%
\begin{equation}
\varepsilon _{s}=\frac{\hbar ^{2}k^{2}}{2m}+sf_{X}.
\end{equation}%
Time-reversal symmetry is defined by%
\begin{equation}
T\varepsilon _{s}\left( \mathbf{k}\right) T^{-1}=\varepsilon _{-s}\left( -%
\mathbf{k}\right) .
\end{equation}%
Inversion symmetry is defined by%
\begin{equation}
P\varepsilon _{s}\left( \mathbf{k}\right) P^{-1}=\varepsilon _{s}\left( -%
\mathbf{k}\right) .
\end{equation}%
When there are both symmetries, we have%
\begin{equation}
PT\varepsilon _{s}\left( \mathbf{k}\right) \left( PT\right)
^{-1}=\varepsilon _{-s}\left( \mathbf{k}\right) .
\end{equation}%
Hence, the band with up and down spins are degenerate perfectly. In this
sense, the breaking of time-reversal symmetry or inversion symmetry is
necessary for spin-splitting band structure. In the $d$-wave, the $g$-wave
and the $i$-wave altermagnets, time-reversal symmetry is broken but
inversion symmetry is preserved. On the other hand, in the $p$-wave and the $%
f$-wave magnets, inversion symmetry is broken but time-reversal symmetry is
preserved. Hence, the spin splitting occurs in all $X$-wave magnets. These
properties are summarized in the following table,

\begin{equation}
\begin{tabular}{|l|l|l|l|l|l|}
\hline
& $p$ & $d$ & $f$ & $g$ & $i$ \\ \hline
Time-reversal & Yes & No & Yes & No & No \\ \hline
Inversion symmetry & No & Yes & No & Yes & Yes \\ \hline
Spin splitting & Yes & Yes & Yes & Yes & Yes \\ \hline
Aletermagnet & No & Yes & No & Yes & Yes \\ \hline
\end{tabular}%
.
\end{equation}

\subsubsection{Spin nondiagonal case}

In the presence of the Rashba interaction, the system is not diagonal with
respect to spin $\sigma _{z}$. In this case, time-reversal symmetry is
defined by%
\begin{equation}
TH\left( \mathbf{k}\right) T^{-1}=H\left( -\mathbf{k}\right)
\end{equation}%
with the time-reversal symmetry operator $T=i\sigma _{y}K$, where $K$ is the
complex conjugation operator. Thus, $T$ is an anti-unitary operator.
Time-reversal symmetry is broken%
\begin{equation}
TH\left( \mathbf{k}\right) T^{-1}=-H\left( -\mathbf{k}\right)
\end{equation}%
for the $d$-wave, the $g$-wave and the $i$-wave altermagnets. On the other
hand, time-reversal symmetry is preserved for the $p$-wave and the $f$-wave
magnets.

Inversion symmetry is defined by%
\begin{equation}
PH\left( \mathbf{k}\right) P^{-1}=H\left( -\mathbf{k}\right)
\end{equation}%
with the inversion symmetry operator $P=\sigma _{z}$, which is a unitary
operator. Inversion symmetry is preserved for $s$-wave, $d$-wave and $g$%
-wave magnets. On the other hand, it is broken%
\begin{equation}
PH\left( \mathbf{k}\right) P^{-1}=-H\left( -\mathbf{k}\right)
\end{equation}%
for $p$-wave, $f$-wave and $i$-wave magnets.

PT symmetry is a combination operator of inversion symmetry and
time-reversal symmetry. If there is PT symmetry%
\begin{equation}
PTH\left( \mathbf{k}\right) \left( PT\right) ^{-1}=H\left( \mathbf{k}\right)
,
\end{equation}%
the bands are two-fold degenerate. All $X$-wave magnets break PT symmetry,
and hence, the band structure is spin split.

The $X$-wave symmetric magnets with $N_{X}$ are characterized by the spin
group, which a combination group of spatial group and spin. The $\frac{2\pi 
}{N_{X}}$-rotation along the $z$ axis of the momentum is given by%
\begin{equation}
R_{z}\left( \frac{\pi }{N_{X}}\right) :\left( 
\begin{array}{c}
k_{x} \\ 
k_{y}%
\end{array}%
\right) \mapsto \left( 
\begin{array}{cc}
\cos \frac{\pi }{N_{X}} & \sin \frac{\pi }{N_{X}} \\ 
-\sin \frac{\pi }{N_{X}} & \cos \frac{\pi }{N_{X}}%
\end{array}%
\right) \left( 
\begin{array}{c}
k_{x} \\ 
k_{y}%
\end{array}%
\right) ,
\end{equation}%
where $N_{X}=1,2,3,4,6$ for $X=p,d,f,g,i$, respectively. The $X$-wave
symmetric magnets with $N_{X}$ in two dimensions have a combinational
symmetry of $\frac{2\pi }{N_{X}}$-rotation and time-reversal symmetry%
\begin{equation}
\left[ R_{z}\left( \frac{\pi }{N_{X}}\right) T\right] H\left( \mathbf{k}%
\right) \left[ R_{z}\left( \frac{\pi }{N_{X}}\right) T\right] ^{-1}=H\left( 
\mathbf{k}\right) .
\end{equation}%
The $X$-wave symmetric magnets in three dimensions also have a combinational
symmetry of spatial group and time-reversal symmetry.

\subsection{Quantum geometry of $X$-wave magnets}

We take $\mathbf{n}=\left( 0,0,1\right) $ and $\mathbf{B}=\left(
0,0,B\right) $ in the Hamiltonian (\ref{ModelHamil}). The Berry curvature is
given by%
\begin{equation}
\Omega _{\pm }^{xy}=\mp \frac{\lambda ^{2}\left( B+Jf_{X}-Jk\partial
_{k}f_{X}\right) }{2\left( \lambda ^{2}k^{2}+\left( B+Jf_{X}\right) \right)
^{3/2}},
\end{equation}%
while the quantum metrices are given by%
\begin{align}
g_{\pm }^{xx}=& \mp \frac{\lambda ^{2}\left( \lambda ^{2}k_{y}^{2}+\left(
B+Jf_{X}\right) ^{2}\right) }{2\left( \lambda ^{2}k^{2}+\left(
B+Jf_{X}\right) ^{2}\right) ^{2}}, \\
g_{\pm }^{xy}=& g_{\pm }^{yx}=\mp \frac{-\lambda ^{4}k_{x}k_{y}}{2\left(
\lambda ^{2}k^{2}+\left( B+Jf_{X}\right) ^{2}\right) ^{2}}, \\
g_{\pm }^{yy}=& \mp \frac{\lambda ^{2}\left( \lambda ^{2}k_{x}^{2}+\left(
B+Jf_{X}\right) ^{2}\right) }{2\left( \lambda ^{2}k^{2}+\left(
B+Jf_{X}\right) ^{2}\right) ^{2}}.
\end{align}

\subsection{Zeeman quantum geometry of $X$-wave magnets}

By inserting Eq.(\ref{ModelHamil}) to Eq.(\ref{Z}), the Zeeman Berry
curvatures are calculated as%
\begin{align}
\mathcal{Z}_{+-}^{xx}=& \frac{\partial n_{x}}{\partial k_{x}}=\lambda k_{y}%
\frac{\lambda ^{2}k_{x}+\left( B+Jf_{X}\right) J\partial _{k_{x}}f_{X}}{%
\left( \lambda ^{2}k^{2}+\left( B+Jf_{X}\right) ^{2}\right) ^{3/2}}, \\
\mathcal{Z}_{+-}^{yy}=& \frac{\partial n_{y}}{\partial k_{y}}=-\lambda k_{x}%
\frac{\lambda ^{2}k_{y}+\left( B+Jf_{X}\right) J\partial _{k_{y}}f_{X}}{%
\left( \lambda ^{2}k^{2}+\left( B+Jf_{X}\right) ^{2}\right) ^{3/2}}, \\
\mathcal{Z}_{+-}^{xy}=& \frac{\partial n_{y}}{\partial k_{x}}=\frac{\lambda 
}{\sqrt{\lambda ^{2}k^{2}+\left( B+Jf_{X}\right) ^{2}}}  \notag \\
& -\lambda k_{x}\frac{2\lambda ^{2}k_{x}+2\left( B+f_{X}\right) J\partial
_{k_{x}}f_{X}}{2\left( \lambda ^{2}k^{2}+\left( B+Jf_{X}\right) ^{2}\right)
^{3/2}}, \\
\mathcal{Z}_{+-}^{yx}=& \frac{\partial n_{x}}{\partial k_{y}}=-\frac{\lambda 
}{\sqrt{\lambda ^{2}k^{2}+\left( B+Jf_{X}\right) ^{2}}}  \notag \\
& +\lambda k_{y}\frac{2\lambda ^{2}k_{y}+2\left( B+f_{X}\right) J\partial
_{k_{y}}f_{X}}{2\left( \lambda ^{2}k^{2}+\left( B+Jf_{X}\right) ^{2}\right)
^{3/2}}, \\
\mathcal{Z}_{+-}^{xz}=& \frac{\partial n_{z}}{\partial k_{x}}=\lambda ^{2}%
\frac{-k_{x}\left( B+Jf_{X}\right) +Jk^{2}J\partial _{k_{x}}f_{X}}{2\left(
\lambda ^{2}k^{2}+\left( B+Jf_{X}\right) ^{2}\right) ^{3/2}}, \\
\mathcal{Z}_{+-}^{yz}=& \frac{\partial n_{z}}{\partial k_{y}}=\lambda ^{2}%
\frac{-k_{y}\left( B+Jf_{X}\right) +Jk^{2}J\partial _{k_{y}}f_{X}}{2\left(
\lambda ^{2}k^{2}+\left( B+Jf_{X}\right) ^{2}\right) ^{3/2}}.
\end{align}

By inserting Eq.(\ref{ModelHamil}) to Eq.(\ref{Q}), the Zeeman quantum
metrices are calculated as%
\begin{align}
\mathcal{Q}_{+-}^{xx}=& \frac{n_{y}\frac{\partial n_{z}}{\partial k_{x}}%
-n_{z}\frac{\partial n_{y}}{\partial k_{x}}}{2}=-\frac{\lambda \left(
B+Jf_{X}-Jk_{x}\partial _{k_{x}}f_{X}\right) }{2\left( \lambda
^{2}k^{2}+\left( B+Jf_{X}\right) ^{2}\right) }, \\
\mathcal{Q}_{+-}^{yy}=& \frac{n_{z}\frac{\partial n_{x}}{\partial k_{y}}%
-n_{x}\frac{\partial n_{z}}{\partial k_{y}}}{2}=-\frac{\lambda \left(
B+Jf_{X}-Jk_{y}\partial _{k_{y}}f_{X}\right) }{2\left( \lambda
^{2}k^{2}+\left( B+Jf_{X}\right) ^{2}\right) }, \\
\mathcal{Q}_{+-}^{xy}=& \frac{n_{z}\frac{\partial n_{x}}{\partial k_{x}}%
-n_{x}\frac{\partial n_{z}}{\partial k_{x}}}{2}=\frac{\lambda k_{y}J\partial
_{k_{x}}f_{X}}{2\left( \lambda ^{2}k^{2}+\left( B+Jf_{X}\right) ^{2}\right) }%
, \\
\mathcal{Q}_{+-}^{yx}=& \frac{n_{y}\frac{\partial n_{z}}{\partial k_{y}}%
-n_{z}\frac{\partial n_{y}}{\partial k_{y}}}{2}=\frac{\lambda k_{x}J\partial
_{k_{y}}f_{X}}{2\left( \lambda ^{2}k^{2}+\left( B+Jf_{X}\right) ^{2}\right) }%
, \\
\mathcal{Q}_{+-}^{xz}=& \frac{n_{x}\frac{\partial n_{y}}{\partial k_{x}}%
-n_{y}\frac{\partial n_{x}}{\partial k_{x}}}{2}=-\frac{\lambda ^{2}k_{y}}{%
2\left( \lambda ^{2}k^{2}+\left( B+Jf_{X}\right) ^{2}\right) }, \\
\mathcal{Q}_{+-}^{yz}=& \frac{n_{x}\frac{\partial n_{y}}{\partial k_{y}}%
-n_{y}\frac{\partial n_{x}}{\partial k_{y}}}{2}=\frac{\lambda ^{2}k_{x}}{%
2\left( \lambda ^{2}k^{2}+\left( B+Jf_{X}\right) ^{2}\right) }.
\end{align}

The diagonal spin quantum metrices are calculated as%
\begin{align}
\mathcal{S}_{+-}^{xx} =&\frac{\lambda ^{2}k_{x}^{2}+\left( B+Jf_{X}\right)
^{2}}{\lambda ^{2}k^{2}+\left( B+Jf_{X}\right) ^{2}}, \\
\mathcal{S}_{+-}^{yy} =&\frac{\lambda ^{2}k_{y}^{2}+\left( B+Jf_{X}\right)
^{2}}{\lambda ^{2}k^{2}+\left( B+Jf_{X}\right) ^{2}}, \\
\mathcal{S}_{+-}^{zz} =&\frac{\lambda ^{2}k^{2}}{\lambda ^{2}k^{2}+\left(
B+Jf_{X}\right) ^{2}}.
\end{align}

The off-diagonal spin quantum metrices are calculated as%
\begin{align}
\mathcal{S}_{+-}^{xy}=& \frac{\lambda ^{2}k_{x}k_{y}}{\sqrt{\lambda
^{2}k^{2}+\left( B+Jf_{X}\right) ^{2}}}, \\
\mathcal{S}_{+-}^{yz}=& -\frac{\left( B+Jf_{X}\right) k_{x}}{\sqrt{\lambda
^{2}k^{2}+\left( B+Jf_{X}\right) ^{2}}}, \\
\mathcal{S}_{+-}^{zx}=& \frac{\left( B+f_{X}\right) k_{y}}{\sqrt{\lambda
^{2}k^{2}+\left( B+Jf_{X}\right) ^{2}}}.
\end{align}%
The spin Berry curvatures are calculated as%
\begin{align}
\mathcal{A}_{+-}^{xy}=& -2\frac{B+Jf_{X}}{\sqrt{\lambda ^{2}k^{2}+\left(
B+Jf_{X}\right) ^{2}}}, \\
\mathcal{A}_{+-}^{yz}=& \frac{2\lambda k_{y}}{\sqrt{\lambda ^{2}k^{2}+\left(
B+Jf_{X}\right) ^{2}}}, \\
\mathcal{A}_{+-}^{zx}=& -\frac{2\lambda k_{x}}{\sqrt{\lambda
^{2}k^{2}+\left( B+Jf_{X}\right) ^{2}}}.
\end{align}

\subsection{Zeeman quantum geometry induced cross response}

We study the X-wave magnet coupled with the Rashba interaction without
applying magnetic field\cite{Chak}, where we set $J\neq 0$ and $B=0$. It is
possible to determine whether there is a response by integrating the quantum
geometric tensors $\mathcal{Z}_{+-}^{\mu \nu }$ and $\mathcal{Q}_{+-}^{\mu
\nu }$ over the angle $\phi $. Most of them vanishes by integration over the
angle $\phi $.

First, we study the spin polarization $S^{\mu ;\nu }$ induced by electric
field $E^{\nu }$. It is determined by $\mathcal{Q}_{+-}^{\mu \nu }$ as in
Eq.(\ref{SE}).

1) $S^{x;x}/E^{x}$

In the absence of the $X$-wave magnet, there is no response in the absence
of magnetic field, while there is nonzero response in the presence of
magnetic field. Only in the $d_{x^{2}-y^{2}}$ altermagnet, there emerges the
diagonal spin polarization $S^{\mu ;\mu }$ by applying electric field $%
E^{\mu }$ with $\mu =x,y$ in the absence of magnetic field. Once magnetic
field is turned on, there is nonzero response for all $X$-wave magnets.

2) $S^{x;y}/E^{y}$

In addition, only in the $d_{xy}$ altermagnet, there emerges the
off-diagonal spin polarization $S^{x}$ ($S^{y}$) by applying electric field $%
E^{y}$ ($E^{x}$) due to the contribution from the Zeeman quantum metric $%
\mathcal{Z}_{+-}^{xy}$ ($\mathcal{Z}_{+-}^{yx}$). Diagonal spin polarization 
$S^{\mu }$ is induced by electric field $E^{\mu }$ for all $X$-wave and $%
X^{\prime }$-wave magnets, which originate from the contribution of the
Rashba interaction.

Next, we study current $J^{\mu ;\nu }$ induced by magnetic field $B^{\nu }$.
It is determined by $\mathcal{Z}_{+-}^{\mu \nu }$ as in Eq.(\ref{JB}).

3) $J^{x;x}/B^{x}$

The diagonal current $J^{\mu ;\mu }$ is not induced by magnetic field $%
B^{\mu }$.

4) $J^{x;y}/B^{y}$

On the other hand, the off-diagonal current $J^{x;y}$ ($J^{y;x}$) is induced
by magnetic field $B^{y}$ ($B^{x}$) when there is nonzero Rashba interaction
irrespective of the presence of the $X$-wave magnet.

They are summarized in the following table. Detailed derivations are shown
in Appendix.\ref{ApCross},

\begin{equation}
\begin{tabular}{|l|l|l|l|l|}
\hline
& $S^{x;x}/E^{x}$ & $S^{x;y}/E^{y}$ & $J^{x;x}/B^{x}$ & $J^{x;y}/B^{y}$ \\ 
\hline
Quantum geometry & $\mathcal{Q}_{+-}^{xx}$ & $\mathcal{Q}_{+-}^{xy}$ & $%
\mathcal{Z}_{+-}^{xx}$ & $\mathcal{Z}_{+-}^{xy}$ \\ \hline
Rashba ($B=0$) & Zero & Zero & Zero & Nonzero \\ \hline
Rashba ($B\neq 0$) & Nonzero & Zero & Zero & Nonzero \\ \hline
X-wave ($B=0$) & $d_{x^{2}-y^{2}}$ & $d_{xy}$ & Zero & Nonzero \\ \hline
X-wave ($B\neq 0$) & Nonzero & $d_{xy}$ & Zero & Nonzero \\ \hline
\end{tabular}%
.
\end{equation}

\subsection{Materials}

We summarize materials realizing $X$-wave magnets\cite{BaiRev,BhoRev}.
Altermagnets have a collinear spin texture\cite{SmejX,SmejX2}, while $p$%
-wave magnets have a spiral spin texture\cite{Hayami2020B,Okumura}.

\subsubsection{$p$-wave magnet}

It was theoretically proposed that CeNiAsO is a $p$-wave magnet\cite{pwave}
and experimentally realized\cite{HZhou}. They were recently realized
experimentally in\cite{Yamada} Gd$_{3}$Ru$_{4}$Al$_{12}$ and in\cite{Comin}
NiI$_{2}$. It was also theoretically proposed that a $p$-wave magnet is
realized in graphene by introducing spin nematic order\cite{BitanRoy}.

\subsubsection{$d$-wave altermagnet}

The $d$-wave magnet in two dimensions was theoretically proposed in organic
materials\cite{Naka}, perovskite materials\cite{NakaB}, and twisted magnetic
Van der Waals bilayers\cite{YLiu}. The $d$-wave altermagnet in three
dimensions is experimentally realized in RuO$_{2}$\cite%
{Ahn,SmeRuO,Tsch,Fed,Lin}, Mn$_{5}$Si$_{3}$\cite{Leiv}, FeSb$_{2}$\cite%
{Mazin}, KV$_{2}$Se$_{2}$ \cite{KVSe}.

\subsubsection{$f$-wave magnet}

It was theoretically proposed that an $f$-wave magnet is theoretically
proposed in Ba$_{3}$MnNb$_{2}$O$_{9}$\cite{Hayami2020}, FePO$_{4}$\cite%
{Hayami2020B} and in graphene by introducing spin nematic order\cite%
{BitanRoy}.

\subsubsection{$g$-wave altermagnet}

A $g$-wave altermagnet in two-dimensions was theoretically proposed in
twisted magnetic Van der Waals bilayers\cite{YLiu}. The Fermi surface
splitting of the $g$-wave altermagnet in three dimensions is experimentally
observed in MnTe\cite{Krem,LeeG,Osumi,Haj,Lee,Masuda}, CrSb\cite%
{Reim,GYang,Ding,Cli,WLu}, V$_{1/3}$NbS$_{2}$\cite{Ray} and FeS\cite{Takagi}.

\subsubsection{$i$-wave altermagnet}

An $i$-wave altermagnet in two dimensions was theoretically proposed in
twisted magnetic Van der Waals bilayers\cite{YLiu} and in MnP(S,Se)$_{3}$%
\cite{MazinIwave}.

\section{Transport properties of $X$-wave magnets}

\label{SecXTra}

\subsection{Without Rashba interaction}

\subsubsection{Spin current generation}

One of the key feature of the $d$-wave altermagnet is that spin current is
generated without using the\ Rashba interaction\cite{Naka}. We analytically
show it.

The current is given by%
\begin{equation}
\mathbf{j}=-e\int dkf\mathbf{v},  \label{currentJ}
\end{equation}%
where $f$ is the Fermi distribution function in the presence of $\mathbf{E}$%
; $\mathbf{v}$ is the velocity,%
\begin{equation}
\mathbf{v}=\frac{1}{\hbar }\frac{\partial \varepsilon }{\partial \mathbf{k}},
\label{Velocity}
\end{equation}%
where $\varepsilon $ is the energy of the Hamiltonian, and we have dropped
the anomalous velocity term proportional to $-e\mathbf{E}\times \mathbf{%
\Omega }/\hbar $ because the Berry curvature $\mathbf{\Omega }$ is zero due
to the single band condition.

We expand the current in terms of electric field $E$ as%
\begin{equation}
j_{b}=\sum_{\ell _{1},\ell _{2}=0}\sigma ^{x^{\ell _{1}}y^{\ell
_{2}};b}(E_{x})^{\ell _{1}}(E_{y})^{\ell _{2}}.  \label{Expansion}
\end{equation}%
Then, the $(\ell _{1}+\ell _{2})$-th order conductivity is defined by

\begin{equation}
\sigma ^{x^{\ell _{1}}y^{\ell _{2}};b}=\frac{1}{\ell _{1}!\ell _{2}!}\frac{%
\partial ^{\ell _{1}+\ell _{2}}j_{b}}{\partial E_{x}^{\ell _{1}}\partial
E_{y}^{\ell _{2}}}.
\end{equation}%
The semi-classical Boltzmann equation in the presence of electric field $%
\mathbf{E}$ is given by%
\begin{equation}
\partial _{t}f-\frac{e\mathbf{E}}{\hbar }\cdot \nabla _{\mathbf{k}}f=-\frac{%
f-f^{\left( 0\right) }}{\tau },
\end{equation}%
where $\tau $\ is the relaxation time, and $f^{\left( 0\right) }$ is the
Fermi distribution function at the equilibrium with the chemical potential $%
\mu $,%
\begin{equation}
f^{\left( 0\right) }=1/\left( \exp \left( \varepsilon -\mu \right) +1\right)
.
\end{equation}%
Corresponding to Eq.(\ref{Expansion}), we expand the Fermi distribution in
powers of $\mathbf{E}$,%
\begin{equation}
f=f^{\left( 0\right) }+f^{\left( 1\right) }+\cdots .  \label{f}
\end{equation}%
The recursive solution gives\cite{YFang}%
\begin{equation}
f^{\left( \ell _{1}+\ell _{2}\right) }=\left( \frac{e/\hbar }{i\omega
+1/\tau }\right) ^{\ell _{1}+\ell _{2}}\frac{\partial ^{\ell _{1}+\ell
_{2}}f^{\left( 0\right) }}{\partial k_{x}^{\ell _{1}}\partial k_{y}^{\ell
_{2}}}(E_{x})^{\ell _{1}}(E_{x})^{\ell _{2}},
\end{equation}%
where $\omega $ is the frequency of the applied electric field $\mathbf{E}$.
The current (\ref{currentJ}) is expanded as in Eq.(\ref{f}) with 
\begin{align}
j_{b}^{\left( \ell _{1}+\ell _{2}\right) }& =-e\int d\mathbf{k}f^{\left(
\ell _{1}+\ell _{2}\right) }v_{b}  \notag \\
& =-\frac{e}{\hbar }\left( \frac{e/\hbar }{i\omega +1/\tau }\right) ^{\ell
_{1}+\ell _{2}}  \notag \\
& \qquad \quad \times \int d\mathbf{k}\,\frac{\partial \varepsilon }{%
\partial k_{b}}\frac{\partial ^{\ell _{1}+\ell _{2}}f^{\left( 0\right) }}{%
\partial k_{x}^{\ell _{1}}\partial k_{y}^{\ell _{2}}}(E_{x})^{\ell
_{1}}(E_{x})^{\ell _{2}}.
\end{align}%
The $(\ell _{1}+\ell _{2})$-th nonlinear Drude conductivity is defined by%
\begin{equation}
\sigma ^{x^{\ell _{1}}y^{\ell _{2}};b}=\frac{\left( -e/\hbar \right) ^{\ell
_{1}+\ell _{2}+1}}{\left( i\omega +1/\tau \right) ^{\ell _{1}+\ell _{2}}}%
\int d\mathbf{k}f^{\left( 0\right) }\frac{\partial ^{\ell _{1}+\ell
_{2}+1}\varepsilon }{\partial k_{x}^{\ell _{1}}\partial k_{y}^{\ell
_{2}}\partial k_{b}}.
\end{equation}%
The static limit is obtained simply by setting $\omega =0$ in this equation.

We define the $\ell $-th order spin-dependent Drude conductivity for each
spin $s$ by the formula%
\begin{equation}
\sigma _{s}^{x^{\ell _{1}}y^{\ell _{2}};b}=\frac{\left( -e/\hbar \right)
^{\ell _{1}+\ell _{2}+1}}{\left( i\omega +1/\tau \right) ^{\ell _{1}+\ell
_{2}}}\int d\mathbf{k}\,f_{s}^{\left( 0\right) }\frac{\partial ^{\ell
_{1}+\ell _{2}+1}\varepsilon _{s}}{\partial k_{x}^{\ell _{1}}\partial
k_{y}^{\ell _{2}}\partial k_{b}}.  \label{Drude}
\end{equation}%
This formula is nontrivial only when%
\begin{equation}
\frac{\partial ^{\ell _{1}+\ell _{2}+1}\varepsilon _{s}}{\partial
k_{x}^{\ell _{1}}\partial k_{y}^{\ell _{2}}\partial k_{b}}\neq 0,
\label{BasicCond}
\end{equation}%
which leads \ to a conclusion that there is no $\ell $-th order nonlinear
spin-Drude conductivity for $\ell \geq \ell _{1}+\ell _{2}$. It is necessary
to calculate explicitly the $\ell $-th order nonlinear spin-Drude
conductivity for $\ell =0,1,\cdots ,\ell _{1}+\ell _{2}-1$. In particular,
the choice of $\ell =0$\ and $1$\ yield the persistent spin current without
electric field and the linear spin conductivity, respectively.

We define the $(\ell _{1}+\ell _{2})$-th order nonlinear spin-Drude
conductivity by 
\begin{equation}
\sigma _{\text{spin}}^{x^{\ell _{1}}y^{\ell _{2}};b}=\frac{\sigma _{\uparrow
}^{x^{\ell _{1}}y^{\ell _{2}};b}-\sigma _{\downarrow }^{x^{\ell _{1}}y^{\ell
_{2}};b}}{2}.  \label{spinDr}
\end{equation}%
On the other hand, the charge conductivity is given by%
\begin{equation}
\sigma _{\text{charge}}^{x^{\ell _{1}}y^{\ell _{2}};b}=\sigma _{\uparrow
}^{x^{\ell _{1}}y^{\ell _{2}};b}+\sigma _{\downarrow }^{x^{\ell _{1}}y^{\ell
_{2}};b}.  \label{ChargeCurre}
\end{equation}

Spin current is generated in linear response only for the $d$-wave
altermagnet. The second-order nonlinear spin current is generated in the $f$%
-wave magnet, the third-order nonlinear spin current is generated in the $g$%
-wave altermagnet and the fifth-order nonlinear spin current is generated in
the $i$-wave altermagnet\cite{GI}. The results are summarized in the
following table.

\begin{equation}
\begin{tabular}{|c|c|c|c|c|c|c|}
\hline
& $s$ & $p$ & $d$ & $f$ & $g$ & $i$ \\ \hline
nodes & 0 & 1 & 2 & 3 & 4 & 6 \\ \hline
$\ell $ &  &  & linear & 2nd NL & 3rd NL & 5th NL \\ \hline
2D & None & None & $\sigma _{\text{spin}}^{y;x}$ & $\sigma _{\text{spin}%
}^{xx;y}$ & $\sigma _{\text{spin}}^{yyy;x}$ & $\sigma _{\text{spin}%
}^{yyyyy;x}$ \\ \hline
3D & None & None & $\sigma _{\text{spin}}^{y;x}$ & $\sigma _{\text{spin}%
}^{zy;x}$ & $\sigma _{\text{spin}}^{xxx;z}$ & $\sigma _{\text{spin}%
}^{yyyyx;x}$ \\ \hline
\end{tabular}%
.
\end{equation}

\subsubsection{Spin Nernst effects}

The spin Nernst effect is an effect that spin current is generated
perpendicularly to the thermal gradient. It is discussed in the $d$-wave
altermagnet\cite{Naka}. So far, there is no analytic result on it, which we
derive.

Tts momentum derivative reads%
\begin{equation}
\frac{\partial f}{\partial \mathbf{k}}=\frac{\partial \varepsilon \left( 
\mathbf{k}\right) }{\partial \mathbf{k}}\frac{\partial f}{\partial
\varepsilon },
\end{equation}%
while its spatial derivative reads%
\begin{equation}
\frac{\partial f}{\partial \mathbf{r}}=\frac{\partial T\left( \mathbf{r}%
\right) }{\partial \mathbf{r}}\frac{\partial f}{\partial T}.
\end{equation}%
We rewrite the Boltzmann equation as%
\begin{align}
\frac{df}{dt}=& \frac{\partial f}{\partial t}+\frac{\partial \mathbf{k}}{%
\partial t}\cdot \frac{\partial f}{\partial \mathbf{k}}+\frac{\partial 
\mathbf{r}}{\partial t}\cdot \frac{\partial f}{\partial \mathbf{r}}  \notag
\\
=& \frac{\partial \mathbf{k}}{\partial t}\cdot \left( \frac{\partial
\varepsilon \left( \mathbf{k}\right) }{\partial \mathbf{k}}\frac{\partial f}{%
\partial \varepsilon }\right) +\frac{\partial \mathbf{r}}{\partial t}\cdot
\left( \frac{\partial T\left( \mathbf{r}\right) }{\partial \mathbf{r}}\frac{%
\partial f}{\partial T}\right)  \notag \\
=& -\frac{f-f^{\left( 0\right) }}{\tau }.
\end{align}%
We use the kinetic equation%
\begin{align}
\frac{\partial \mathbf{r}}{\partial t}=& \frac{1}{\hbar }\frac{\partial
\varepsilon \left( \mathbf{k}\right) }{\partial \mathbf{k}}, \\
\frac{\partial \mathbf{k}}{\partial t}=& 0,
\end{align}%
and obtain%
\begin{equation}
\frac{1}{\hbar }\left( \frac{\partial \varepsilon \left( \mathbf{k}\right) }{%
\partial \mathbf{k}}\cdot \frac{\partial T\left( \mathbf{r}\right) }{%
\partial \mathbf{r}}\right) \frac{\partial f}{\partial T}=-\frac{f-f^{\left(
0\right) }}{\tau }.
\end{equation}%
The first order expansion reads%
\begin{equation}
\frac{1}{\hbar }\left( \frac{\partial \varepsilon \left( \mathbf{k}\right) }{%
\partial \mathbf{k}}\cdot \frac{\partial T\left( \mathbf{r}\right) }{%
\partial \mathbf{r}}\right) \frac{\partial f^{\left( 0\right) }}{\partial T}%
=-\frac{f^{\left( 1\right) }}{\tau },
\end{equation}%
whose solution is given by%
\begin{equation}
f^{\left( 1\right) }=-\frac{\tau }{\hbar }\left( \frac{\partial \varepsilon
\left( \mathbf{k}\right) }{\partial \mathbf{k}}\cdot \frac{\partial T\left( 
\mathbf{r}\right) }{\partial \mathbf{r}}\right) \frac{\partial f^{\left(
0\right) }}{\partial T}.
\end{equation}%
The current driven by temperature gradient is given by%
\begin{align}
j_{b}=& -e\int d\mathbf{k}f^{\left( 1\right) }v_{b}  \notag \\
=& \frac{e\tau }{\hbar ^{2}}\int d\mathbf{k}\left( \frac{\partial
\varepsilon \left( \mathbf{k}\right) }{\partial \mathbf{k}}\cdot \frac{%
\partial T\left( \mathbf{r}\right) }{\partial \mathbf{r}}\right) \frac{%
\partial \varepsilon }{\partial k_{b}}\frac{\partial f^{\left( 0\right) }}{%
\partial T}.
\end{align}%
Especially, current $j_{x}$ and $j_{y}$ flow when the temperature gradient
is along the $x$ axis,%
\begin{equation}
j_{x}=e\tau \int d\mathbf{k}\left( \frac{1}{\hbar }\frac{\partial
\varepsilon }{\partial k_{x}}\right) ^{2}\frac{\partial T\left( \mathbf{r}%
\right) }{\partial x}\frac{\partial f^{\left( 0\right) }}{\partial T},
\end{equation}%
which is known as the Seebeck effect, and 
\begin{align}
j_{y} =&-e\int d\mathbf{k}f^{\left( 1\right) }v_{y}  \notag \\
=&\frac{e\tau }{\hbar ^{2}}\int d\mathbf{k}\frac{\partial \varepsilon }{%
\partial k_{x}}\cdot \frac{\partial T\left( \mathbf{r}\right) }{\partial x}%
\frac{\partial \varepsilon }{\partial k_{y}}\frac{\partial f^{\left(
0\right) }}{\partial T},
\end{align}%
which is known as the Nernst effect. We assume that the temperature gradient
is linear,%
\begin{equation}
T\left( \mathbf{r}\right) =ax.
\end{equation}%
When $\varepsilon \left( \mathbf{k}\right) -\mu \gg k_{\text{B}}T\left( 
\mathbf{r}\right) $, we approximate the Fermi distribution by the Boltzmann
distribution,%
\begin{equation}
f=\frac{1}{\exp \left( \frac{\varepsilon \left( \mathbf{k}\right) -\mu }{k_{%
\text{B}}T\left( \mathbf{r}\right) }\right) +1}\simeq \exp \left( -\frac{%
\varepsilon \left( \mathbf{k}\right) -\mu }{k_{\text{B}}T\left( \mathbf{r}%
\right) }\right) .
\end{equation}%
Then, the spin Hall conductivity is obtained as%
\begin{equation}
j_{s}=\frac{e\tau }{\hbar ^{2}}\frac{8e^{\mu /k_{\text{B}}T}Jm\pi \left(
2T-\mu \right) }{\sqrt{1-4J^{2}m^{2}}}
\end{equation}%
for the $d$-wave altermagnet. It is proportional to the coupling constant $J$
of the $X$-wave magnet in the Hamiltonian (\ref{ModelHamil}). Hence, it is
possible to detect the sign of $J$ by the spin Nernst effect. On the other
hand, there is no spin current generation for the other $X$-wave magnets.

\begin{figure}[t]
\centerline{\includegraphics[width=0.48\textwidth]{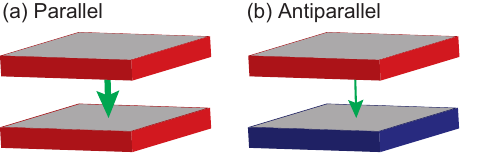}}
\caption{Illustration of a bilayer magnetic tunneling junction made of
magnets. (a) Parallel configuration, where the spin directions are identical
at each lattice site between the two layers. (b) Antiparallel configuration,
where the spin directions are opposite at each lattice site between the two
layers. The green arrow indicates the tunneling current, which is larger in
the parallel configuration.}
\label{FigIllustMTJ}
\end{figure}

\begin{figure*}[t]
\centerline{\includegraphics[width=0.88\textwidth]{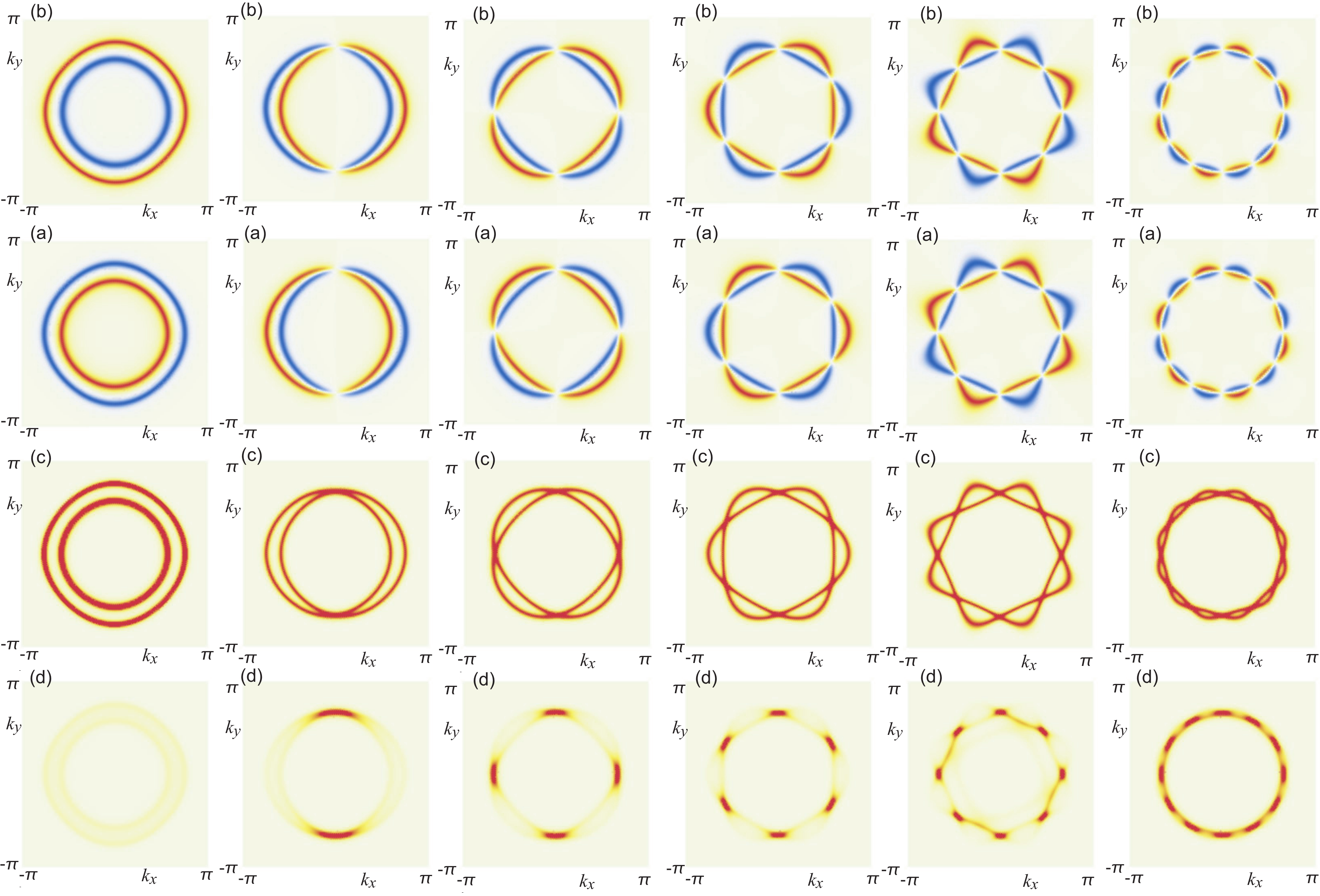}}
\caption{{}(a) Spin density $\left\langle S_{z}\right\rangle $ when $J>0$,
and (b) that when $J<0$ in one layer. Red (blue) color indicates up (down)
spin. (c) Overlap $\mathcal{O}_{\text{P}}$ for the parallel configuration,
and (d) overlap $\mathcal{O}_{\text{AP}}$ for the antiparallel
configuration. Red color indicates a large overlap. }
\label{FigMTJX}
\end{figure*}

\subsubsection{Tunneling magnetoresistance}

Magnetic tunneling junction is a most successful spintronic device\cite%
{Julli,Mood}. It consists of the bilayer ferromagnets spaced by an insulator
as shown in Fig.\ref{FigIllustMTJ}, where the resistance is low (large) if
the directions of spins are identical (opposite). It is called the tunneling
magnetoresistance (TMR). The spin direction of the memory can be readout by
using the TMR. Tunneling magnetoresistance is discussed in the $d$-wave
altermagnet\cite{SmejX2,FLiu,Chi,Yao,ZYan,ZYang,Volo,YSun} and the $p$-wave
magnet\cite{Brek}.

The differential conductance $G=dI/dV$ is calculated based on the Green
function\cite{Brek},%
\begin{equation}
\frac{G}{4e\pi ^{3}}=\sum_{s=\pm 1}\sum_{\mathbf{k}_{1},\mathbf{k}%
_{2}}\left\vert T_{\mathbf{k}_{1},\mathbf{k}_{2}}\right\vert ^{2}\text{Tr}%
\left[ \text{Im}\mathcal{G}_{s}^{\text{T}}\left( 0;\mathbf{k}_{1}\right) 
\text{Im}\mathcal{G}_{s}^{\text{B}}\left( 0;\mathbf{k}_{2}\right) \right] ,
\label{G}
\end{equation}%
where $\mathcal{G}_{s}^{\text{T}}$ ($\mathcal{G}_{s}^{\text{B}}$) is the
retarded Green function of the top (bottom) layer defined by%
\begin{equation}
\mathcal{G}_{s}^{\text{T}}\left( \omega ;\mathbf{k}\right) \equiv \mathcal{G}%
_{s}^{\text{B}}\left( \omega ;\mathbf{k}\right) \equiv \mathcal{G}_{s}\left(
\omega ;\mathbf{k}\right)
\end{equation}%
for the parallel configuration, and%
\begin{equation}
\mathcal{G}_{s}^{\text{T}}\left( \omega ;\mathbf{k}\right) \equiv \mathcal{G}%
_{s}\left( \omega ;\mathbf{k}\right) ,\quad \mathcal{G}_{s}^{\text{B}}\left(
\omega ;\mathbf{k}\right) \equiv \mathcal{G}_{-s}\left( \omega ;\mathbf{k}%
\right)  \label{GreenAnti}
\end{equation}%
for the antiparallel configuration, where we have defined%
\begin{equation}
\mathcal{G}_{s}\left( \omega ;\mathbf{k}\right) \equiv \frac{1}{\hbar \omega
-\varepsilon _{s}+i\Gamma }
\end{equation}%
with the self-energy $\Gamma $ and the energy $\varepsilon _{s}$. Im$%
\mathcal{G}_{s}\left( 0;\mathbf{k}_{1}\right) $ represents the density of
states depending on the spin $s$, which is shown in Fig.\ref{FigMTJX}(a) and
(b).

The differential conductivity is determined by the overlap of the density of
states Im$\mathcal{G}_{s}\left( 0;\mathbf{k}_{1}\right) $, which is shown in
Fig.\ref{FigMTJX}(c) and (d).

The differential conductivity for the parallel configuration is analytically
given by\cite{MTJ}%
\begin{equation}
\lim_{\Gamma \rightarrow 0}\frac{G_{\text{P}}}{4e\pi ^{3}}=\frac{2m\pi ^{2}}{%
\hbar ^{2}\Gamma },
\end{equation}%
while that for the antiparallel configuration is analytically given by%
\begin{equation}
\lim_{\Gamma \rightarrow 0}\frac{G_{\text{AP}}}{4e\pi ^{3}}=\sqrt{\frac{m}{%
2\mu }}\frac{N_{X}\pi ^{2}}{\hbar a\left\vert J\right\vert },
\end{equation}%
where $J$ is the strength of the $X$-wave magnet, $\Gamma $ is the
self-energy, $\mu $ is the chemical potential, $a$ is the lattice constant
and $m$ is the mass of electrons. Hence, the TMR ratio is given by%
\begin{equation}
\lim_{\Gamma \rightarrow 0}\text{TMR}_{\text{ratio}}\equiv \lim_{\Gamma
\rightarrow 0}\frac{G_{\text{P}}-G_{\text{AP}}}{G_{\text{AP}}}=\frac{%
2a\left\vert J\right\vert \sqrt{2m\mu }}{N_{X}\hbar \Gamma }.
\end{equation}

It is to be contrasted with the TMR ratio based on ferromagnets, where it is
given by%
\begin{equation}
\lim_{\Gamma \rightarrow 0}\text{TMR}_{\text{ratio}}=\frac{2J^{2}}{\Gamma
^{2}}.
\end{equation}%
Therefore, the TMR ratio is larger in ferromagnets for $\left\vert
J\right\vert >\Gamma $. However, the $X$-wave magnets are expected to
achieve high-speed and ultra-dense memory owing to the zero net
magnetization.

\subsection{With Rashba interaction}

\subsubsection{Anomalous Hall effects}

One of the motivation of studying altermangnets is that the N\'{e}el vector
can be readout by measuring anomalous Hall conductivity because
time-reversal symmetry is broken\cite{Fak,Tsch,Sato,Leiv}. Actually, it is
verified by the density-functional theory and experiments.

The Berry curvature is obtained as%
\begin{equation}
\Omega _{X}=-\frac{\lambda ^{2}\left( B+Jf_{X}-Jk\partial _{k}f_{X}\right) }{%
2\left( \left( B+Jf_{X}\right) ^{2}+\lambda ^{2}k^{2}\right) ^{3/2}}.
\end{equation}%
The Hall conductivity is calculated as%
\begin{equation}
\frac{1}{2\pi }\int \Omega _{X}d\mathbf{k}=-\frac{1}{2}\text{sgn}B.
\end{equation}%
It does not depend on $J$. Hence, it is impossible to detect the sign of $J$
by measuring the anomalous Hall conductivity based on the two-band model. It
is necessary to introduce the orbital degrees of freedom to the two-band
model. 
\begin{figure}[t]
\centerline{\includegraphics[width=0.48\textwidth]{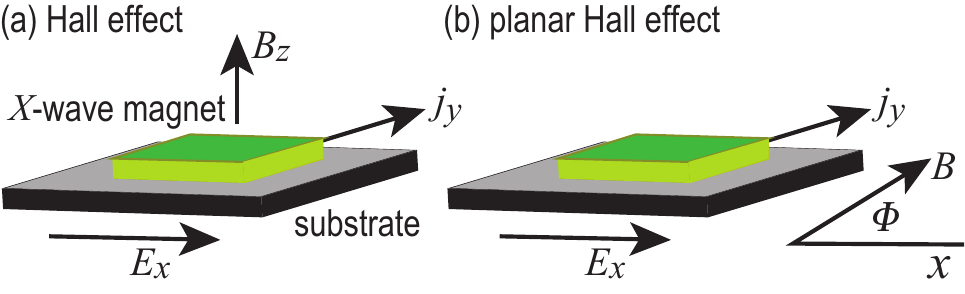}}
\caption{Illustration for (a) Hall effect and (b) planar Hall effect. In the
planar Hall effect, when the electric field is applied along the $x$ axis
and the magnetic field $(B\cos \Phi ,B\sin \Phi ,0)$ is applied parallel to
the system, the Hall current flows along the $y$ axis. The Hall conductivity
is predicted to be given by the formula (\protect\ref{pHall}). }
\label{FigIllust}
\end{figure}

\subsubsection{Planar Hall effects}

The Hall effect is a prominent phenomenon in two-dimensional materials. A
current flows into a direction perpendicular to the applied electric field,
when the magnetic field is applied perpendicular to the plane, as
illustrated in Fig.\ref{FigIllust}(a). Similarly, a current flows into a
direction perpendicular to the applied electric field, when the magnetic
field is applied parallel to the plane, as illustrated in Fig.\ref{FigIllust}%
(b). It is the planar Hall effect\cite%
{Tang,XLiu,Nandy,Taskin,Burkov,Kumar,Zliu}. The planar Hall effect is
discussed in $d$-wave altermagnets\cite{Korra,BFChen}.

The planar Hall effect is discussed in the $X$-wage magnet\cite{Planar}. The
Dirac point shifts in the presence of the in-plane magnetic field. The
shifted Dirac point is given by $(k_{x}^{\prime },k_{y}^{\prime })=0$, where%
\begin{equation}
k_{x}^{\prime }=k_{x}+\frac{B_{y}}{\lambda },\qquad k_{y}^{\prime }=k_{y}-%
\frac{B_{x}}{\lambda },
\end{equation}%
by solving the equation%
\begin{equation}
\lambda \left( k_{x}\sigma _{y}-k_{y}\sigma _{x}\right) +B_{x}\sigma
_{x}+B_{y}\sigma _{y}=0.
\end{equation}

The Dirac gap is explicitly determined as%
\begin{equation}
\Delta =\left( -1\right) ^{s_{X}}\frac{B^{N_{X}}}{\lambda ^{N_{X}}}\sin
N_{X}\Phi ,  \label{sDirac2}
\end{equation}%
which leads to the Hall conductance%
\begin{equation}
\sigma _{xy}=\frac{e^{2}}{h}\frac{\left( -1\right) ^{s_{X}}}{2}\text{sgn}%
\left( J\frac{B^{N_{X}}}{\lambda ^{N_{X}}}\sin N_{X}\Phi \right) ,
\label{pHall}
\end{equation}%
where $s_{X}=1,1,-1,-1,1$ for $X=p,d,f,g,i$, respectively.

\section{Quantum Hall effects}

\label{SecLL}

\subsection{Landau levels}

If we apply magnetic field perpendicular to the sample, Landau levels are
formed. Landau levels are obtained for the $d$-wave altermagnet\cite%
{WangChen25,XiaLi}. We apply a homogeneous magnetic field $\mathbf{B}=%
\mathbf{\nabla }\times \mathbf{A}=\left( 0,0,-B\right) $ with $B>0$ along
the $z$ axis to the sample. The Hamiltonian under magnetic field is obtained
by replacing the momentum $k_{i}$ to the covariant momentum $P_{i}\equiv
k_{i}+eA_{i}$. We introduce a pair of Landau-level ladder operators, 
\begin{equation}
\hat{a}=\frac{\ell _{B}(P_{x}+iP_{y})}{\sqrt{2}\hbar },\quad \hat{a}%
^{\dagger }=\frac{\ell _{B}(P_{x}-iP_{y})}{\sqrt{2}\hbar },  \label{G-OperaA}
\end{equation}%
satisfying the bosonic commutation relation $[\hat{a},\hat{a}^{\dag }]=1$,
where $\ell _{B}=\sqrt{\hbar /eB}$ is the magnetic length. The inverse
relations read 
\begin{equation}
P_{x}=\frac{\hbar }{\sqrt{2}\ell _{B}}\left( \hat{a}+\hat{a}^{\dagger
}\right) ,\quad P_{y}=\frac{\hbar }{i\sqrt{2}\ell _{B}}\left( \hat{a}-\hat{a}%
^{\dagger }\right) .  \label{PxPy}
\end{equation}%
The Hamiltonian for free electrons reads%
\begin{equation}
H_{0}=\hbar \omega _{0}\hat{a}^{\dagger }\hat{a}.
\end{equation}

\subsubsection{$p$-wave magnets and coherent states}

By inserting Eq.(\ref{PxPy}) to Eq.(\ref{f2p}), the $p$-wave term under
magnetic field reads%
\begin{equation}
f_{p}^{2\text{D}}\left( \mathbf{k}\right) =\frac{a\hbar }{\sqrt{2}\ell _{B}}%
\left( \hat{a}+\hat{a}^{\dagger }\right) .
\end{equation}%
The Hamiltonian is identical to that of the coherent state%
\begin{equation}
\hat{H}=\hbar \omega _{0}\hat{a}^{\dagger }\hat{a}+c^{\ast }\hat{a}+c\hat{a}%
^{\dagger }  \label{HamilCoherent}
\end{equation}%
provided%
\begin{equation}
c=c^{\ast }=\frac{a\hbar }{\sqrt{2}\ell _{B}}.
\end{equation}%
It is diagonalized as%
\begin{equation}
\hat{H}=\hbar \omega \hat{b}^{\dagger }\hat{b}-c^{2}/\hbar \omega ,
\end{equation}%
where we have introduced displaced operators%
\begin{equation}
\hat{b}\equiv \hat{a}+c/\hbar \omega ,\qquad \hat{b}^{\dagger }\equiv \hat{a}%
^{\dagger }+c/\hbar \omega .  \label{pb}
\end{equation}%
The energy of the landau level is obtained as%
\begin{equation}
\hat{H}=\hbar \omega N-\frac{a^{2}\hbar }{2\ell _{B}^{2}\omega },
\end{equation}%
while the eigenstate is determined as%
\begin{equation}
\left\vert N\right\rangle _{b}\equiv \frac{1}{\sqrt{N!}}\left( \hat{b}%
^{\dagger }\right) ^{N}\left\vert 0\right\rangle .
\end{equation}

\subsubsection{$d$-wave altermagnets and squeezed states}

By inserting Eq.(\ref{PxPy}) to Eq.(\ref{f2d}), the $d$-wave term under
magnetic field reads%
\begin{align}
f_{d^{\prime }}^{2\text{D}}\left( \mathbf{k}\right) =& \frac{\hbar ^{2}a^{2}%
}{2\ell _{B}^{2}}\left( \left( \hat{a}+\hat{a}^{\dagger }\right) ^{2}+\left( 
\hat{a}-\hat{a}^{\dagger }\right) ^{2}\right) \\
& =\frac{\hbar ^{2}a^{2}}{\ell _{B}^{2}}\left( \left( \hat{a}^{\dagger
}\right) ^{2}+\hat{a}^{2}\right) .
\end{align}%
The Hamiltonian is identical to that of the squeezed state,%
\begin{equation}
\mathcal{\hat{H}}\equiv \hbar \omega _{0}+\frac{c}{2}\left( \hat{a}^{\dagger
}\right) ^{2}+\frac{c^{\ast }}{2}\hat{a}^{2}  \label{HPDC}
\end{equation}%
provided%
\begin{equation}
c=c^{\ast }=\frac{2\hbar ^{2}a^{2}}{\ell _{B}^{2}}.
\end{equation}%
It is diagonalized as%
\begin{equation}
H=\hbar \omega \hat{b}^{\dagger }\hat{b}-\hbar \frac{\omega _{0}-\omega }{2}
\end{equation}%
with%
\begin{equation}
\omega =\sqrt{\omega _{0}^{2}-c^{2}},
\end{equation}%
where we have made the Bogoliubov transformation of bosons%
\begin{equation}
\left( 
\begin{array}{c}
\hat{b} \\ 
\hat{b}^{\dagger }%
\end{array}%
\right) =\left( 
\begin{array}{cc}
\cosh r & -\sinh r \\ 
-\sinh r & \cosh r%
\end{array}%
\right) \left( 
\begin{array}{c}
\hat{a} \\ 
\hat{a}^{\dagger }%
\end{array}%
\right)  \label{bd}
\end{equation}%
with%
\begin{equation}
\tanh 2r=c/\hbar \omega .
\end{equation}%
The energy of the landau level is obtained as%
\begin{equation}
E_{n}=\hbar \omega N-\hbar \frac{\omega _{0}-\omega }{2},
\end{equation}%
while the eigenstate is determined as%
\begin{equation}
\left\vert N\right\rangle _{b}\equiv \frac{1}{\sqrt{N!}}(\hat{b}^{\dagger
})^{N}\left\vert 0\right\rangle .
\end{equation}

\subsubsection{$X$-wave magnets}

In general, it is hard to exactly diagonalize the Hamiltonian except for the 
$p$-wave magnet and the $d$-wave altermagnet because the Hamiltonian
contains more than quadratic order of the creation and annihilation
operators. However, it is possible to numerically obtain the energy of the
Landau levels by using the states%
\begin{equation}
\left\vert N\right\rangle =\frac{1}{\sqrt{N!}}(a^{\dagger })^{N}\left\vert
0\right\rangle .
\end{equation}

\subsection{Magneto-optical conductivity}

Optical absorption between Landau levels are known as magneto-optical
conductivity, which is calculated by the Kubo formula\cite%
{Ando,Koshino,Nikol,Carbo},

\begin{align}
& \sigma _{\mu \nu }\left( \omega \right) =-i\hbar e^{2}\int d\mathbf{k} 
\notag \\
& \times \sum_{n\neq m}\frac{f\left( \varepsilon _{n}\left( \mathbf{k}%
\right) \right) -f\left( \varepsilon _{m}\left( \mathbf{k}\right) \right) }{%
\left( \varepsilon _{n}\left( \mathbf{k}\right) -\varepsilon _{m}\left( 
\mathbf{k}\right) \right) \left( \varepsilon _{n}\left( \mathbf{k}\right)
-\varepsilon _{m}\left( \mathbf{k}\right) +\hbar \omega +i\Gamma \right) } 
\notag \\
& \times \lbrack \left\langle \psi _{n}\left( \mathbf{k}\right) \right\vert
\upsilon _{\mu }\left\vert \psi _{m}\left( \mathbf{k}\right) \right\rangle
\left\langle \psi _{m}\left( \mathbf{k}\right) \right\vert \upsilon _{\nu
}\left\vert \psi _{n}\left( \mathbf{k}\right) \right\rangle ]
\end{align}%
with the velocity operators%
\begin{equation}
\upsilon _{x}\equiv \frac{\partial H}{\partial P_{x}},\qquad \upsilon
_{y}\equiv \frac{\partial H}{\partial P_{y}}.
\end{equation}%
Especially, the magneto-optical conductivity for the $d$-wave altermagnet is
studied\cite{WangChen25}.

\subsubsection{$p$-wave magnets}

We first study magneto-optical conductivity for the $p$-wave magnet. The
velocity operators are explicitly obtained as%
\begin{align}
\upsilon _{x}=& \frac{\partial H}{\partial P_{x}}=\frac{\partial \hat{b}%
^{\dagger }}{\partial P_{x}}\frac{\partial H}{\partial \hat{b}^{\dagger }}+%
\frac{\partial \hat{b}}{\partial P_{x}}\frac{\partial H}{\partial \hat{b}} 
\notag \\
=& \frac{\ell _{B}\omega }{\sqrt{2}}\left( \hat{b}+\hat{b}^{\dagger }\right)
,
\end{align}%
and%
\begin{align}
\upsilon _{y}=& \frac{\partial H}{\partial P_{y}}=\frac{\partial \hat{b}%
^{\dagger }}{\partial P_{y}}\frac{\partial H}{\partial \hat{b}^{\dagger }}+%
\frac{\partial \hat{b}}{\partial P_{y}}\frac{\partial H}{\partial \hat{b}} 
\notag \\
=& i\frac{\ell _{B}\omega }{\sqrt{2}}\left( -\hat{b}+\hat{b}^{\dagger
}\right) ,
\end{align}%
where we have used%
\begin{align}
\hat{b}=& \frac{\ell _{B}}{\sqrt{2}\hbar }\left( P_{x}+iP_{y}\right) +\frac{c%
}{\hbar \omega }, \\
\hat{b}^{\dagger }=& \frac{\ell _{B}}{\sqrt{2}\hbar }\left(
P_{x}-iP_{y}\right) +\frac{c^{\ast }}{\hbar \omega },
\end{align}%
derived from Eq.(\ref{pb}). The interband matrix elements $\left\langle \psi
_{n}\left( \mathbf{k}\right) \right\vert \upsilon _{\mu }\left\vert \psi
_{m}\left( \mathbf{k}\right) \right\rangle $ and $\left\langle \psi
_{m}\left( \mathbf{k}\right) \right\vert \upsilon _{\nu }\left\vert \psi
_{n}\left( \mathbf{k}\right) \right\rangle $\ are nonzero for $m=n\pm 1$.
Then, the optical absorption occurs between the adjacent Landau levels.
There are nonzero $\text{Re}\sigma _{xx}$, $\text{Re}\sigma _{yy}$ and $%
\text{Im}\sigma _{xy}$ because $\upsilon _{x}$ is real and $\upsilon _{y}$
is imaginary.

\subsubsection{$d$-wave altermagnets}

We next study magneto-optical conductivity for the $d$-wave altermagnet. The
velocity operators are explicitly obtained as%
\begin{align}
\upsilon _{x} =&\frac{\partial H}{\partial P_{x}}=\frac{\partial \hat{b}%
^{\dagger }}{\partial P_{x}}\frac{\partial H}{\partial \hat{b}^{\dagger }}+%
\frac{\partial \hat{b}}{\partial P_{x}}\frac{\partial H}{\partial \hat{b}} 
\notag \\
=&\frac{\ell _{B}\omega }{\sqrt{2}\left( \cosh r+\sinh r\right) }\left( \hat{%
b}+\hat{b}^{\dagger }\right) ,
\end{align}%
and%
\begin{align}
\upsilon _{y} =&\frac{\partial H}{\partial P_{y}}=\frac{\partial \hat{b}%
^{\dagger }}{\partial P_{y}}\frac{\partial H}{\partial \hat{b}^{\dagger }}+%
\frac{\partial \hat{b}}{\partial P_{y}}\frac{\partial H}{\partial \hat{b}} 
\notag \\
=&\frac{-i\ell _{B}\omega }{\sqrt{2}\left( \cosh r-\sinh r\right) }\left( 
\hat{b}-\hat{b}^{\dagger }\right)
\end{align}%
where we have used%
\begin{align}
\hat{b} =&\frac{\ell _{B}}{\sqrt{2}}\left( \frac{P_{x}}{\cosh r+\sinh r}+%
\frac{iP_{y}}{\cosh r-\sinh r}\right) , \\
\hat{b}^{\dagger } =&\frac{\ell _{B}}{\sqrt{2}}\left( \frac{P_{x}}{\cosh
r+\sinh r}-\frac{iP_{y}}{\cosh r-\sinh r}\right) ,
\end{align}%
derived from Eq.(\ref{bd}). The interband matrix elements $\left\langle \psi
_{n}\left( \mathbf{k}\right) \right\vert \upsilon _{\mu }\left\vert \psi
_{m}\left( \mathbf{k}\right) \right\rangle $ and $\left\langle \psi
_{m}\left( \mathbf{k}\right) \right\vert \upsilon _{\nu }\left\vert \psi
_{n}\left( \mathbf{k}\right) \right\rangle $\ are nonzero for $m=n\pm 1$.
Then, the optical absorption occurs between the adjacent Landau levels.
There are nonzero $\text{Re}\sigma _{xx}$, $\text{Re}\sigma _{yy}$ and $%
\text{Im}\sigma _{xy}$ because $\upsilon _{x}$ is real and $\upsilon _{y}$
is imaginary.

\subsection{Magnetic circular dichroism}

Circular dichroism is a phenomenon that the optical absorption is different
between the right-hand and left-hand polarized light,%
\begin{equation}
\sigma _{\pm }=\sigma _{xx}\pm i\sigma _{xy}.
\end{equation}%
Its real part is given by%
\begin{equation}
\text{Re}\sigma _{\pm }=\text{Re}\sigma _{xx}\mp \text{Im}\sigma _{xy}.
\end{equation}%
By using it, the degree of the circular dichroism is evaluated by the
quantity%
\begin{equation}
I_{\text{CD}}\equiv \frac{\text{Re}\sigma _{+}-\text{Re}\sigma _{-}}{\text{Re%
}\sigma _{+}+\text{Re}\sigma _{-}}=-\frac{\text{Im}\sigma _{xy}}{\text{Re}%
\sigma _{xx}}.
\end{equation}%
Hence, circular dichroism occurs when $\text{Im}\sigma _{xy}\neq 0$. Indeed,
it occurs in the case of the $p$-wave magnet and the $d$-wave altermagnet,
where $\text{Im}\sigma _{xy}\neq 0$ as we have seen.

\section{Friedel oscillation}

\label{SecFriedel}

We study the effect of an impurity. Friedel oscillation is that the local
the density of sates (LDOS) oscillates in the presence of the impurity. We
show that this LDOS has the $X$-wave symmetry for the $X$-wave magnet. The
energy-dependent LDOS\ is determined by 
\begin{equation}
\rho \left( \omega ,\mathbf{r}\right) =-\frac{1}{\pi }\text{Im}G^{\text{R}%
}\left( \omega ,\mathbf{r},\mathbf{r}\right) ,
\end{equation}%
where $G\left( \omega ,\mathbf{r},\mathbf{r}\right) $ is the real space
representation of the retarded Green function in the presence of the
impurity defined by%
\begin{equation}
G^{\text{R}}\left( \omega ,\mathbf{r},\mathbf{r}^{\prime }\right) \equiv
\left\langle \mathbf{r}\right\vert \frac{1}{\hbar \omega -H+i\eta }%
\left\vert \mathbf{r}^{\prime }\right\rangle .
\end{equation}%
The Green function is obtained by using the Dyson equation%
\begin{align}
& G^{\text{R}}\left( \omega ,\mathbf{r},\mathbf{r}^{\prime }\right)  \notag
\\
=& G_{0}^{\text{R}}\left( \omega ,\mathbf{r},\mathbf{r}^{\prime }\right) 
\notag \\
& +\int d\mathbf{r}_{1}d\mathbf{r}_{2}G_{0}^{\text{R}}\left( \omega ,\mathbf{%
r},\mathbf{r}_{1}\right) T\left( \omega ,\mathbf{r}_{1},\mathbf{r}%
_{2}\right) G_{0}^{\text{R}}\left( \omega ,\mathbf{r}_{2},\mathbf{r}^{\prime
}\right)
\end{align}%
with the T matrix%
\begin{equation}
T\left( \omega ,\mathbf{r},\mathbf{r}_{1}\right) \equiv V\left( 1-G_{0}^{%
\text{R}}\left( \omega ,\mathbf{r},\mathbf{r}_{1}\right) V\right) ^{-1}.
\end{equation}%
We assume that the impurity is described by the delta function potential%
\begin{equation}
\left( V_{0}\sigma _{0}+V_{s}\sigma _{z}\right) \delta \left( r\right) ,
\end{equation}%
where $V_{0}$ describes a non-magnetic impurity and $V_{s}$ describes a
magnetic impurity. It is enough to study spin-dependent potential%
\begin{equation}
\left( V_{0}+V_{s}s\right) \delta \left( r\right)
\end{equation}%
with $s=\pm 1$ because the spin operator is diagonal. In this case, the T
matrix has a form%
\begin{equation}
T\left( \omega ,\mathbf{r},\mathbf{r}_{1}\right) =T\left( \omega \right)
\delta \left( \mathbf{r}\right) \delta \left( \mathbf{r}^{\prime }\right) ,
\end{equation}%
which leads to the Dyson equation%
\begin{align}
& G^{\text{R}}\left( \omega ,\mathbf{r},\mathbf{r}^{\prime }\right)  \notag
\\
=& G_{0}^{\text{R}}\left( \omega ,\mathbf{r},\mathbf{r}^{\prime }\right)
+G_{0}^{\text{R}}\left( \omega ,\mathbf{r},\mathbf{0}\right) T\left( \omega
\right) G_{0}^{\text{R}}\left( \omega ,\mathbf{0},\mathbf{r}^{\prime
}\right) .
\end{align}%
For small $V$, the difference of the local density of states is simply given
by%
\begin{align}
\delta \rho \equiv & -\frac{1}{\pi }\text{Im}\left[ G^{\text{R}}\left(
\omega ,\mathbf{r},\mathbf{r}\right) -G^{\text{R}}\left( \omega ,\mathbf{r},%
\mathbf{r}\right) \right]  \notag \\
=& -\frac{1}{\pi }\text{Im}\left[ VG_{0}^{\text{R}}\left( \omega ,\mathbf{r},%
\mathbf{0}\right) G_{0}^{\text{R}}\left( \omega ,\mathbf{0},\mathbf{r}%
^{\prime }\right) \right] ^{2}.
\end{align}%
The real space Green function is obtained by the Fourier transformation of
the momentum space Green function as%
\begin{align}
G_{0}^{\text{R}}\left( \omega ,\mathbf{r},\mathbf{r}^{\prime }\right) =&
G_{0}^{\text{R}}\left( \omega ,\mathbf{r}-\mathbf{r}^{\prime },\mathbf{0}%
\right)  \notag \\
\equiv & \frac{1}{\left( 2\pi \right) ^{2}}\int e^{-\mathbf{k}\cdot \left( 
\mathbf{r}-\mathbf{r}^{\prime }\right) }G_{0}^{\text{R}}\left( \omega ,%
\mathbf{k}\right) d\mathbf{r,}
\end{align}%
with%
\begin{equation}
G_{0}^{\text{R}}\left( \omega ,\mathbf{k}\right) \equiv \frac{1}{i\omega
-H+i\eta }
\end{equation}%
where we have used the translational symmetry of the Green function in the
absence of the impurity.

\subsection{Free electrons}

First, we review the Friedel oscillation of free electrons. We derive the
real space representation of the Green function in the presence of the
impurity and determine the symmetry of the LDOS. The Green function for free
electrons is given by 
\begin{equation}
G_{0}^{\text{R}}\left( \omega ,\mathbf{k}\right) =\frac{1}{\frac{\hbar ^{2}}{%
2m}\left( k_{0}^{2}-k^{2}\right) +i\eta },
\end{equation}%
where we have set%
\begin{equation}
\omega =\frac{\hbar ^{2}k_{0}^{2}}{2m}.
\end{equation}%
We make a Fourier transformation%
\begin{align}
G_{0}^{\text{R}}\left( r\right) =& \frac{1}{\left( 2\pi \right) ^{2}}%
\int_{0}^{\infty }k\frac{e^{ikr\cos \left( \theta -\phi \right) }}{\frac{%
\hbar ^{2}}{2m}\left( k_{0}^{2}-k^{2}\right) +i\eta }dkd\theta  \notag \\
=& \frac{1}{2\pi }\int_{0}^{\infty }\frac{kJ_{0}\left( kr\right) }{\frac{%
\hbar ^{2}}{2m}\left( k_{0}^{2}-k^{2}\right) +i\eta }dk,
\end{align}%
where $J_{0}$ is the Bessel function and we have used the formula%
\begin{equation}
\int_{0}^{2\pi }e^{ikr\cos \theta }d\theta =2\pi J_{0}\left( kr\right) .
\end{equation}%
The Bessel function is decomposed as the Hankel function%
\begin{equation}
J_{0}\left( kr\right) =\frac{H_{0}^{\left( 1\right) }\left( kr\right)
+H_{0}^{\left( 2\right) }\left( kr\right) }{2},
\end{equation}%
where $H_{0}^{\left( 1\right) }\left( kr\right) $ ($H_{0}^{\left( 2\right)
}\left( kr\right) $) is the Hankel function the first (second) kind. The
asymptotic forms of the Hankel functions are given by%
\begin{align}
& H_{0}^{\left( 1\right) }\left( kr\right) \sim \sqrt{\frac{2}{\pi kr}}%
e^{i\left( kr-\frac{\pi }{4}\right) }, \\
& H_{0}^{\left( 2\right) }\left( kr\right) \sim \sqrt{\frac{2}{\pi kr}}%
e^{-i\left( kr-\frac{\pi }{4}\right) }.
\end{align}%
We make a complex integral%
\begin{equation}
G_{0}\left( r\right) =\frac{1}{16\pi }\int_{-\infty }^{\infty }\frac{k\left(
H_{0}^{\left( 1\right) }\left( kr\right) +H_{0}^{\left( 2\right) }\left(
kr\right) \right) }{\frac{\hbar ^{2}}{2m}\left( k_{0}^{2}-k^{2}\right)
+i\eta }dk,
\end{equation}%
where poles exist at%
\begin{equation}
k=\pm k_{0}+i\eta .
\end{equation}%
We take the upper half circle for $H_{0}^{\left( 1\right) }\left( kr\right) $
and the lower half circle for $H_{0}^{\left( 2\right) }\left( kr\right) $ so
that the complex integral converges. Only the contribution from $%
H_{0}^{\left( 1\right) }\left( kr\right) $ is nonzero and the Green function
is calculated as%
\begin{align}
G_{0}\left( r\right) =& \frac{1}{16\pi }\int_{-\infty }^{\infty }\frac{%
kH_{0}^{\left( 1\right) }\left( kr\right) }{\frac{\hbar ^{2}}{2m}\left(
k_{0}^{2}-k^{2}\right) +i\eta }dk  \notag \\
=& \frac{i}{8}\left( H_{0}^{\left( 1\right) }\left( k_{0}r\right)
+H_{0}^{\left( 1\right) }\left( -k_{0}r\right) \right)  \notag \\
=& \frac{i}{4}H_{0}^{\left( 1\right) }\left( k_{0}r\right) .
\end{align}%
Then, the LDOS has the $s$-wave symmetry.

\subsection{$p$-wave magnets}

Second, we study the Friedel oscillation of the $p$-wave magnet\cite{Suk}.
We calculate the real space representation of the Green function%
\begin{align}
G\left( r\right) =&\frac{1}{\left( 2\pi \right) ^{2}}\int_{0}^{\infty }k%
\frac{e^{ikr\cos \left( \theta -\phi \right) }}{\frac{\hbar ^{2}}{2m}\left(
k_{0}^{2}-k^{2}\right) -sJk\cos \phi +i\eta }dkd\theta  \notag \\
=&\frac{1}{2\pi }\int_{0}^{\infty }\frac{kJ_{0}\left( kr\right) }{\frac{%
\hbar ^{2}}{2m}\left( k_{0}^{2}-k^{2}\right) -sJk\cos \phi +i\eta }dk.
\end{align}%
Poles exist at%
\begin{equation}
k_{p}=-\frac{sJm\cos \phi }{\hbar ^{2}}\pm \sqrt{k_{0}^{2}+\frac{J^{2}m^{2}}{%
\hbar ^{4}}\cos ^{2}\phi }+i\eta ,
\end{equation}%
and we obtain%
\begin{equation}
G\left( r\right) =\frac{i}{4}H_{0}^{\left( 1\right) }\left( k_{p}r\right) .
\end{equation}%
Then, the LDOS has the $p$-wave symmetry and spin dependence.

\subsection{$d$-wave altermagnets}

Third, we study the Friedel oscillation of the $d$-wave altermagnet\cite%
{WangChen24,Suk}. We calculate the real space representation of the Green
function 
\begin{align}
G\left( r\right) =&\frac{1}{\left( 2\pi \right) ^{2}}\int_{0}^{\infty }k%
\frac{e^{ikr\cos \left( \theta -\phi \right) }}{k_{0}^{2}-k^{2}-sJk^{2}\cos
\phi +i\eta }dkd\theta  \notag \\
=&\frac{1}{2\pi }\int_{0}^{\infty }\frac{kJ_{0}\left( kr\right) }{\frac{%
\hbar ^{2}}{2m}\left( k_{0}^{2}-k^{2}\right) -sJk^{2}\cos \phi +i\eta }dk
\end{align}%
Poles exist at%
\begin{equation}
k_{d}=\pm \frac{k_{0}}{\sqrt{1+\frac{2msJ}{\hbar ^{2}}\cos \phi }}+i\eta ,
\end{equation}%
and we obtain%
\begin{equation}
G\left( r\right) =\frac{i}{4}H_{0}^{\left( 1\right) }\left( k_{d}r\right) .
\end{equation}%
Then, the LDOS has the $d$-wave symmetry.

\subsection{$X$-wave magnets}

Finally, we study the Friedel oscillation of the $X$-wave magnet. We
calculate the real space representation of the Green function%
\begin{align}
G\left( r\right) =&\frac{1}{\left( 2\pi \right) ^{2}}\int_{0}^{\infty }k%
\frac{e^{ikr\cos \left( \theta -\phi \right) }}{k_{0}^{2}-k^{2}-f_{X,X^{%
\prime }}^{\text{2D}}\left( \mathbf{k}\right) +i\eta }dkd\theta  \notag \\
=&\frac{1}{2\pi }\int_{0}^{\infty }\frac{kJ_{0}\left( kr\right) }{\frac{%
\hbar ^{2}}{2m}\left( k_{0}^{2}-k^{2}\right) -f_{X,X^{\prime }}^{\text{2D}%
}\left( \mathbf{k}\right) +i\eta }dk
\end{align}%
The pole is determined by the solution of%
\begin{equation}
\frac{\hbar ^{2}}{2m}\left( k_{0}^{2}-k_{X,X^{\prime }}^{2}\right)
-f_{X,X^{\prime }}^{\text{2D}}\left( \mathbf{k}_{X,X^{\prime }}\right)
+i\eta =0,
\end{equation}%
which has the $X$-wave symmetry although it is hard to obtain analytic
formula except for the $p$-wave magnet and the $d$-wave altermagnet. By
using it, we obtain%
\begin{equation}
G\left( r\right) =\frac{i}{4}H_{0}^{\left( 1\right) }\left( k_{X,X^{\prime
}}r\right) .
\end{equation}%
Then, the LDOS has the $X$-wave symmetry and spin dependence.

\section{Summaries, discussions and outlooks}

\label{SecSummary}

We reviewed recent progress on quantum geometry and $X$-wave magnets. We
systematically formulated quantum geometry based on the quantum distance.
Quantum geometry is generalized to the Zeemann quantum geometry,
non-Hermitian quantum geometry and quantum information geometry. Then, we
reviewed the $X$-wave magnets including altermagnets and $p$-wave magnets in
a universal manner. Universal transport properties were discussed for the $X$%
-wave magnets.

In this paper, we have derived various analytic formulas based on the
two-band Hamiltonian. However, actual materials are more complicated. For
example, four-band models including the orbital degrees of freedom have been
studied\cite{Roig,pwave}. Nevertheless, if there are only two bands at the
Fermi energy, the effective Hamiltonian is reduced to be a two-band
Hamiltonian\cite{EzawaPNeel}. In addition, optical absorption is well
described by the two-band Hamiltonian consisting of the valence and
conduction bands. The two-band Hamiltonian reveals the basic structure of
the relevant phenomena in these cases. It is an interesting problem to study
based on more realistic models.

In passing, we give some outlooks. The general relativistic quantities such
as the Christoffel symbols are defined based on the quantum metric as shown
in Sec.\ref{SecQGR}. However, it is not clear whether there are physical
observable quantities such as nonlinear conductivities or nonlinear optical
conductivities directly relating to them. Quantum geometry is mainly studied
in the context of condensed matter physics. Applications of the notion of
quantum geometry to quantum information and quantum computing are still
rare, where quantum information geometry for density matrices will be
useful. There will be developments in this direction. Zeeman quantum
geometry will be useful for spintronics.

Magnonic properties of the $X$-wave magnets will also be interesting, which
are mainly studied so far for $d$-wave altermagnets\cite{SmeMag,Cui} and $g$%
-wave altermagnets\cite{LiuMag}. The magnon spectrum is given by\cite{Eto}%
\begin{equation}
E_{\pm }\left( \mathbf{k}\right) =vk+\kappa k^{3}\pm Jk^{N_{x}}\sin N_{X}\phi
\end{equation}%
for the $X$-wave magnet in general. With respect to material realization of
the $X$-wave magnets, there are no experiments on the $f$-wave, $g$-wave and 
$i$-wave magnets in two dimensions and the $f$-wave and $i$-wave magnets in
three dimensions. In this paper, we have mainly focused on electronic and
optical properties coupled with the $X$-wave magnet. On the other hand, the
control of the magnetism of the $X$-wave magnet is also important. It is a
nontrivial problem to switch the direction of the spins of the $X$-wave
magnet. In order to switch the direction of spins, multiferroic altermagnets%
\cite{SmeMulti,GuPRL} are promising because the spin direction is reversed
by applying electric field. It is a fascinating problem both theoretically
and experimentally to search universal physics in the viewpoint of the $X$%
-wave magnets.

\section*{Acknowledgements}

The author is grateful to N. Nagaosa, M. Hirschberger, S. Okumura, Y.
Motome, T. Morimoto, H. Seo, J. Wang, B. J. Yang and W. Chen for helpful
discussions on the subject. This work is supported by CREST, JST (Grants No.
JPMJCR20T2) and Grants-in-Aid for Scientific Research from MEXT KAKENHI
(Grant No. 23H00171).

\section*{Appendices}

\appendix

\section{Quantum distance and Abelian quantum geometric tensor}

\label{ApQ}

The Hilbert-Schmidt distance is defined by%
\begin{equation}
\left( ds_{\text{HS}}\right) ^{2}\equiv 1-\left\vert \left\langle \psi
_{n}\left( \mathbf{k}\right) \left\vert \psi _{n}\left( \mathbf{k}\right) +d%
\mathbf{k}\right\rangle \right. \right\vert ^{2}.
\end{equation}%
The wave function is expanded as%
\begin{align}
& \left\vert \psi _{n}\left( \mathbf{k}+d\mathbf{k}\right) \right\rangle 
\notag \\
=& \left\vert \psi _{n}\left( \mathbf{k}\right) \right\rangle +\left\vert
\partial _{k_{\mu }}\psi _{n}\left( \mathbf{k}\right) \right\rangle dk_{\mu }
\notag \\
& +\frac{1}{2}\left\vert \partial _{k_{\mu }}\partial _{k_{\nu }}\psi
_{n}\left( \mathbf{k}\right) \right\rangle dk_{\mu }dk_{\nu }+\cdots .
\end{align}

The inner product reads%
\begin{align}
& \left\vert \left\langle \psi _{n}\left( \mathbf{k}\right) \left\vert \psi
_{n}\left( \mathbf{k}^{\prime }\right) \right\rangle \right. \right\vert ^{2}
\notag \\
=& \left\langle \psi _{n}\left( \mathbf{k}^{\prime }\right) \left\vert \psi
_{n}\left( \mathbf{k}\right) \right\rangle \right. \left\langle \psi
_{n}\left( \mathbf{k}\right) \left\vert \psi _{n}\left( \mathbf{k}^{\prime
}\right) \right\rangle \right.  \notag \\
=& \left\langle \psi _{n}\left( \mathbf{k}\right) \left\vert \psi _{n}\left( 
\mathbf{k}\right) \right\rangle \right.  \notag \\
& +\left( \left\langle \partial _{k_{\mu }}\psi _{n}\left( \mathbf{k}\right)
\left\vert \psi _{n}\left( \mathbf{k}\right) \right\rangle \right.
+\left\langle \psi _{n}\left( \mathbf{k}\right) \left\vert \partial _{k_{\mu
}}\psi _{n}\left( \mathbf{k}\right) \right\rangle \right. \right) dk_{\mu } 
\notag \\
& +(\frac{1}{2}\left\langle \partial _{k_{\mu }}\partial _{k_{\nu }}\psi
_{n}\left( \mathbf{k}\right) \left\vert \psi _{n}\left( \mathbf{k}\right)
\right\rangle \right. +\frac{1}{2}\left\langle \psi _{n}\left( \mathbf{k}%
\right) \left\vert \partial _{k_{\mu }}\partial _{k_{\nu }}\psi _{n}\left( 
\mathbf{k}\right) \right\rangle \right.  \notag \\
& +\left\langle \partial _{k_{\mu }}\psi _{n}\left( \mathbf{k}\right)
\left\vert \psi _{n}\left( \mathbf{k}\right) \right\rangle \right.
\left\langle \psi _{n}\left( \mathbf{k}\right) \left\vert \partial _{k_{\mu
}}\psi _{n}\left( \mathbf{k}\right) \right\rangle \right. )dk_{\mu }dk_{\nu
}.
\end{align}%
\newline
The term proportional to $dk_{\mu }$ is zero because%
\begin{align}
& \left( \left\langle \partial _{k_{\mu }}\psi _{n}\left( \mathbf{k}\right)
\left\vert \psi _{n}\left( \mathbf{k}\right) \right\rangle \right.
+\left\langle \psi _{n}\left( \mathbf{k}\right) \left\vert \partial _{k_{\mu
}}\psi _{n}\left( \mathbf{k}\right) \right\rangle \right. \right)  \notag \\
=& \partial _{k_{\mu }}\left( \left\langle \psi _{n}\left( \mathbf{k}\right)
\left\vert \psi _{n}\left( \mathbf{k}\right) \right\rangle \right. \right)
=0,
\end{align}%
where we have used the normalization condition $\left\langle \psi _{n}\left( 
\mathbf{k}\right) \left\vert \psi _{n}\left( \mathbf{k}\right) \right\rangle
\right. =1$. By using the relation%
\begin{align}
& \partial _{k_{\mu }}\partial _{k_{\nu }}\left( \left\langle \psi
_{n}\left( \mathbf{k}\right) \left\vert \psi _{n}\left( \mathbf{k}\right)
\right\rangle \right. \right)  \notag \\
=& \left\langle \partial _{k_{\mu }}\partial _{k_{\nu }}\psi _{n}\left( 
\mathbf{k}\right) \left\vert \psi _{n}\left( \mathbf{k}\right) \right\rangle
\right. +\left\langle \psi _{n}\left( \mathbf{k}\right) \left\vert \partial
_{k_{\mu }}\partial _{k_{\nu }}\psi _{n}\left( \mathbf{k}\right)
\right\rangle \right.  \notag \\
& +\left\langle \partial _{k_{\mu }}\psi _{n}\left( \mathbf{k}\right)
\left\vert \partial _{k_{\nu }}\psi _{n}\left( \mathbf{k}\right)
\right\rangle \right. +\left\langle \partial _{k_{\nu }}\psi _{n}\left( 
\mathbf{k}\right) \left\vert \partial _{k_{\mu }}\psi _{n}\left( \mathbf{k}%
\right) \right\rangle \right. ,
\end{align}%
we have%
\begin{align}
& \frac{1}{2}\left\langle \partial _{k_{\mu }}\partial _{k_{\nu }}\psi
_{n}\left( \mathbf{k}\right) \left\vert \psi _{n}\left( \mathbf{k}\right)
\right\rangle \right. +\frac{1}{2}\left\langle \psi _{n}\left( \mathbf{k}%
\right) \left\vert \partial _{k_{\mu }}\partial _{k_{\nu }}\psi _{n}\left( 
\mathbf{k}\right) \right\rangle \right.  \notag \\
=& -\frac{1}{2}(\left\langle \partial _{k_{\mu }}\psi _{n}\left( \mathbf{k}%
\right) \left\vert \partial _{k_{\nu }}\psi _{n}\left( \mathbf{k}\right)
\right\rangle \right.  \notag \\
& +\left\langle \partial _{k_{\nu }}\psi _{n}\left( \mathbf{k}\right)
\left\vert \partial _{k_{\mu }}\psi _{n}\left( \mathbf{k}\right)
\right\rangle \right. )dk_{\mu }dk_{\nu }  \notag \\
=& -\left\langle \partial _{k_{\mu }}\psi _{n}\left( \mathbf{k}\right)
\left\vert \partial _{k_{\nu }}\psi _{n}\left( \mathbf{k}\right)
\right\rangle \right. dk_{\mu }dk_{\nu }.
\end{align}%
Then, the inner product is written as%
\begin{align}
& \left\vert \left\langle \psi _{n}\left( \mathbf{k}\right) \left\vert \psi
_{n}\left( \mathbf{k}^{\prime }\right) \right\rangle \right. \right\vert ^{2}
\notag \\
=& 1+(\left\langle \partial _{k_{\mu }}\psi _{n}\left( \mathbf{k}\right)
\left\vert \psi _{n}\left( \mathbf{k}\right) \right\rangle \right.
\left\langle \psi _{n}\left( \mathbf{k}\right) \left\vert \partial _{k_{\mu
}}\psi _{n}\left( \mathbf{k}\right) \right\rangle \right.  \notag \\
& -\left\langle \partial _{k_{\mu }}\psi _{n}\left( \mathbf{k}\right)
\left\vert \partial _{k_{\nu }}\psi _{n}\left( \mathbf{k}\right)
\right\rangle \right. )dk_{\mu }dk_{\nu }  \notag \\
=& 1+\left\langle \partial _{k_{\mu }}\psi _{n}\left( \mathbf{k}\right)
\right\vert \left( \left\vert \psi _{n}\left( \mathbf{k}\right)
\right\rangle \left\langle \psi _{n}\left( \mathbf{k}\right) \right\vert
-1\right) \left\vert \partial _{k_{\mu }}\psi _{n}\left( \mathbf{k}\right)
\right\rangle dk_{\mu }dk_{\nu },
\end{align}%
and the quantum distance reads%
\begin{align}
& \left( ds_{\text{HS}}\right) ^{2}  \notag \\
\equiv & \left\langle \partial _{k_{\mu }}\psi _{n}\left( \mathbf{k}\right)
\right\vert \left( 1-\left\vert \psi _{n}\left( \mathbf{k}\right)
\right\rangle \left\langle \psi _{n}\left( \mathbf{k}\right) \right\vert
^{2}\right) \left\vert \partial _{k_{\mu }}\psi _{n}\left( \mathbf{k}\right)
\right\rangle dk_{\mu }dk_{\nu }  \notag \\
=& \mathcal{F}_{n}^{\mu \nu }dk_{\mu }dk_{\nu }.
\end{align}%
This is Eq.(\ref{F12}) in the main text.

\section{Quantum distance and non-Abelian quantum geometric tensor}

\label{ApQ2}

We summarize non-Abelian quantum geometry for the $N$-fold degenerate wave
functions\cite{Ma}.

The Hilbert-Schmidt distance is defined by%
\begin{equation}
\left( ds_{\text{HS}}\right) ^{2}\equiv 1-\left\vert \left\langle u\left( 
\mathbf{k}\right) \left\vert u\left( \mathbf{k}+d\mathbf{k}\right)
\right\rangle \right. \right\vert ^{2}.
\end{equation}%
where $\left\vert u\left( \mathbf{k}\right) \right\rangle $ is the wave
function of $N$-fold degenerate bands%
\begin{equation}
\left\vert u\left( \mathbf{k}\right) \right\rangle
=\sum_{n=1}^{N}c_{n}\left\vert \psi _{n}\left( \mathbf{k}\right)
\right\rangle ,
\end{equation}%
where 
\begin{equation}
\sum_{n=1}^{N}\left\vert c_{n}\right\vert ^{2}=1.
\end{equation}%
The inner product is calculated as%
\begin{align}
& \left\vert \left\langle u\left( \mathbf{k}\right) \left\vert u\left( 
\mathbf{k}+d\mathbf{k}\right) \right\rangle \right. \right\vert ^{2}  \notag
\\
=& \left\langle u\left( \mathbf{k}\right) \left\vert u\left( \mathbf{k}%
\right) \right\rangle \right.  \notag \\
& +\left( \left\langle \partial _{k_{\mu }}u\left( \mathbf{k}\right)
\left\vert u\left( \mathbf{k}\right) \right\rangle \right. +\left\langle
u\left( \mathbf{k}\right) \left\vert \partial _{k_{\mu }}u\left( \mathbf{k}%
\right) \right\rangle \right. \right) dk_{\mu }  \notag \\
& +(\frac{1}{2}\left\langle \partial _{k_{\mu }}\partial _{k_{\nu }}u\left( 
\mathbf{k}\right) \left\vert u\left( \mathbf{k}\right) \right\rangle \right.
+\frac{1}{2}\left\langle u\left( \mathbf{k}\right) \left\vert \partial
_{k_{\mu }}\partial _{k_{\nu }}u\left( \mathbf{k}\right) \right\rangle
\right.  \notag \\
& +\left\langle \partial _{k_{\mu }}u\left( \mathbf{k}\right) \left\vert
u\left( \mathbf{k}\right) \right\rangle \right. \left\langle u\left( \mathbf{%
k}\right) \left\vert \partial _{k_{\mu }}u\left( \mathbf{k}\right)
\right\rangle \right. )dk_{\mu }dk_{\nu }  \notag \\
=& 1+(-\frac{1}{2}\left\langle \partial _{k_{\mu }}u\left( \mathbf{k}\right)
\left\vert \partial _{k_{\nu }}u\left( \mathbf{k}\right) \right\rangle
\right. -\frac{1}{2}\left\langle \partial _{k_{\nu }}u\left( \mathbf{k}%
\right) \left\vert \partial _{k_{\mu }}u\left( \mathbf{k}\right)
\right\rangle \right.  \notag \\
& +\left\langle \partial _{k_{\mu }}u\left( \mathbf{k}\right) \left\vert
u\left( \mathbf{k}\right) \right\rangle \right. \left\langle u\left( \mathbf{%
k}\right) \left\vert \partial _{k_{\mu }}u\left( \mathbf{k}\right)
\right\rangle \right. )dk_{\mu }dk_{\nu }  \notag \\
=& 1+\left\langle \partial _{k_{\mu }}u\left( \mathbf{k}\right) \right\vert
\left( \left\vert u\left( \mathbf{k}\right) \right\rangle \left\langle
u\left( \mathbf{k}\right) \right\vert -1\right) \left\vert \partial _{k_{\mu
}}u\left( \mathbf{k}\right) \right\rangle dk_{\mu }dk_{\nu }.
\end{align}%
Then, we have%
\begin{equation}
\left( ds_{\text{HS}}\right) ^{2}=\left\langle \partial _{k_{\mu }}u\left( 
\mathbf{k}\right) \right\vert \left( 1-P\left( \mathbf{k}\right) \right)
\left\vert \partial _{k_{\mu }}u\left( \mathbf{k}\right) \right\rangle
dk_{\mu }dk_{\nu },
\end{equation}%
where $P\left( \mathbf{k}\right) \equiv \left\vert u\left( \mathbf{k}\right)
\right\rangle \left\langle u\left( \mathbf{k}\right) \right\vert $ is the
projection operator to the $N$-fold degenerate bands satisyfing $P\left( 
\mathbf{k}\right) ^{2}=P\left( \mathbf{k}\right) $. By diagonalizing it, we
obtain%
\begin{equation}
P\left( \mathbf{k}\right) \equiv \sum_{n=1}^{N}\left\vert \psi _{n}\left( 
\mathbf{k}\right) \right\rangle \left\langle \psi _{n}\left( \mathbf{k}%
\right) \right\vert .
\end{equation}%
This is Eq.(\ref{F12Mu}) in the main text.

\section{Hellmann-Feynman theorem}

\label{ApHF}

By differentiating the eigenvalue equation%
\begin{equation}
H\left\vert \psi _{n}\left( \mathbf{k}\right) \right\rangle =\varepsilon
_{n}\left( \mathbf{k}\right) \left\vert \psi _{n}\left( \mathbf{k}\right)
\right\rangle  \label{EG}
\end{equation}%
with respect to $k_{\mu }$, we obtain%
\begin{align}
& \partial _{k_{\mu }}H\left( \mathbf{k}\right) \left\vert \psi _{n}\left( 
\mathbf{k}\right) \right\rangle +H\left( \mathbf{k}\right) \partial _{k_{\mu
}}\left\vert \psi _{n}\left( \mathbf{k}\right) \right\rangle  \notag \\
=& \partial _{k_{\mu }}\varepsilon _{n}\left( \mathbf{k}\right) \left\vert
\psi _{n}\left( \mathbf{k}\right) \right\rangle +\varepsilon _{n}\left( 
\mathbf{k}\right) \partial _{k_{\mu }}\left\vert \psi _{n}\left( \mathbf{k}%
\right) \right\rangle .
\end{align}%
Applying $\left\langle \psi _{m}\left( \mathbf{k}\right) \right\vert $ to
the above equation, we obtain%
\begin{align}
& \left\langle \psi _{m}\left( \mathbf{k}\right) \right\vert \partial
_{k_{\mu }}H\left( \mathbf{k}\right) \left\vert \psi _{n}\left( \mathbf{k}%
\right) \right\rangle  \notag \\
& +\left\langle \psi _{m}\left( \mathbf{k}\right) \right\vert H\left( 
\mathbf{k}\right) \partial _{k_{\mu }}\left\vert \psi _{n}\left( \mathbf{k}%
\right) \right\rangle  \notag \\
=& \left\langle \psi _{m}\left( \mathbf{k}\right) \right\vert \partial
_{k_{\mu }}\varepsilon _{n}\left( \mathbf{k}\right) \left\vert \psi
_{n}\left( \mathbf{k}\right) \right\rangle  \notag \\
& +\left\langle \psi _{m}\left( \mathbf{k}\right) \right\vert \varepsilon
_{n}\left( \mathbf{k}\right) \partial _{k_{\mu }}\left\vert \psi _{n}\left( 
\mathbf{k}\right) \right\rangle .
\end{align}%
Using (\ref{EG}) and defining the velocity 
\begin{equation}
v_{\mu }=\frac{\partial H\left( \mathbf{k}\right) }{\hbar \partial k_{\mu }},
\end{equation}%
we obtain%
\begin{align}
& \hbar \left\langle \psi _{m}\left( \mathbf{k}\right) \right\vert v_{\mu
}\left\vert \psi _{n}\left( \mathbf{k}\right) \right\rangle  \notag \\
& +\left\langle \psi _{m}\left( \mathbf{k}\right) \right\vert \varepsilon
_{m}\left( \mathbf{k}\right) \partial _{k_{\mu }}\left\vert \psi _{n}\left( 
\mathbf{k}\right) \right\rangle  \notag \\
=& \left\langle \psi _{m}\left( \mathbf{k}\right) \right\vert \partial
_{k_{\mu }}\varepsilon _{n}\left( \mathbf{k}\right) \left\vert \psi
_{n}\left( \mathbf{k}\right) \right\rangle  \notag \\
& +\varepsilon _{n}\left( \mathbf{k}\right) \left\langle \psi _{m}\left( 
\mathbf{k}\right) \right\vert \partial _{k_{\mu }}\left\vert \psi _{n}\left( 
\mathbf{k}\right) \right\rangle ,
\end{align}%
where we have used the eigenequation%
\begin{equation}
\left\langle \psi _{m}\left( \mathbf{k}\right) \right\vert H\left( \mathbf{k}%
\right) =\left\langle \psi _{m}\left( \mathbf{k}\right) \right\vert
\varepsilon _{m}\left( \mathbf{k}\right) .
\end{equation}%
This equation is rewritten as%
\begin{align}
& \left\langle \psi _{m}\left( \mathbf{k}\right) \right\vert v_{\mu
}\left\vert \psi _{n}\left( \mathbf{k}\right) \right\rangle  \notag \\
=& \frac{1}{\hbar }\partial _{k_{\mu }}\varepsilon _{n}\left( \mathbf{k}%
\right) \langle \psi _{m}\left( \mathbf{k}\right) |\psi _{n}\left( \mathbf{k}%
\right) \rangle  \notag \\
& +\frac{1}{\hbar }\left( \varepsilon _{n}\left( \mathbf{k}\right)
-\varepsilon _{m}\left( \mathbf{k}\right) \right) \left\langle \psi
_{m}\left( \mathbf{k}\right) \right\vert \partial _{k_{\mu }}\left\vert \psi
_{n}\left( \mathbf{k}\right) \right\rangle .  \label{Helmann}
\end{align}%
It is Eq.(\ref{HF}) in the main text. Assuming that the states satisfy the
orthonormalization condition $\langle \psi _{m}\left( \mathbf{k}\right)
|\psi _{n}\left( \mathbf{k}\right) \rangle =\delta _{mn}$ we find%
\begin{align}
& \left\langle \psi _{m}\left( \mathbf{k}\right) \right\vert v_{\mu
}\left\vert \psi _{n}\left( \mathbf{k}\right) \right\rangle  \notag \\
=& \frac{1}{\hbar }\left( \varepsilon _{n}\left( \mathbf{k}\right)
-\varepsilon _{m}\left( \mathbf{k}\right) \right) \left\langle \psi
_{m}\left( \mathbf{k}\right) \right\vert \partial _{k_{\mu }}\left\vert \psi
_{n}\left( \mathbf{k}\right) \right\rangle ,\quad \text{for}\quad m\neq n.
\end{align}%
Indeed, when $n$ is the band index, it is obvious that $\langle \psi
_{m}\left( \mathbf{k}\right) |\psi _{n}\left( \mathbf{k}\right) \rangle =0$
for $m\neq n$. Furthermore, it is always possible to normalized as $\langle
\psi _{n}\left( \mathbf{k}\right) |\psi _{n}\left( \mathbf{k}\right) \rangle
=1$ at each $\mathbf{k}$.

By applying $\left\vert \psi _{m}\left( \mathbf{k}\right) \right\rangle $ to
Eq.(\ref{Helmann}) and taking a sum of $m$, we have%
\begin{align}
& \sum_{m}\left\vert \psi _{m}\left( \mathbf{k}\right) \right\rangle
\left\langle \psi _{m}\left( \mathbf{k}\right) \right\vert v_{\mu
}\left\vert \psi _{n}\left( \mathbf{k}\right) \right\rangle  \notag \\
=& \frac{1}{\hbar }\partial _{k_{\mu }}\varepsilon _{n}\left( \mathbf{k}%
\right) \sum_{m}\left\vert \psi _{m}\left( \mathbf{k}\right) \right\rangle
\langle \psi _{m}\left( \mathbf{k}\right) |\psi _{n}\left( \mathbf{k}\right)
\rangle  \notag \\
& +\frac{1}{\hbar }\left( \varepsilon _{n}\left( \mathbf{k}\right)
-\varepsilon _{m}\left( \mathbf{k}\right) \right)  \notag \\
& \times \sum_{m}\left\vert \psi _{m}\left( \mathbf{k}\right) \right\rangle
\left\langle \psi _{m}\left( \mathbf{k}\right) \right\vert \partial _{k_{\mu
}}\left\vert \psi _{n}\left( \mathbf{k}\right) \right\rangle  \notag \\
=& \frac{1}{\hbar }\partial _{k_{\mu }}\varepsilon _{n}\left( \mathbf{k}%
\right) \sum_{m}\left\vert \psi _{m}\left( \mathbf{k}\right) \right\rangle
\delta _{mn}  \notag \\
& +\frac{1}{\hbar }\left( \varepsilon _{n}\left( \mathbf{k}\right)
-\varepsilon _{m}\left( \mathbf{k}\right) \right) \sum_{m}\left\vert \psi
_{m}\left( \mathbf{k}\right) \right\rangle \left\langle \psi _{m}\left( 
\mathbf{k}\right) \right\vert .
\end{align}%
We have%
\begin{align}
& \sum_{m}\left\vert \psi _{m}\left( \mathbf{k}\right) \right\rangle
\left\langle \psi _{m}\left( \mathbf{k}\right) \right\vert \partial _{k_{\mu
}}\left\vert \psi _{n}\left( \mathbf{k}\right) \right\rangle  \notag \\
=& \sum_{m\neq n}\left\vert \psi _{m}\left( \mathbf{k}\right) \right\rangle 
\frac{\left\langle \psi _{m}\left( \mathbf{k}\right) \right\vert \hbar
v_{\mu }\left\vert \psi _{n}\left( \mathbf{k}\right) \right\rangle }{%
\varepsilon _{n}\left( \mathbf{k}\right) -\varepsilon _{m}\left( \mathbf{k}%
\right) }.
\end{align}%
for $m\neq n$. By using the complete condition $\sum_{m}\left\vert \psi
_{m}\left( \mathbf{k}\right) \right\rangle \left\langle \psi _{m}\left( 
\mathbf{k}\right) \right\vert =1$, we have%
\begin{equation}
\partial _{k_{\mu }}\left\vert \psi _{n}\left( \mathbf{k}\right)
\right\rangle =\sum_{m\neq n}\left\vert \psi _{m}\left( \mathbf{k}\right)
\right\rangle \frac{\left\langle \psi _{m}\left( \mathbf{k}\right)
\right\vert \hbar v_{\mu }\left\vert \psi _{n}\left( \mathbf{k}\right)
\right\rangle }{\varepsilon _{n}\left( \mathbf{k}\right) -\varepsilon
_{m}\left( \mathbf{k}\right) }.  \label{dPsi}
\end{equation}%
We use Eq.(\ref{dPsi}) to analyze the diagonal component of the quantum
metric%
\begin{align}
\mathcal{F}_{nn}^{\mu \nu }\left( \mathbf{k}\right) =& \left\langle \partial
_{k_{\mu }}\psi _{n}\left( \mathbf{k}\right) \right\vert \mathcal{Q}\left( 
\mathbf{k}\right) \left\vert \partial _{k_{\nu }}\psi _{n}\left( \mathbf{k}%
\right) \right\rangle  \notag \\
=& \sum_{m^{\prime }\neq n}\frac{\left\langle \psi _{n}\left( \mathbf{k}%
\right) \right\vert \hbar v_{\mu }\left\vert \psi _{m^{\prime }}\left( 
\mathbf{k}\right) \right\rangle }{\varepsilon _{n}\left( \mathbf{k}\right)
-\varepsilon _{m^{\prime }}\left( \mathbf{k}\right) }\left\langle \psi
_{m^{\prime }}\left( \mathbf{k}\right) \right\vert  \notag \\
& \times \left( 1-\sum_{m^{\prime \prime }=1}^{N}\left\vert \psi _{m^{\prime
\prime }}\left( \mathbf{k}\right) \right\rangle \left\langle \psi
_{m^{\prime \prime }}\left( \mathbf{k}\right) \right\vert \right)  \notag \\
& \times \sum_{n^{\prime }\neq n}\left\vert \psi _{n^{\prime }}\left( 
\mathbf{k}\right) \right\rangle \frac{\left\langle \psi _{n^{\prime }}\left( 
\mathbf{k}\right) \right\vert \hbar v_{\nu }\left\vert \psi _{n}\left( 
\mathbf{k}\right) \right\rangle }{\varepsilon _{n}\left( \mathbf{k}\right)
-\varepsilon _{n^{\prime }}\left( \mathbf{k}\right) }.
\end{align}%
We have%
\begin{align}
& \left\langle \psi _{m^{\prime }}\left( \mathbf{k}\right) \right\vert
\left( 1-\sum_{m^{\prime \prime }=1}^{N}\left\vert \psi _{m^{\prime \prime
}}\left( \mathbf{k}\right) \right\rangle \left\langle \psi _{m^{\prime
\prime }}\left( \mathbf{k}\right) \right\vert \right) \left\vert \psi
_{n^{\prime }}\left( \mathbf{k}\right) \right\rangle  \notag \\
=& \delta _{m^{\prime }n^{\prime }}-\sum_{m^{\prime \prime }=1}^{N}\delta
_{m^{\prime }m^{\prime \prime }}\delta _{m^{\prime \prime }n^{\prime }}=0, 
\notag
\end{align}%
when $m^{\prime \prime }$ is the occupied band, while%
\begin{align}
& \left\langle \psi _{m^{\prime }}\left( \mathbf{k}\right) \right\vert
\left( 1-\sum_{m^{\prime \prime }=1}^{N}\left\vert \psi _{m^{\prime \prime
}}\left( \mathbf{k}\right) \right\rangle \left\langle \psi _{m^{\prime
\prime }}\left( \mathbf{k}\right) \right\vert \right) \left\vert \psi
_{n^{\prime }}\left( \mathbf{k}\right) \right\rangle  \notag \\
=& \delta _{m^{\prime }n^{\prime }},
\end{align}%
when $m^{\prime \prime }$ is the unoccupied band.%
\begin{align}
& \mathcal{F}_{nn}^{\mu \nu }\left( \mathbf{k}\right)  \notag \\
=& \sum_{m\in \text{Unoccupied}}\frac{\left\langle \psi _{n}\left( \mathbf{k}%
\right) \right\vert \hbar v_{x}\left\vert \psi _{m}\left( \mathbf{k}\right)
\right\rangle \left\langle \psi _{m}\left( \mathbf{k}\right) \right\vert
\hbar v_{y}\left\vert \psi _{n}\left( \mathbf{k}\right) \right\rangle }{%
\left[ \varepsilon _{n}\left( \mathbf{k}\right) -\varepsilon _{m}\left( 
\mathbf{k}\right) \right] ^{2}}.
\end{align}%
The quantum metric reads%
\begin{equation}
g_{n}^{\mu \nu }=\text{Re}\sum_{m\in \text{Unoccupied}}\frac{\left\langle
\psi _{n}\left( \mathbf{k}\right) \right\vert \hbar v_{x}\left\vert \psi
_{m}\left( \mathbf{k}\right) \right\rangle \left\langle \psi _{m}\left( 
\mathbf{k}\right) \right\vert \hbar v_{y}\left\vert \psi _{n}\left( \mathbf{k%
}\right) \right\rangle }{\left[ \varepsilon _{n}\left( \mathbf{k}\right)
-\varepsilon _{m}\left( \mathbf{k}\right) \right] ^{2}},
\end{equation}%
while the Berry curvature reads%
\begin{equation}
\Omega _{n}^{\mu \nu }=2\text{Im}\sum_{m\in \text{Unoccupied}}\frac{%
\left\langle \psi _{n}\left( \mathbf{k}\right) \right\vert \hbar
v_{x}\left\vert \psi _{m}\left( \mathbf{k}\right) \right\rangle \left\langle
\psi _{m}\left( \mathbf{k}\right) \right\vert \hbar v_{y}\left\vert \psi
_{n}\left( \mathbf{k}\right) \right\rangle }{\left[ \varepsilon _{n}\left( 
\mathbf{k}\right) -\varepsilon _{m}\left( \mathbf{k}\right) \right] ^{2}}.
\end{equation}

\section{Bulk photovoltaic effects}

We apply alternating\ and monochromatic electric fields,%
\begin{equation}
E_{x}\left( t\right) =E_{x}\left( \omega \right) e^{-i\omega t}+E_{x}\left(
-\omega \right) e^{i\omega t}  \label{monochro}
\end{equation}%
with $\omega >0$. We assume that the electric field is real,%
\begin{equation}
E_{x}\left( t\right) =E_{x}^{\ast }\left( t\right) ,
\end{equation}%
or%
\begin{equation}
E_{x}\left( \omega \right) =E_{x}^{\ast }\left( -\omega \right) .
\label{Exw}
\end{equation}%
Excited electrons from the valence band and the conduction band contribute
to the bulk photovoltaic current.

\subsection{Injection currents}

\label{ApInject}

We derive the injection current. {We consider an acceleration due to the
static electric field and an optical excitation from the valence band to the
conduction band due to the oscillatory electric field, separately}\cite%
{Jerk,JerkComment,JerkReply,snap}. The equation of motion of electrons is%
\begin{equation}
\hbar \frac{d\mathbf{k}}{dt}=-e\frac{\partial \mathbf{A}_{0}}{\partial t},
\end{equation}%
where $\mathbf{A}_{0}=\left( A_{0},0,0\right) $ is the vector potential. The
static electric field is obtained from the vector potential as%
\begin{equation}
E_{x}\left( 0\right) \equiv -\frac{\partial A_{0}}{\partial t},
\end{equation}%
or $A_{0}=-E_{x}\left( 0\right) t+$constant. The Bloch velocity under
electric field is given by the minimal substitution,%
\begin{equation}
v_{n}\left( k_{x}\right) \rightarrow v_{n}\left( k_{x}-eA_{0}/\hbar \right) .
\label{vk}
\end{equation}%
A comment is in order. We have only considered the contribution from the
static electric field in the vector potential. Even if we take into account
the contribution from the oscillatory electric field in the vector
potential, the time evolution of the mean momentum does not change. It is
understood as follows. The equation of motion under the oscillatory electric
field reads%
\begin{equation}
\hbar \frac{dk}{dt}=eE_{x}\left( 0\right) +eE_{x}\left( \omega \right) \cos
\omega t,
\end{equation}%
whose solution is given by%
\begin{equation}
k=\frac{e}{\hbar }\left( tE_{x}\left( 0\right) +\frac{E_{x}\left( \omega
\right) }{\omega }\sin \omega t\right) +k_{0}.
\end{equation}%
Its mean for one period with respect to the time $t$\ is zero.\ Hence, it is
enough only to consider the static electric field in the vector potential.

We expand Eq.(\ref{vk}) in powers of $E_{x}\left( 0\right) $, and obtain%
\begin{equation}
\frac{d^{2}v_{n}^{x}}{dt^{2}}=\left( \frac{e}{\hbar }\right) ^{2}\frac{%
\partial ^{3}\omega _{n}}{\partial k_{x}^{3}}\left[ E_{x}\left( 0\right) %
\right] ^{2},
\end{equation}%
where we have used%
\begin{equation}
\frac{dv_{n}^{x}}{dt}=\frac{dk_{x}}{dt}\frac{dv_{n}^{x}}{dk_{x}}=-\left( 
\frac{e}{\hbar }\right) \frac{\partial A_{0}}{\partial t}\frac{d^{2}\omega
_{n}}{dk_{x}^{2}}=\frac{e}{\hbar }\frac{d^{2}\omega _{n}}{dk_{x}^{2}}%
E_{x}\left( 0\right) .
\end{equation}%
The current is given by%
\begin{equation}
J=\frac{e}{V}\sum_{n,\mathbf{k}}f_{n}v_{n}^{x},
\end{equation}%
with the velocity along the $x$ direction%
\begin{equation}
v_{n}^{x}\equiv \frac{\partial \omega _{n}}{\partial k_{x}}.
\end{equation}%
The injection current originates in $dJ^{x;x^{2}}/dt$. The injection current
is obtained as%
\begin{equation}
\frac{dJ_{\text{inject}}^{x;x^{2}}}{dt}=\frac{e}{V}\sum_{n}\sum_{s=0}^{1}%
\left( 
\begin{array}{c}
\ell -1 \\ 
s%
\end{array}%
\right) \frac{d^{s}f_{n}}{dt^{s}}\frac{d^{1-s}v_{n}^{x}}{dt^{1-s}}.
\label{JEll}
\end{equation}%
The term proportional to $E_{x}\left( \omega \right) E_{x}\left( -\omega
\right) $ is given by taking the terms with $s=1$ from Eq.(\ref{JEll}) as%
\begin{equation}
\frac{dJ_{\text{inject}}^{x;x^{2}}}{dt}=\frac{e}{V}\sum_{n}\frac{df_{n}}{dt}%
v_{n}^{x}.  \label{JElll}
\end{equation}%
The Fermi golden rule reads%
\begin{equation}
\frac{df_{\pm }}{dt}=\pm \frac{2\pi e^{2}}{\hbar ^{2}}\left\vert \mathbf{E}%
\left( \omega \right) \cdot \mathbf{a}_{+-}\right\vert ^{2}\delta \left(
\omega _{+-}-\omega \right) ,
\end{equation}%
{where} 
\begin{align}
\mathbf{E}\left( \omega \right) \cdot \mathbf{r}_{+-}^{2}=& \left\vert
E_{x}\left( \omega \right) a_{+-}^{x}\right\vert ^{2}=E_{x}\left( \omega
\right) a_{+-}^{x}E_{x}^{\ast }\left( \omega \right) a_{+-}^{x\ast }  \notag
\\
=& -E_{x}\left( \omega \right) E_{x}\left( -\omega \right)
a_{+-}^{x}a_{-+}^{x},
\end{align}%
{with the use of} the Hermitian condition Eq.(\ref{AHermitian}).

Inserting it into Eq.(\ref{JElll}), we have%
\begin{align}
& J_{\text{inject}}^{x;x^{2}}  \notag \\
& =\frac{2\pi e^{2}}{\hbar ^{2}}\frac{e}{V}a_{-+}^{x}a_{+-}^{x}\delta \left(
\omega _{+-}-\omega \right)  \notag \\
& \times \frac{\partial ^{3}\omega _{+-}}{\partial k_{x}^{3}}E_{x}\left(
\omega \right) E_{x}\left( -\omega \right) .
\end{align}%
By assuming a monochromic oscillation $J^{x;x^{2}}\propto e^{-i\omega _{0}t}$%
, we obtain the $\ell $-th order current $J^{x;x^{\ell }}$,%
\begin{align}
J_{\text{inject}}^{x;x^{\ell }}=& \frac{1}{i\omega _{0}+1/\tau }\frac{2\pi
e^{3}}{\hbar ^{2}V}a_{-+}^{x}a_{+-}^{x}\delta \left( \omega _{+-}-\omega
\right)  \notag \\
& \times \frac{\partial \omega _{+-}}{\partial k_{x}}E_{x}\left( \omega
\right) E_{x}\left( -\omega \right) \left[ E_{x}\left( 0\right) \right]
^{\ell -2},
\end{align}%
where we have introduced a cutoff by the relaxation time $\tau $. When we
concentrate on the direct current component $\omega _{0}=0$, it is
proportional to $\tau $. The longitudinal component is 
\begin{equation}
\frac{\sigma _{\text{inject}}^{x;x^{2}}}{\sigma _{\text{inject}}}=\sum_{%
\mathbf{k}}f_{-+}\frac{\partial \omega _{+-}}{\partial k_{x}}%
a_{-+}^{x}a_{+-}^{x}\delta \left( \omega _{+-}-\omega \right) ,
\end{equation}%
with 
\begin{equation}
\sigma _{\text{inject}}\equiv 2\pi V\frac{e^{3}}{\hbar ^{2}}\tau .
\label{SlInject}
\end{equation}%
This is Eq.(\ref{Inject}) in the main text.

\subsection{Shift currents}

\label{ApShift}

The shift current is induced by the difference of the position between the
valence band and the conduction band when electrons are excited from the
valence band to the conduction band. It induces the electric dipole $%
E_{x}\left( t\right) a_{+-;x}^{x}$. Hence, the shift current originates in a
quantum interference\cite{snap} between the oscillation of $\rho _{mn}$\ and
the oscillation of the dipole velocity $E_{x}a_{nm;x}^{x}$,%
\begin{equation}
J_{\text{shift}}^{x;x^{2}}=-\frac{e^{2}}{\hbar V}\sum_{\mathbf{k}%
}E_{x}a_{+-;x}^{x}\rho _{+-}^{\left( 1\right) },
\end{equation}%
which is defined by the interband contributions a$_{+-;x}^{x}$ and $\rho
_{+-}^{\left( 1\right) }\left( t\right) $.

The density matrix is obtained by solving the von-Neumann equation\cite%
{Sipe,Ave,snap}, \ 
\begin{align}
& \frac{\partial \rho _{mn}}{\partial t}+i\omega _{mn}\rho _{mn}  \notag \\
=& \frac{e}{i\hbar }\sum_{s}E_{x}\left( \rho _{ml}a_{sn}^{x}-a_{ms}^{x}\rho
_{sn}\right) -\frac{e}{\hbar }E_{x}\rho _{mn;x}.
\end{align}%
For the two-band system, it is rewritten as%
\begin{equation}
i\left( \omega _{+-}-\omega _{0}\right) \rho _{+-}=\frac{e}{i\hbar }%
E_{x}a_{+-}^{x}\left( \rho _{++}-\rho _{--}\right) -\frac{e}{\hbar }%
E_{x}\rho _{+-;x},
\end{equation}%
where we have used a$_{nn}^{x}=0$ and assumed a monochromatic solution $\rho
_{mn}\propto e^{-i\omega _{0}t}$. This definition recovers properly the
definition of the shift current for $\ell =2$.

The zeroth order solution is given by%
\begin{equation}
\rho _{\pm }^{\left( 0\right) }=f_{\pm },
\end{equation}%
because electric field is not applied. The first order solution is given by%
\begin{align}
\rho _{+-}^{\left( 1\right) }& =\frac{ie}{\hbar \left( \omega _{+-}-\omega
_{0}\right) }E_{x}a_{+-}^{x}\left( \rho _{++}^{\left( 0\right) }-\rho
_{--}^{\left( 0\right) }\right) e^{-i\omega _{0}t}  \notag \\
& =\frac{ie}{\hbar \left( \omega _{+-}-\omega _{0}\right) }%
E_{x}a_{+-}^{x}f_{-+}e^{-i\omega _{0}t}.
\end{align}%
Especially, we have $\rho _{++}^{\left( 1\right) }=0$ because $f_{++}\equiv
f_{+}-f_{+}=0$.

Hence, the shift current is obtained as%
\begin{align}
J_{\text{shift}}^{x;x^{2}}=& -\frac{e^{2}}{\hbar V}\sum_{\mathbf{k}%
}E_{x}a_{+-;x}^{x}\rho _{+-}^{\left( 1\right) }  \notag \\
=& -\frac{e^{2}}{\hbar V}e^{-i\omega _{0}t}\sum_{\mathbf{k}%
}a_{+-;x}^{x}f_{-+}a_{+-}^{x}E_{x}^{l}.
\end{align}%
We use the monochromatic condition (\ref{monochro}) and study the direct
current component,%
\begin{align}
J_{\text{shift}}^{x;x^{2}}=& \sigma _{\text{shift}}^{x;x^{2}}\left( 0;\omega
,-\omega ,0,0,\cdots ,0\right)  \notag \\
& \times E_{x}\left( \omega \right) E_{x}\left( -\omega \right) .
\end{align}%
Then, the shift current is obtained as 
\begin{align}
J_{\text{shift}}^{x;x^{2}}=& -\frac{e^{2}}{\hbar V}\sum_{\mathbf{k}%
}a_{+-;x}^{x}a_{+-}^{x}  \notag \\
& \times f_{-+}E_{x}\left( \omega \right) E_{x}\left( -\omega \right) \delta
\left( \omega _{+-}-\omega \right) .
\end{align}%
It is independent of $\tau $. The shift conductivity is given by 
\begin{equation}
\frac{\sigma _{\text{shift}}^{x;x^{2}}}{\sigma _{\text{shift}}}=\frac{1}{V}%
\sum_{\mathbf{k}}f_{-+}a_{+-;x}^{x}a_{+-}^{x}\delta \left( \omega
_{+-}-\omega \right) ,
\end{equation}%
with 
\begin{equation}
\sigma _{\text{shift}}\equiv -\frac{e^{3}}{\hbar ^{2}}.  \label{SlShift}
\end{equation}%
This is Eq.(\ref{Shift}) in the main text.

\section{Quantum distance and Zeeman quantum geometric tensor}

\label{ApZQ}

The Zeeman quantum geometric tensor $g_{nm}^{\mu \nu }$ is defined by the
quantum distance $ds_{\text{HS}}$ for the infinitesimal translation $d%
\mathbf{k}$ of the momentum as\cite{Xiang2,Xiang25,Chak,Xiang3} 
\begin{equation}
ds_{\text{HS}}\left( \mathbf{k}\right) \equiv \sqrt{1-\left\vert
\left\langle \psi _{n}\left( \mathbf{k}\right) \right\vert U_{d\mathbf{%
\theta }}U_{d\mathbf{k}}\left\vert \psi _{n}\left( \mathbf{k}\right)
\right\rangle \right\vert ^{2}},
\end{equation}%
where 
\begin{equation}
U_{d\mathbf{\theta }}\equiv e^{-\frac{i}{2}d\mathbf{\theta }\cdot \mathbf{%
\sigma }},\quad U_{d\mathbf{k}}\equiv e^{-id\mathbf{k}\cdot \mathbf{r}}
\end{equation}%
are the generators of the spin angular momentum $d\mathbf{\theta }$ and the
momentum $d\mathbf{k}$. By using the relation%
\begin{align}
& U_{d\mathbf{\theta }}U_{d\mathbf{k}}\left\vert \psi _{n}\right\rangle 
\notag \\
=& \left( 1-\frac{i\sigma _{\mu }}{2}d\theta _{\nu }-\frac{\sigma _{\mu
}\sigma _{\nu }}{8}d\theta _{\mu }d\theta _{\mu }\cdots \right)  \notag \\
& \times \left( 1+\partial _{k_{\mu }}dk_{\mu }+\frac{1}{2}\left\vert
\partial _{k_{\mu }}\partial _{k_{\nu }}\psi _{n}\left( \mathbf{k}\right)
\right\rangle dk_{\mu }dk_{\nu }\cdots \right) \left\vert \psi
_{n}\right\rangle  \notag \\
=& 1+\frac{1}{2}\left\vert \partial _{k_{\mu }}\partial _{k_{\nu }}\psi
_{n}\left( \mathbf{k}\right) \right\rangle dk_{\mu }dk_{\nu }  \notag \\
& -\frac{i\sigma _{\mu }}{2}\left\vert \psi _{n}\right\rangle d\theta ^{\mu
}-\frac{1}{8}\left\vert \psi _{n}\right\rangle d\theta ^{\mu }d\theta ^{\mu
}-\frac{i\sigma _{\nu }}{2}\partial _{k_{\mu }}\left\vert \psi
_{n}\right\rangle dk_{\mu }d\theta ^{\nu }+\cdots ,
\end{align}%
the quantum distance is given by%
\begin{align}
\left( ds_{\text{HS}}\right) ^{2}=& \sum_{n\neq m}g_{nm}^{\mu \nu }dk_{\mu
}dk_{\nu }+\sum_{n\neq m}s_{nm}^{\mu \nu }\frac{d\theta _{\mu }d\theta _{\nu
}}{4}  \notag \\
& +\sum_{n\neq m}\left( z_{nm}^{\mu \nu }+z_{mn}^{\mu \nu }\right) dk_{\mu }%
\frac{d\theta _{\nu }}{2},  \label{dsZeeman}
\end{align}%
where $g_{nm}^{\mu \nu }$ is the quantum metric, $s_{nm}^{\mu \nu }$ is the
spin quantum geometric tensor and $z_{nm}^{\mu \nu }$ is the Zeeman quantum
geometric tensor.

We determine $z_{nm}^{\mu \nu }$ by calculating the coefficient of $dk_{\mu }%
\frac{d\theta _{\nu }}{2}$. The fidelity is calculated as%
\begin{align}
& \left\vert \left\langle \psi _{n}\left( \mathbf{k}\right) \left\vert \psi
_{n}\left( \mathbf{k}+d\mathbf{k}\right) \right\rangle \right. \right\vert
^{2}  \notag \\
=& \left\langle \psi _{n}\left( \mathbf{k}+d\right) \left\vert \psi
_{n}\left( \mathbf{k}\right) \right\rangle \right. \left\langle \psi
_{n}\left( \mathbf{k}\right) \left\vert \psi _{n}\left( \mathbf{k}+d\right)
\right\rangle \right.  \notag \\
=& \left\langle \psi _{n}\left( \mathbf{k}\right) \left\vert \psi _{n}\left( 
\mathbf{k}\right) \right\rangle \right. +(\left\langle \partial _{k_{\mu
}}\psi _{n}\left( \mathbf{k}\right) \left\vert \psi _{n}\left( \mathbf{k}%
\right) \right\rangle \right.  \notag \\
& +\left\langle \psi _{n}\left( \mathbf{k}\right) \left\vert \partial
_{k_{\mu }}\psi _{n}\left( \mathbf{k}\right) \right\rangle \right. )dk_{\mu }
\notag \\
& +(-\frac{i}{2}\left\langle \partial _{k_{\mu }}\psi _{n}\left( \mathbf{k}%
\right) \left\vert \psi _{n}\left( \mathbf{k}\right) \right\rangle \right.
\left\langle \psi _{n}\left( \mathbf{k}\right) \right\vert \sigma _{\nu
}\left\vert \psi _{n}\left( \mathbf{k}\right) \right\rangle  \notag \\
& +\frac{i}{2}\left\langle \psi _{n}\left( \mathbf{k}\right) \right\vert
\sigma _{\nu }\left\vert \psi _{n}\left( \mathbf{k}\right) \right\rangle
\left\langle \psi _{n}\left( \mathbf{k}\right) \left\vert \partial _{k_{\mu
}}\psi _{n}\left( \mathbf{k}\right) \right\rangle \right.  \notag \\
& +\frac{i}{2}\left\langle \partial _{k_{\mu }}\psi _{n}\left( \mathbf{k}%
\right) \right\vert \sigma _{\nu }\left\vert \psi _{n}\left( \mathbf{k}%
\right) \right\rangle  \notag \\
& -\frac{i}{2}\left\langle \psi _{n}\left( \mathbf{k}\right) \right\vert
\sigma _{\nu }\left\vert \partial _{k_{\mu }}\psi _{n}\left( \mathbf{k}%
\right) \right\rangle )dk_{\mu }d\theta _{\nu }  \notag \\
=& 1+\frac{i}{2}(\left\langle \partial _{k_{\mu }}\psi _{n}\left( \mathbf{k}%
\right) \right\vert \left( 1-P\left( \mathbf{k}\right) \right) \sigma _{\nu
}\left\vert \psi _{n}\left( \mathbf{k}\right) \right\rangle  \notag \\
& -\frac{i}{2}\left\langle \psi _{n}\left( \mathbf{k}\right) \right\vert
\left( 1-P\left( \mathbf{k}\right) \right) \sigma _{\nu }\left\vert \partial
_{k_{\mu }}\psi _{n}\left( \mathbf{k}\right) \right\rangle )dk_{\mu }\frac{%
d\theta _{\nu }}{2}  \notag \\
=& 1+\frac{i}{2}(\sum_{m\neq n}\left\langle \partial _{k_{\mu }}\psi
_{n}\left( \mathbf{k}\right) \left\vert \psi _{m}\left( \mathbf{k}\right)
\right\rangle \right. \left\langle \psi _{m}\left( \mathbf{k}\right)
\right\vert \sigma _{\nu }\left\vert \psi _{n}\left( \mathbf{k}\right)
\right\rangle  \notag \\
& -\left\langle \psi _{n}\left( \mathbf{k}\right) \right\vert \sigma _{\nu
}\left\vert \psi _{m}\left( \mathbf{k}\right) \right\rangle \left\langle
\psi _{m}\left( \mathbf{k}\right) \left\vert \partial _{k_{\mu }}\psi
_{n}\left( \mathbf{k}\right) \right\rangle \right. )dk_{\mu }\frac{d\theta
_{\nu }}{2}  \notag \\
=& 1-\frac{1}{2}\sum_{m\neq n}\left\langle \psi _{n}\left( \mathbf{k}\right)
\left\vert \partial _{k_{\mu }}\psi _{m}\left( \mathbf{k}\right)
\right\rangle \right. \left\langle \psi _{m}\left( \mathbf{k}\right)
\right\vert \sigma _{\nu }\left\vert \psi _{n}\left( \mathbf{k}\right)
\right\rangle  \notag \\
& +\left\langle \psi _{n}\left( \mathbf{k}\right) \right\vert \sigma _{\nu
}\left\vert \psi _{m}\left( \mathbf{k}\right) \right\rangle \left\langle
\psi _{m}\left( \mathbf{k}\right) \left\vert \partial _{k_{\mu }}\psi
_{n}\left( \mathbf{k}\right) \right\rangle \right. dk_{\mu }\frac{d\theta
_{\nu }}{2}  \notag \\
=& 1-\frac{1}{2}\sum_{m\neq n}\left( a_{nm}^{\mu }s_{mn}^{\nu }+s_{nm}^{\nu
}a_{mn}^{\mu }\right) dk_{\mu }\frac{d\theta _{\nu }}{2},
\end{align}%
where we have used Eq.(\ref{WZa}). By comparing it to Eq.(\ref{dsZeeman}),
we obtain%
\begin{equation}
z_{nm}^{\mu \nu }=a_{nm}^{\mu }s_{mn}^{\nu },
\end{equation}
which is the Zeeman geometric tensor. This is Eq.(\ref{Zz12}) in the main
text.

We calculate $s_{nm}^{\mu \nu }$ as%
\begin{align}
& \left\vert \left\langle \psi _{n}\left( \mathbf{k}\right) \left\vert \psi
_{n}\left( \mathbf{k}^{\prime }\right) \right\rangle \right. \right\vert ^{2}
\notag \\
=& \left\langle \psi _{n}\left( \mathbf{k}^{\prime }\right) \left\vert \psi
_{n}\left( \mathbf{k}\right) \right\rangle \right. \left\langle \psi
_{n}\left( \mathbf{k}\right) \left\vert \psi _{n}\left( \mathbf{k}^{\prime
}\right) \right\rangle \right.  \notag \\
=& \left( \left\langle \psi _{n}\left( \mathbf{k}\right) \right\vert
+\left\langle \psi _{n}\left( \mathbf{k}\right) \right\vert \frac{i\sigma
_{\mu }}{2}d\theta ^{\mu }+\cdots \right) \left\vert \psi _{n}\left( \mathbf{%
k}\right) \right\rangle  \notag \\
& \left\langle \psi _{n}\left( \mathbf{k}\right) \left\vert \left(
\left\vert \psi _{n}\left( \mathbf{k}\right) \right\rangle +\frac{i\sigma
_{\mu }}{2}\left\vert \psi _{n}\left( \mathbf{k}\right) \right\rangle
d\theta ^{\mu \nu }+\cdots \right) \right\rangle \right.  \notag \\
=& \left\langle \psi _{n}\left( \mathbf{k}\right) \left\vert \psi _{n}\left( 
\mathbf{k}\right) \right\rangle \right.  \notag \\
& +\left( \frac{i}{2}\left\langle \psi _{n}\left( \mathbf{k}\right)
\right\vert \sigma _{\mu }\left\vert \psi _{n}\left( \mathbf{k}\right)
\right\rangle -\frac{i\sigma _{\mu }}{2}\left\langle \psi _{n}\left( \mathbf{%
k}\right) \right\vert \sigma _{\mu }\left\vert \psi _{n}\left( \mathbf{k}%
\right) \right\rangle \right) d\theta ^{\mu }  \notag \\
& +(-\frac{1}{4}\left\langle \psi _{n}\left( \mathbf{k}\right) \right\vert
\sigma _{\mu }\left\vert \psi _{n}\left( \mathbf{k}\right) \right\rangle
\left\langle \psi _{n}\left( \mathbf{k}\right) \right\vert \sigma _{\nu
}\left\vert \psi _{n}\left( \mathbf{k}\right) \right\rangle )d\theta ^{\mu
}d\theta ^{\nu }  \notag \\
=& 1+\frac{1}{4}\left\langle \psi _{n}\left( \mathbf{k}\right) \right\vert
\sigma _{\mu }\left\vert \psi _{n}\left( \mathbf{k}\right) \right\rangle
\left\langle \psi _{n}\left( \mathbf{k}\right) \right\vert \sigma _{\nu
}\left\vert \psi _{n}\left( \mathbf{k}\right) \right\rangle )d\theta ^{\mu
}d\theta ^{\nu }
\end{align}%
Then, we have%
\begin{equation}
s_{nm}^{\mu \nu }\equiv s_{nm}^{\mu }s_{mn}^{\nu }
\end{equation}%
with%
\begin{equation}
s_{nm}^{\mu }\equiv \left\langle \psi _{n}\left( \mathbf{k}\right)
\right\vert \sigma _{\mu }\left\vert \psi _{n}\left( \mathbf{k}\right)
\right\rangle .
\end{equation}%
is the spin geometric tensor. This is Eq.(\ref{SGT}) in the main text.

\section{Zeeman quantum geometry induced intrinsic cross responses}

\label{ApCrossRes}

\subsection{Hall current and AC Hall current}

Inserting the density matrix (\ref{Rho1}) to Eq.(\ref{Jab}) and assuming $%
\omega \ll \varepsilon _{mn}$, the current is calculated as%
\begin{align}
J^{\mu ;\nu }=& \frac{1}{2}\sum_{nm}\sum_{\omega _{1}=\pm \omega }\int d%
\mathbf{k}v_{nm}^{\mu }f_{nm}a_{mn}^{\nu }\frac{E^{\nu }e^{-i\omega _{1}t}}{%
\hbar \omega _{1}-\varepsilon _{mn}+i\eta }  \notag \\
=& \frac{1}{2}\sum_{nm}\sum_{\omega _{1}=\pm \omega }\int d\mathbf{k}%
f_{nm}i\varepsilon _{nm}a_{nm}^{\mu }a_{mn}^{\nu }\frac{E^{\nu }e^{-i\omega
_{1}t}}{\hbar \omega _{1}-\varepsilon _{mn}+i\eta }.  \label{JabA}
\end{align}%
We assume the off-resonant condition 
\begin{equation}
\hbar \omega _{1}-\varepsilon _{mn}\neq 0.
\end{equation}%
Then, we can expand the denominator in Eq.(\ref{JabA}) 
\begin{equation}
\frac{1}{\hbar \omega _{1}-\varepsilon _{mn}+i\eta }=-\sum_{j=1}^{\infty }%
\frac{\omega ^{j-1}}{\varepsilon _{mn}^{j}}.
\end{equation}%
By using it, we obtain%
\begin{align}
J^{\mu ;\nu }=& -\frac{1}{2}\sum_{nm}\sum_{\omega _{1}=\pm \omega }\int d%
\mathbf{k}f_{nm}i\varepsilon _{nm}a_{nm}^{\mu }a_{mn}^{\nu }  \notag \\
& \times E^{\nu }e^{-i\omega _{1}t}\sum_{j=1}^{\infty }\frac{\omega
_{1}^{j-1}}{\varepsilon _{mn}^{j}}  \notag \\
=& \frac{1}{2}\sum_{nm}\sum_{\omega _{1}=\pm \omega }\int d\mathbf{k}%
f_{nm}ia_{nm}^{\mu }a_{mn}^{\nu }E^{\nu }e^{-i\omega
_{1}t}\sum_{j=0}^{\infty }\frac{\omega _{1}^{j}}{\varepsilon _{mn}^{j}} 
\notag \\
=& \frac{1}{2}\int d\mathbf{k}\sum_{n>m}\sum_{\omega _{1}=\pm \omega
}\sum_{j=0}^{\infty }f_{nm}  \notag \\
& \times i\left[ a_{nm}^{\mu }a_{mn}^{\nu }-a_{mn}^{\mu }a_{nm}^{\nu }\right]
\frac{\omega _{1}^{2j}}{\varepsilon _{mn}^{2j}}E^{\nu }e^{-i\omega _{1}t} 
\notag \\
& +i\left[ a_{nm}^{\mu }a_{mn}^{\nu }+a_{mn}^{\mu }a_{nm}^{\nu }\right] 
\frac{\omega _{1}^{2j+1}}{\varepsilon _{mn}^{2j+1}}E^{\nu }e^{-i\omega
_{1}t},
\end{align}%
where we have used the relations 
\begin{equation}
f_{nm}\equiv f_{n}-f_{m}=-f_{mn},
\end{equation}%
and%
\begin{equation}
\varepsilon _{nm}\equiv \varepsilon _{n}-\varepsilon _{m}=-\varepsilon _{mn}.
\end{equation}%
By taking the sum of $\omega _{1}=\pm \omega $, we obtain%
\begin{align}
J^{\mu ;\nu }=& \int d\mathbf{k}\sum_{n>m}\sum_{j=0}^{\infty }f_{nm}\left[
\Omega _{nm}^{\mu \nu }\frac{\omega ^{2j}}{\varepsilon _{mn}^{2j}}E^{\nu
}\cos \omega t\right.  \notag \\
& \left. +2g_{nm}^{\mu \nu }\frac{\omega ^{2j+1}}{\varepsilon _{mn}^{2j+1}}%
E^{\nu }\sin \omega t\right]  \notag \\
=& \int d\mathbf{k}\sum_{n>m}f_{nm}\left[ \Omega _{nm}^{\mu \nu }\frac{1}{%
1-\left( \omega /\varepsilon _{mn}\right) ^{2}}E^{\nu }\cos \omega t\right. 
\notag \\
& \left. +2g_{nm}^{\mu \nu }\frac{\omega /\varepsilon _{mn}}{1-\left( \omega
/\varepsilon _{mn}\right) ^{2}}E^{\nu }\sin \omega t\right] .
\end{align}%
Only the order of $\Omega ^{0}$ and $\Omega $ are valid in the linear
response theory. Then, we obtain the current%
\begin{equation}
\frac{J^{\mu ;\nu }}{E^{\nu }}=\int d\mathbf{k}\sum_{n>m}f_{nm}\left[ \Omega
_{nm}^{\mu \nu }+2g_{nm}^{\mu \nu }\frac{\omega }{\varepsilon _{mn}}\sin
\omega t\right] ,
\end{equation}%
which is Eq.(\ref{JE}) in the main text.

In the static limit $\Omega =0$, it recovers the TKNN formula%
\begin{equation}
J^{\mu \nu }=\int d\mathbf{k}\sum_{n>m}f_{nm}\Omega _{nm}^{\mu \nu }E^{\nu }.
\end{equation}

\subsection{Intrinsic gyrotropic magnetic current}

The current under external magnetic field is calculated as%
\begin{align}
J^{\mu ;\nu }=& -\frac{g\mu _{\text{B}}}{2}\sum_{nm}\sum_{\omega _{1}=\pm
\omega }\int d\mathbf{k}v_{nm}^{\mu }\frac{f_{nm}s_{mn}^{\nu }B^{\nu
}e^{-i\omega _{1}t}}{\hbar \omega _{1}-\varepsilon _{mn}+i\eta }  \notag \\
=& g\mu _{\text{B}}\int d\mathbf{k}\sum_{n>m}f_{nm}\left[ \mathcal{Z}%
_{nm}^{\mu \nu }\frac{1}{1-\left( \omega /\varepsilon _{mn}\right) ^{2}}%
\right.  \notag \\
& \left. +2\mathcal{Q}_{nm}^{\mu \nu }\frac{\omega /\varepsilon _{mn}}{%
1-\left( \omega /\varepsilon _{mn}\right) ^{2}}\sin \omega t\right] B^{\nu }.
\end{align}%
It is%
\begin{equation}
J^{\mu ;\nu }=g\mu _{\text{B}}\int d\mathbf{k}\sum_{n>m}f_{nm}\left[ 
\mathcal{Z}_{nm}^{\mu \nu }\cos \omega t+2\mathcal{Q}_{nm}^{\mu \nu }\frac{%
\omega }{\varepsilon _{mn}}\sin \omega t\right] B^{\nu }.
\end{equation}%
up to the linear order in $\omega $, which is Eq.(\ref{JB}) in the main text.

\subsection{Intrinsic electric field induced spin density}

The spin polarization under external electric field is calculated as%
\begin{align}
S^{\mu }=& \frac{1}{2}\sum_{nm}\sum_{\omega _{1}=\pm \omega }\int d\mathbf{k}%
s_{nm}^{\mu }f_{nm}r_{mn}^{\nu }\frac{E^{\nu }e^{-i\omega _{1}t}}{\hbar
\omega _{1}-\varepsilon _{mn}+i\eta }  \notag \\
=& -\int d\mathbf{k}\sum_{n>m}f_{nm}\left[ \frac{2\mathcal{Q}_{nm}^{\nu \mu }%
}{\varepsilon _{mn}}\frac{1}{1-\left( \omega /\varepsilon _{mn}\right) ^{2}}%
\right.  \notag \\
& \left. +\mathcal{Z}_{nm}^{\nu \mu }\frac{\omega /\varepsilon _{mn}^{2}}{%
1-\left( \omega /\varepsilon _{mn}\right) ^{2}}\sin \omega t\right] E^{\nu }.
\end{align}%
It is%
\begin{align}
& S^{\mu }  \notag \\
=& -\int d\mathbf{k}\sum_{n>m}f_{nm}\left[ \frac{2\mathcal{Q}_{nm}^{\nu \mu }%
}{\varepsilon _{mn}}\cos \omega t+\mathcal{Z}_{nm}^{\nu \mu }\frac{\omega }{%
\varepsilon _{mn}^{2}}\sin \omega t\right] E^{\nu }.
\end{align}%
up to the linear order in $\omega $, which is Eq.(\ref{SE}) in the main text.

\subsection{Intrinsic magnetic field induced spin density}

The spin polarization under external magnetic field is calculated as%
\begin{align}
S^{\mu ;\nu }=& -\frac{g\mu _{\text{B}}}{2}\sum_{\omega _{1}=\pm \omega
}\int d\mathbf{k}s_{nm}^{\mu }\frac{f_{nm}s_{mn}^{\nu }B^{\nu }e^{-i\omega
_{1}t}}{\hbar \omega _{1}-\varepsilon _{mn}+i\eta }  \notag \\
=& g\mu _{\text{B}}\int d\mathbf{k}\sum_{n>m}f_{nm}\left[ \frac{\mathcal{S}%
_{nm}^{\mu \nu }}{\varepsilon _{mn}}\frac{1}{1-\left( \omega /\varepsilon
_{mn}\right) ^{2}}B^{\nu }\right.  \notag \\
& \left. +2\mathcal{A}_{nm}^{\mu \nu }\frac{\omega /\varepsilon _{mn}^{2}}{%
1-\left( \omega /\varepsilon _{mn}\right) ^{2}}B^{\nu }\sin \omega t\right] .
\end{align}%
It is 
\begin{align}
& S^{\mu ;\nu }  \notag \\
=& g\mu _{\text{B}}\int d\mathbf{k}\sum_{n>m}f_{nm}\left[ \frac{\mathcal{S}%
_{nm}^{\mu \nu }}{\varepsilon _{mn}}\cos \omega t+2\mathcal{A}_{nm}^{\mu \nu
}\frac{\omega }{\varepsilon _{mn}^{2}}\sin \omega t\right] B^{\nu }
\end{align}%
up to the linear order in $\omega $, which is Eq.(\ref{SB}) in the main text.

\section{Zeeman quantum geometry for two-band systems}

\label{ApZQ2}

\subsection{Zeeman Berry curvature}

By inserting $n=+$ and $m=-$ in Eq.(\ref{Zab}), we obtain%
\begin{align}
\mathcal{Z}_{+-}^{xx}=& i\left(
r_{+-}^{x}s_{-+}^{x}-r_{-+}^{x}s_{+-}^{x}\right)  \notag \\
=& \frac{\partial \theta }{\partial k_{x}}\cos \theta \cos \phi -\frac{%
\partial \phi }{\partial k_{x}}\sin \theta \sin \phi =\frac{\partial n_{x}}{%
\partial k_{x}}, \\
\mathcal{Z}_{+-}^{yy}=& i\left(
r_{+-}^{y}s_{-+}^{y}-r_{-+}^{y}s_{+-}^{y}\right)  \notag \\
=& \frac{\partial \theta }{\partial k_{y}}\cos \theta \sin \phi +\frac{%
\partial \phi }{\partial k_{y}}\cos \theta \sin \phi =\frac{\partial n_{y}}{%
\partial k_{y}}, \\
\mathcal{Z}_{+-}^{xy}=& i\left(
r_{+-}^{x}s_{-+}^{y}-r_{-+}^{x}s_{+-}^{y}\right)  \notag \\
=& \frac{\partial \theta }{\partial k_{x}}\cos \theta \sin \phi +\frac{%
\partial \phi }{\partial k_{x}}\sin \theta \cos \phi =\frac{\partial n_{y}}{%
\partial k_{x}}, \\
\mathcal{Z}_{+-}^{yx}=& i\left(
r_{+-}^{y}s_{-+}^{x}-r_{-+}^{y}s_{+-}^{x}\right)  \notag \\
=& \frac{\partial \theta }{\partial k_{y}}\cos \theta \cos \phi -\frac{%
\partial \phi }{\partial k_{y}}\sin \theta \sin \phi =\frac{\partial n_{x}}{%
\partial k_{y}}, \\
\mathcal{Z}_{+-}^{xz}=& i\left(
r_{+-}^{x}s_{-+}^{z}-r_{-+}^{x}s_{+-}^{z}\right)  \notag \\
=& -\frac{\partial \theta }{\partial k_{x}}\sin \theta =\frac{\partial n_{z}%
}{\partial k_{x}}, \\
\mathcal{Z}_{+-}^{yz}=& i\left(
r_{+-}^{y}s_{-+}^{z}-r_{-+}^{y}s_{+-}^{z}\right)  \notag \\
=& -\frac{\partial \theta }{\partial k_{y}}\sin \theta =\frac{\partial n_{z}%
}{\partial k_{y}}.
\end{align}%
They are summarized as%
\begin{equation}
\mathcal{Z}_{+-}^{\mu \nu }=-\mathcal{Z}_{-+}^{\mu \nu }=\frac{\partial
n_{\nu }}{\partial k_{\mu }}.
\end{equation}%
This is Eq.(\ref{Z}) in the main text.

\subsection{Zeeman quantum metric}

The expectation value of the spin%
\begin{equation}
s_{nm}^{\mu }\equiv \left\langle \psi _{n}\left( \mathbf{k}\right)
\right\vert \sigma _{\mu }\left\vert \psi _{m}\left( \mathbf{k}\right)
\right\rangle
\end{equation}%
is calculated as%
\begin{align}
s_{-+}^{x}=& -i\sin \phi +\cos \theta \cos \phi , \\
s_{-+}^{y}=& i\cos \phi +\cos \theta \sin \phi , \\
s_{-+}^{z}=& -\sin \theta .
\end{align}%
By inserting $n=+$ and $m=-$ in Eq.(\ref{Qab}), we obtain%
\begin{align}
\mathcal{Q}_{+-}^{xx}=& \frac{r_{+-}^{x}s_{-+}^{x}+r_{-+}^{x}s_{+-}^{x}}{2} 
\notag \\
=& -\frac{\sin \phi }{2}\frac{\partial \theta }{\partial k_{x}}-\frac{\cos
\phi }{2}\sin \theta \cos \theta \frac{\partial \phi }{\partial k_{x}} 
\notag \\
=& \frac{1}{2}\left( n_{y}\frac{\partial n_{z}}{\partial k_{x}}-n_{z}\frac{%
\partial n_{y}}{\partial k_{x}}-\right) , \\
\mathcal{Q}_{+-}^{yy}=& \frac{r_{+-}^{y}s_{-+}^{y}+r_{-+}^{y}s_{+-}^{y}}{2} 
\notag \\
=& \frac{\cos \phi }{2}\frac{\partial \theta }{\partial k_{y}}-\frac{\sin
\phi }{2}\sin \theta \cos \theta \frac{\partial \phi }{\partial k_{y}} 
\notag \\
=& \frac{1}{2}\left( n_{z}\frac{\partial n_{x}}{\partial k_{y}}-n_{x}\frac{%
\partial n_{z}}{\partial k_{y}}\right) , \\
\mathcal{Q}_{+-}^{xy}=& \frac{r_{+-}^{x}s_{-+}^{y}+r_{-+}^{x}s_{+-}^{y}}{2} 
\notag \\
=& \frac{\cos \phi }{2}\frac{\partial \theta }{\partial k_{x}}-\frac{\sin
\phi }{2}\sin \theta \cos \theta \frac{\partial \phi }{\partial k_{x}} 
\notag \\
=& \frac{1}{2}\left( n_{z}\frac{\partial n_{x}}{\partial k_{x}}-n_{x}\frac{%
\partial n_{z}}{\partial k_{x}}\right) , \\
\mathcal{Q}_{+-}^{yx}=& \frac{r_{+-}^{y}s_{-+}^{x}+r_{-+}^{y}s_{+-}^{x}}{2} 
\notag \\
=& \frac{\cos \phi }{2}\frac{\partial \theta }{\partial k_{y}}-\frac{\sin
\phi }{2}\sin \theta \cos \theta \frac{\partial \phi }{\partial k_{y}} 
\notag \\
=& \frac{1}{2}\left( n_{y}\frac{\partial n_{z}}{\partial k_{y}}-n_{z}\frac{%
\partial n_{y}}{\partial k_{y}}\right) , \\
\mathcal{Q}_{+-}^{xz}=& \frac{\sin ^{2}\theta }{2}\frac{\partial \phi }{%
\partial k_{x}}=\frac{1}{2}\left( n_{x}\frac{\partial n_{y}}{\partial k_{x}}%
-n_{y}\frac{\partial n_{x}}{\partial k_{x}}\right) , \\
\mathcal{Q}_{+-}^{yz}=& \frac{\sin ^{2}\theta }{2}\frac{\partial \phi }{%
\partial k_{y}}=\frac{1}{2}\left( n_{x}\frac{\partial n_{y}}{\partial k_{y}}%
-n_{y}\frac{\partial n_{x}}{\partial k_{y}}\right) ,
\end{align}%
where we have used the relations%
\begin{align}
\frac{\partial n_{x}}{\partial k_{\mu }}=& \cos \theta \cos \phi \frac{%
\partial \theta }{\partial k_{\mu }}-\sin \theta \sin \phi \frac{\partial
\phi }{\partial k_{\mu }}, \\
\frac{\partial n_{y}}{\partial k_{\mu }}=& \cos \theta \sin \phi \frac{%
\partial \theta }{\partial k_{\mu }}+\sin \theta \cos \phi \frac{\partial
\phi }{\partial k_{\mu }}, \\
\frac{\partial n_{z}}{\partial k_{\mu }}=& -\sin \theta \frac{\partial
\theta }{\partial k_{\mu }}.
\end{align}%
They are summarized as%
\begin{equation}
\mathcal{Q}_{+-}^{\mu \nu }=\frac{1}{2}\varepsilon _{\nu \rho \sigma
}n_{\rho }\frac{\partial n_{\sigma }}{\partial k_{\mu }}.
\end{equation}%
This is Eq.(\ref{Q}) in the main text.

\subsection{Spin quantum geometry}

By inserting $n=+$ and $m=-$ in Eq.(\ref{SGT}), we obtain the diagonal
components,%
\begin{align}
s_{+-}^{xx}=& s_{+-}^{x}s_{-+}^{x}=\frac{1}{4}\left( 3+\cos 2\theta -2\cos
2\phi \sin ^{2}\theta \right)  \notag \\
=& n_{y}^{2}+n_{z}^{2}=1-n_{x}^{2}, \\
s_{+-}^{yy}=& s_{+-}^{y}s_{-+}^{y}=\frac{1}{4}\left( 3+\cos 2\theta +2\cos
2\phi \sin ^{2}\theta \right)  \notag \\
=& n_{x}^{2}+n_{z}^{2}=1-n_{y}^{2}, \\
s_{+-}^{zz}=& s_{+-}^{z}s_{-+}^{z}=\sin ^{2}\theta =1-n_{z}^{2}.
\end{align}%
They are summarized as%
\begin{equation}
s_{+-}^{\mu \mu }=1-n_{\mu }^{2}.  \label{Sigma1}
\end{equation}%
We also obtain the off-diagonal components, 
\begin{align}
s_{+-}^{xy}=& i\cos \theta -\sin ^{2}\theta \sin \phi \cos \phi  \notag \\
=& in_{z}-n_{x}n_{y}, \\
s_{+-}^{yx}=& -i\cos \theta -\sin ^{2}\theta \sin \phi \cos \phi  \notag \\
=& -in_{z}-n_{x}n_{y}=\left( s_{+-}^{xy}\right) ^{\ast }, \\
s_{+-}^{xz}=& -\sin \theta \left( i\sin \phi \cos \theta +\cos \phi \right) 
\notag \\
=& -in_{y}-n_{z}n_{x}, \\
s_{+-}^{yz}=& \sin \theta \left( +i\cos \phi -\cos \theta \sin \phi \right) 
\notag \\
=& in_{x}-n_{z}n_{y}.
\end{align}%
They are summarized as%
\begin{equation}
s_{+-}^{\mu \nu }=i\varepsilon _{\mu \nu \rho }n_{\rho }-n_{\mu }n_{\nu }
\label{Sigma2}
\end{equation}%
for $\mu \neq \nu $. Eqs.(\ref{Sigma1}) and (\ref{Sigma2}) are summarized as%
\begin{equation}
s_{+-}^{\mu \nu }=\delta _{\mu \nu }+i\varepsilon _{\mu \nu \rho }n_{\rho
}-n_{\mu }n_{\nu }.
\end{equation}

The spin quantum metric (\ref{SQGT}) is given by%
\begin{equation}
\mathcal{S}_{+-}^{\mu \nu }=\delta _{\mu \nu }-n_{\mu }n_{\nu }.
\end{equation}%
This is Eq.(\ref{S}) in the main text.

The spin Berry curvature (\ref{SBerry}) is given by%
\begin{equation}
\mathcal{A}_{+-}^{\mu \nu }=-2\varepsilon _{\mu \nu \rho }n_{\rho }.
\end{equation}%
This is Eq.(\ref{A}) in the main text.

\section{Zeeman geometry induced cross reponses in X-wave magnets}

\label{ApCross}

We evaluate whether there is a response by integrating the Zeeman quantum
geometric quantities $\mathcal{Z}_{+-}^{\mu \nu }$ and $\mathcal{Q}%
_{+-}^{\mu \nu }$. It is determined by the integration over $\phi $.

1) $J^{x;x}/E^{x}$

$\mathcal{Z}_{+-}^{xx}$ is expanded as%
\begin{align}
&\mathcal{Z}_{+-}^{xx}  \notag \\
=&\frac{\lambda ^{3}k_{x}k_{y}}{\left( \lambda ^{2}k^{2}+B^{2}\right) ^{3/2}}
\notag \\
&+J\lambda k_{y}\left( -\frac{3B\lambda ^{2}k_{x}f_{X}}{\left( \lambda
^{2}k^{2}+B^{2}\right) ^{5/2}}+\frac{B\partial _{k_{x}}f_{X}}{\left( \lambda
^{2}k^{2}+B^{2}\right) ^{3/2}}\right)
\end{align}%
up to the first order in $J$. The integration is 
\begin{equation}
\int \mathcal{Z}_{+-}^{xx}d\phi =0
\end{equation}%
for $f_{X}=k^{N_{X}}\cos N_{X}\phi $, and $f_{X}=k^{N_{X}}\sin N_{X}\phi $
with all $N_{X}$. Hence there is no response $J^{x;x}/E^{x}$.

2) $J^{y;y}/E^{y}$

$\mathcal{Z}_{+-}^{yy}$ is expanded as%
\begin{align}
&\mathcal{Z}_{+-}^{yy}  \notag \\
=&-\frac{\lambda ^{3}k_{x}k_{y}}{\left( \lambda ^{2}k^{2}+B^{2}\right) ^{3/2}%
}  \notag \\
&+J\lambda k_{x}\left( \frac{3B\lambda ^{2}k_{y}f_{X}}{\left( \lambda
^{2}k^{2}+B^{2}\right) ^{5/2}}-\frac{B\partial _{k_{y}}f_{X}}{\left( \lambda
^{2}k^{2}+B^{2}\right) ^{3/2}}\right)
\end{align}%
up to the first order in $J$. The integration is 
\begin{equation}
\int \mathcal{Z}_{+-}^{yy}d\phi =0
\end{equation}%
for $f_{X}=k^{N_{X}}\cos N_{X}\phi $, and $f_{X}=k^{N_{X}}\sin N_{X}\phi $
with all $N_{X}$. Hence there is no response $J^{y;y}/E^{y}$.

3) $J^{x;y}/E^{y}$

$\mathcal{Z}_{+-}^{xy}$ is expanded as%
\begin{align}
\mathcal{Z}_{+-}^{xy}& =\frac{\lambda \left( \lambda
^{2}k_{y}^{2}+B^{2}\right) }{\left( \lambda ^{2}k^{2}+\left( B+Jf_{X}\right)
^{2}\right) ^{3/2}}  \notag \\
& +J\lambda \left( -\frac{3B\left( \lambda ^{2}k_{y}^{2}+B^{2}\right) f_{X}}{%
\left( \lambda ^{2}k^{2}+B^{2}\right) ^{5/2}}+B\frac{2f_{X}-k_{x}\partial
_{k_{y}}f_{X}}{\left( \lambda ^{2}k^{2}+B^{2}\right) ^{3/2}}\right)
\end{align}%
up to the first order in $J$. It is nonzero for all $N_{X}$ due to the first
term.

4) $\mathcal{S}^{x;x}/E^{x}$

We show that $\mathcal{S}^{x;x}/E^{x}$ is nonzero only for the $%
d_{x^{2}-y^{2}}$-wave altermagnet. We have $\mathcal{S}^{x;x}/E^{x}\neq 0$
only for $d^{\prime }=d_{x^{2}-y^{2}}$ as in shown as follows. $\mathcal{Q}%
_{+-}^{xx}/\varepsilon _{mn}$ is expanded as%
\begin{align}
\frac{\mathcal{Q}_{+-}^{xx}}{\varepsilon _{mn}} =&-\frac{\lambda B}{2\left(
\lambda ^{2}k^{2}+B^{2}\right) ^{3/2}}-\frac{J\lambda ^{2}k^{2}\left(
f_{X}-k_{x}\partial _{k_{x}}f_{X}\right) }{2\left( \lambda
^{2}k^{2}+B^{2}\right) ^{5/2}}  \notag \\
&+\frac{JB^{2}\left( 2f_{X}+k_{x}\partial _{k_{x}}f_{X}\right) }{2\left(
\lambda ^{2}k^{2}+B^{2}\right) ^{5/2}}
\end{align}%
up to the first order in $J$.

First, we consider the case with $J=0$, where%
\begin{equation}
\frac{\mathcal{Q}_{+-}^{xx}}{\varepsilon _{mn}}=-\frac{\lambda B}{2\left(
\lambda ^{2}k^{2}+B^{2}\right) ^{3/2}}.
\end{equation}
We have%
\begin{equation}
\int_{0}^{\infty }kdk\int \frac{\mathcal{Q}_{+-}^{xx}}{\varepsilon _{mn}}%
d\phi =-\frac{B}{2\left\vert B\right\vert \lambda }.
\end{equation}%
Hence, there is nonzero response in the presence of nonzero $B$.

Next, we consider the case with $B=0$, where%
\begin{equation}
\frac{\mathcal{Q}_{+-}^{xx}}{\varepsilon _{mn}}=-\frac{J\lambda
^{2}k^{2}\left( f_{X}-k_{x}\partial _{k_{x}}f_{X}\right) }{2\left( \lambda
^{2}k^{2}+B^{2}\right) ^{5/2}}.
\end{equation}%
By using 
\begin{align}
\partial _{k_{x}} =&\cos \phi \partial _{k}-\frac{\sin \phi }{k}\partial
_{\phi }, \\
\partial _{k_{y}} =&\sin \phi \partial _{k}+\frac{\cos \phi }{k}\partial
_{\phi },
\end{align}%
we have%
\begin{align}
&f_{X}-k_{x}\partial _{k_{x}}f_{X}  \notag \\
=&f_{X}-k\cos \phi \left( \cos \phi \partial _{k}f_{X}-\frac{\sin \phi }{k}%
\partial _{\phi }f_{X}\right) .
\end{align}%
For $f_{X}=k^{N_{X}}\cos N_{X}\phi $, we have%
\begin{align}
&f_{X}-k_{x}\partial _{k_{x}}f_{X} = & -\frac{k^{N_{X}}}{2}(N_{X}\cos \left(
N_{X}-2\right) \phi  \notag \\
&+\left( N_{X}-2\right) \cos N_{X}\phi ). & 
\end{align}%
It is nonzero only for $N_{X}=2$. It is nonzero contribution only for the $%
d_{x^{2}-y^{2}}$-wave altermagnet. For $f_{X}=k^{N_{X}}\sin N_{X}\phi $, we
also have%
\begin{align}
&f_{X}-k_{x}\partial _{k_{x}}f_{X} = & -\frac{k^{N_{X}}}{2}(N_{X}\sin \left(
N_{X}-2\right) \phi  \notag \\
&+\left( N_{X}-2\right) \sin N_{X}\phi ), & 
\end{align}%
which is zero for all $N_{X}$. It leads to%
\begin{align}
\frac{S^{x;x}}{E^{x}} =&-\int d^{2}kf_{+-}\frac{2\mathcal{Q}_{+-}^{xx}}{%
\varepsilon _{+-}}  \notag \\
=&-2\pi Jm\sqrt{m\left( m\lambda ^{2}+2\mu \right) }\text{sgn}\left( \lambda
\right)  \notag \\
&-\frac{\pi B}{2\lambda ^{3}}\left( BJ+\lambda ^{2}\right)  \notag \\
&+B\ln \frac{B^{2}+2m\lambda ^{2}\left\{ m\lambda ^{2}+\mu -M\right\} }{%
B^{2}+2m\lambda ^{2}\left\{ m\lambda ^{2}+\mu +M\right\} }
\end{align}%
with%
\begin{equation}
M\equiv \sqrt{\left( m\lambda ^{2}\right) ^{2}+2\mu m\lambda ^{2}+B^{2}}
\end{equation}%
up to the first order in $J$. For $B=0$, we obtain%
\begin{equation}
\frac{S^{x}}{E^{x}}=-\frac{2mJ\pi }{\lambda }.  \notag
\end{equation}%
It is nonzero even for $B=0$.

5) $\mathcal{S}^{y;y}/E^{y}$

We show that $\mathcal{S}^{y;y}/E^{y}$ is nonzero only for the $d_{xy}$-wave
altermagnet. We have $\mathcal{S}^{y;y}/E^{y}\neq 0$ only for $d^{\prime
}=d_{xy}$ as in shown as follows. $\mathcal{Q}_{+-}^{yy}/\varepsilon _{mn}$
is expanded as%
\begin{align}
\frac{\mathcal{Q}_{+-}^{yy}}{\varepsilon _{mn}} =&-\frac{\lambda B}{2\left(
\lambda ^{2}k^{2}+B^{2}\right) ^{3/2}}+\frac{J\lambda ^{2}k^{2}\left(
f_{X}-k_{y}\partial _{k_{y}}f_{X}\right) }{2\left( \lambda
^{2}k^{2}+B^{2}\right) ^{5/2}}  \notag \\
&+\frac{JB^{2}\left( 2f_{X}+k_{y}\partial _{k_{y}}f_{X}\right) }{2\left(
\lambda ^{2}k^{2}+B^{2}\right) ^{5/2}}
\end{align}%
up to the first order in $J$. As in the case of $\mathcal{Q}_{+-}^{xx}$,
there is nonzero response for $B\neq 0$.

We consider the case with $B=0$, where%
\begin{equation}
\frac{\mathcal{Q}_{+-}^{yy}}{\varepsilon _{mn}}=\frac{J\lambda
^{2}k^{2}\left( f_{X}-k_{y}\partial _{k_{y}}f_{X}\right) }{2\left( \lambda
^{2}k^{2}+B^{2}\right) ^{5/2}}.
\end{equation}%
For $f_{X}=k^{N_{X}}\cos N_{X}\phi $, we have%
\begin{align}
f_{X}-k_{x}\partial _{k_{x}}f_{X} =&\frac{k^{N_{X}}}{2}(N_{X}\sin \left(
N_{X}-2\right) \phi  \notag \\
&+N_{X}\sin N_{X}\phi +2\cos N_{X}\phi ).
\end{align}%
On the other hand, for $f_{X}=k^{N_{X}}\sin N_{X}\phi $, we have 
\begin{align}
f_{X}-k_{x}\partial _{k_{x}}f_{X} =&-\frac{k^{N_{X}}}{2}(N_{X}\cos \left(
N_{X}-2\right) \phi  \notag \\
&+N_{X}\cos N_{X}\phi -2\sin N_{X}\phi ),
\end{align}%
which is nonzero only for $N_{X}=2$ because 
\begin{equation}
\int \left( f_{X}-k_{x}\partial _{k_{x}}f_{X}\right) d\phi =-2k^{2}\pi
\delta _{N_{X},2}.
\end{equation}%
It leads to%
\begin{align}
\frac{S^{y;y}}{E^{x}} =&-\int d^{2}kf_{+-}\frac{2\mathcal{Q}_{+-}^{yy}}{%
\varepsilon _{+-}}  \notag \\
=&-2\pi Jm\sqrt{m\left( m\lambda ^{2}+2\mu \right) }\text{sgn}\left( \lambda
\right)  \notag \\
&-\frac{\pi B}{2\lambda ^{3}}\left( BJ+\lambda ^{2}\right)  \notag \\
&+B\ln \frac{B^{2}+2m\lambda ^{2}\left\{ m\lambda ^{2}+\mu -M\right\} }{%
B^{2}+2m\lambda ^{2}\left\{ m\lambda ^{2}+\mu +M\right\} }
\end{align}%
Hence, we have nonzero contribution only for the $d_{xy}$-wave altermagnet.

6) $\mathcal{S}^{x;y}/E^{y}$

We show that $\mathcal{S}^{x;y}/E^{y}$ is nonzero only for the $d_{xy}$-wave
altermagnet. $\mathcal{Q}_{+-}^{xy}$ is expanded as 
\begin{equation}
\mathcal{Q}_{+-}^{xy}=\frac{\lambda k_{y}J\partial _{k_{x}}f_{X}}{2\left(
\lambda ^{2}k^{2}+B^{2}\right) ^{3/2}}
\end{equation}%
up to the first order in $J$. It is evaluated as follows. The numerator of $%
\mathcal{Q}_{+-}^{xy}$ reads

\begin{equation}
k_{y}\partial _{k_{x}}f_{X}=Nk^{N_{X}}\cos \left( N_{X}-1\right) \phi \sin
\phi
\end{equation}%
for $f_{X}=k^{N_{X}}\cos N_{X}\phi $, whose integration over $\phi $ is zero,%
\begin{equation}
\int k_{y}\partial _{k_{x}}f_{X}d\phi =0.
\end{equation}%
We also have%
\begin{equation}
k_{y}\partial _{k_{x}}f_{X}=Nk^{N_{X}}\sin \left( N_{X}-1\right) \phi \sin
\phi ,
\end{equation}%
for $f_{X}=k^{N_{X}}\sin N_{X}\phi $, whose integration over $\phi $ is
obtained as%
\begin{equation}
\int k_{y}\partial _{k_{x}}f_{X}d\phi =-2k^{2}\pi \delta _{N_{X},2}.
\end{equation}%
Then, we have the static electric-field induced spin polarization%
\begin{align}
\frac{S^{x;y}}{E^{y}} =&-\int d^{2}kf_{+-}\frac{2\mathcal{Q}_{+-}^{xy}}{%
\varepsilon _{+-}}  \notag \\
=&-\frac{J\pi }{\lambda ^{3}}\sum_{\eta =\pm }\eta \frac{2\left(
B^{2}+m\lambda ^{2}\left( m\lambda ^{2}+\mu +\eta M\right) \right) }{\sqrt{%
B^{2}+2m\lambda ^{2}\left( m\lambda ^{2}+\mu +\eta M\right) }}
\end{align}%
up to the first order in $J$. For $B=0$, it is simplified as%
\begin{equation}
\frac{S^{x}}{E^{y}}=-\frac{2mJ\pi }{\lambda }
\end{equation}%
for $\mu >0$. Hence, $\mathcal{S}^{x;y}/E^{y}$ is nonzero only for the $%
d_{xy}$-wave altermagnet.

7) $\mathcal{S}^{y;x}/E^{x}$

We show that $\mathcal{S}^{y;x}/E^{x}$ is nonzero only for the $d_{xy}$-wave
altermagnet. $\mathcal{Q}_{+-}^{yx}$ is expanded as 
\begin{equation}
\mathcal{Q}_{+-}^{yx}=\frac{\lambda k_{x}J\partial _{k_{y}}f_{X}}{2\left(
\lambda ^{2}k^{2}+B^{2}\right) ^{3/2}}
\end{equation}%
up to the first order in $J$. It is evaluated as follows. We study $\mathcal{%
Q}_{+-}^{yx}$, whose numerator is $k_{x}\partial _{k_{y}}f_{X}$.

\begin{equation}
k_{x}\partial _{k_{y}}f_{X}=Nk^{N_{X}}\sin \left( N_{X}-1\right) \phi \cos
\phi
\end{equation}%
for $f_{X}=k^{N_{X}}\cos N_{X}\phi $, whose integration over $\phi $ is 
\begin{equation}
\int k_{y}\partial _{k_{x}}f_{X}d\phi =0.
\end{equation}%
We also have%
\begin{equation}
k_{x}\partial _{k_{y}}f_{X}=Nk^{N_{X}}\cos \left( N_{X}-1\right) \phi \cos
\phi
\end{equation}%
for $f_{X}=k^{N_{X}}\sin N_{X}\phi $, whose integration is obtained as%
\begin{equation}
\int k_{x}\partial _{k_{y}}f_{X}d\phi =2k^{2}\pi \delta _{N_{X},2}.
\end{equation}%
Then, we have the static electric-field induced spin polarization%
\begin{align}
\frac{S^{y;x}}{E^{y}} =&-\int d^{2}kf_{+-}\frac{2\mathcal{Q}_{+-}^{yx}}{%
\varepsilon _{+-}}  \notag \\
=&\frac{J\pi }{\lambda ^{3}}\sum_{\eta =\pm }\eta \frac{2\left(
B^{2}+m\lambda ^{2}\left( m\lambda ^{2}+\mu +\eta M\right) \right) }{\sqrt{%
B^{2}+2m\lambda ^{2}\left( m\lambda ^{2}+\mu +\eta M\right) }}
\end{align}%
up to the first order in $J$. For $B=0$, it is simplified as%
\begin{equation}
\frac{S^{x}}{E^{y}}=\frac{2mJ\pi }{\lambda }
\end{equation}%
for $\mu >0$. Hence, $\mathcal{S}^{y;x}/E^{x}$ is nonzero only for the $%
d_{xy}$-wave altermagnet.

\section{Uhlmann geometry}

\label{ApU}

We prove Eq.(\ref{SLD}) in the main text. We first calculate 
\begin{align}
&\left\langle \psi _{n}\left( \mathbf{k}\right) \right\vert d\rho \left\vert
\psi _{m}\left( \mathbf{k}\right) \right\rangle  \notag \\
=&\sum_{n^{\prime }}\partial _{k_{\mu }}p_{n^{\prime }}\delta _{nn^{\prime
}}\delta _{n^{\prime }m}  \notag \\
&-ip_{n^{\prime }}a_{nn^{\prime }}^{\mu }\left( \mathbf{k}\right) \delta
_{n^{\prime }m}-ip_{n^{\prime }}\delta _{nn^{\prime }}a_{mn^{\prime }}^{\mu
\ast }\left( \mathbf{k}\right)  \notag \\
=&\delta _{nm}\partial _{k_{\mu }}p_{n}-ip_{n}a_{nm}^{\mu }\left( \mathbf{k}%
\right) +ip_{m}a_{nm}^{\mu }\left( \mathbf{k}\right) ,  \label{SLD1}
\end{align}%
where we have used%
\begin{align}
d\rho =&\sum_{n^{\prime }}\partial _{k_{\mu }}p_{n^{\prime }}\left\vert \psi
_{n^{\prime }}\left( \mathbf{k}\right) \right\rangle \left\langle \psi
_{n^{\prime }}\left( \mathbf{k}\right) \right\vert  \notag \\
&+p_{n^{\prime }}\left( \partial _{k_{\mu }}\left\vert \psi _{n^{\prime
}}\left( \mathbf{k}\right) \right\rangle \right) \left\langle \psi
_{n^{\prime }}\left( \mathbf{k}\right) \right\vert  \notag \\
&+p_{n^{\prime }}\left\vert \psi _{n^{\prime }}\left( \mathbf{k}\right)
\right\rangle \left( \partial _{k_{\mu }}\left\langle \psi _{n^{\prime
}}\left( \mathbf{k}\right) \right\vert \right) .
\end{align}%
On the other hand, we have%
\begin{align}
&\left\langle \psi _{n}\left( \mathbf{k}\right) \right\vert \frac{\mathcal{L}%
\rho +\rho \mathcal{L}}{2}\left\vert \psi _{m}\left( \mathbf{k}\right)
\right\rangle  \notag \\
=&\left\langle \psi _{n}\left( \mathbf{k}\right) \right\vert
(\sum_{n^{\prime }}\frac{1}{2}\frac{\partial _{k_{\mu }}p_{n^{\prime }}}{%
p_{n^{\prime }}}\left\vert \psi _{n^{\prime }}\right\rangle \left\langle
\psi _{n^{\prime }}\right\vert  \notag \\
&\times \sum_{m^{\prime }}p_{m^{\prime }}\left\vert \psi _{m^{\prime
}}\left( \mathbf{k}\right) \right\rangle \left\langle \psi _{m^{\prime
}}\left( \mathbf{k}\right) \right\vert  \notag \\
&+i\sum_{n^{\prime }\neq m^{\prime }}\frac{p_{m^{\prime }}-p_{n^{\prime }}}{%
p_{m^{\prime }}+p_{n^{\prime }}}a_{n^{\prime }m^{\prime }}^{\mu }\left\vert
\psi _{n^{\prime }}\right\rangle \left\langle \psi _{m^{\prime }}\right\vert
\notag \\
&\times \sum_{m^{\prime \prime }}p_{m^{\prime \prime }}\left\vert \psi
_{m^{\prime \prime }}\left( \mathbf{k}\right) \right\rangle \left\langle
\psi _{m^{\prime \prime }}\left( \mathbf{k}\right) \right\vert  \notag \\
&+\sum_{m^{\prime \prime }}p_{m^{\prime \prime }}\left\vert \psi _{m^{\prime
\prime }}\left( \mathbf{k}\right) \right\rangle \left\langle \psi
_{m^{\prime \prime }}\left( \mathbf{k}\right) \right\vert  \notag \\
&\times \sum_{n^{\prime }}\frac{1}{2}\frac{\partial _{k_{\mu }}p_{n^{\prime
}}}{p_{n^{\prime }}}\left\vert \psi _{n^{\prime }}\right\rangle \left\langle
\psi _{n^{\prime }}\right\vert  \notag \\
&+\sum_{m^{\prime }}p_{m^{\prime \prime }}\left\vert \psi _{m^{\prime \prime
}}\left( \mathbf{k}\right) \right\rangle \left\langle \psi _{m^{\prime
\prime }}\left( \mathbf{k}\right) \right\vert  \notag \\
&\times i\sum_{n^{\prime }\neq m^{\prime }}\frac{p_{m^{\prime
}}-p_{n^{\prime }}}{p_{m^{\prime }}+p_{n^{\prime }}}a_{n^{\prime }m^{\prime
}}^{\mu }\left\vert \psi _{n^{\prime }}\right\rangle \left\langle \psi
_{m^{\prime }}\right\vert )\left\vert \psi _{m}\left( \mathbf{k}\right)
\right\rangle .
\end{align}%
It is further calculated as%
\begin{align}
&\sum_{n^{\prime }}\delta _{nn^{\prime }}\frac{1}{2}\frac{\partial _{k_{\mu
}}p_{n^{\prime }}}{p_{n^{\prime }}}\delta _{n^{\prime }m^{\prime
}}p_{m^{\prime }}\delta _{m^{\prime }m}  \notag \\
&+i\sum_{n^{\prime }\neq m^{\prime }}\frac{p_{m^{\prime }}-p_{n^{\prime }}}{%
p_{m^{\prime }}+p_{n^{\prime }}}a_{n^{\prime }m^{\prime }}^{\mu }\delta
_{nn^{\prime }}\delta _{m^{\prime }m^{\prime \prime }}\delta _{m^{\prime
\prime }m}  \notag \\
&+\sum_{m^{\prime }}p_{m^{\prime }}\delta _{nm^{\prime }}\sum_{n^{\prime }}%
\frac{1}{2}\frac{\partial _{k_{\mu }}p_{n^{\prime }}}{p_{n^{\prime }}}\delta
_{m^{\prime }n^{\prime }}\delta _{n^{\prime }m}  \notag \\
&+\sum_{n^{\prime }\neq m^{\prime }}\delta _{nm^{\prime \prime
}}p_{m^{\prime \prime }}\frac{p_{m^{\prime }}-p_{n^{\prime }}}{p_{m^{\prime
}}+p_{n^{\prime }}}a_{n^{\prime }m^{\prime }}^{\mu }\delta _{m^{\prime
\prime }n^{\prime }}\delta _{m^{\prime }m}  \notag \\
=&\delta _{nm}\frac{1}{2}\frac{\partial _{k_{\mu }}p_{n}}{p_{n}}p_{n}+i\frac{%
p_{m}-p_{n}}{p_{m}+p_{n}}a_{nm}^{\mu }p_{m}  \notag \\
&+\delta _{nm}p_{n}\frac{1}{2}\frac{\partial _{k_{\mu }}p_{n}}{p_{n}}+ip_{n}%
\frac{p_{m}-p_{n}}{p_{m}+p_{n}}a_{nm}^{\mu }  \notag \\
=&\delta _{nm}\partial _{k_{\mu }}p_{n}+i\left( p_{m}-p_{n}\right)
a_{nm}^{\mu },
\end{align}%
which is identical to Eq.(\ref{SLD1}).

\subsection{Positive Operator-Valued Measure}

Positive Operator-Valued Measure (POVM) $\Pi _{n}$ satisfies\cite{Holevo}

\begin{equation}
p_{n}=\text{Tr}\rho \Pi _{n},
\end{equation}%
and%
\begin{equation}
\sum_{n=1}^{N}\Pi _{n}=1.
\end{equation}%
It is a generalization of the projective measurement. These two conditions
lead to the conservation of the provability,%
\begin{equation}
\sum_{n=1}^{N}p_{n}=\sum_{n=1}^{N}\text{Tr}\rho \Pi _{n}=\text{Tr}\rho
\sum_{n=1}^{N}\Pi _{n}=\text{Tr}\rho =1.
\end{equation}%
By using it, the classical Fisher information (\ref{CFisher}) is rewritten
as 
\begin{align}
\mathcal{F}_{\text{CFisher}}^{\mu \nu } =&\sum_{n=1}^{N}\text{Tr}\rho \Pi
_{n}\frac{\text{Tr}\frac{\partial \rho }{\partial k_{\mu }}\Pi _{n}}{\text{Tr%
}\rho \Pi _{n}}\frac{\text{Tr}\frac{\partial \rho }{\partial k_{\nu }}\Pi
_{n}}{\text{Tr}\rho \Pi _{n}}  \notag \\
=&\sum_{n=1}^{N}\frac{1}{\text{Tr}\rho \Pi _{n}}\text{Tr}\left( \frac{%
\partial \rho }{\partial k_{\mu }}\Pi _{n}\right) \text{Tr}\left( \frac{%
\partial \rho }{\partial k_{\nu }}\Pi _{n}\right) .
\end{align}

\subsection{Quantum Cram\'{e}r-Rao inequality}

\label{ApQCR}

We prove the quantum Cram\'{e}r-Rao inequality (\ref{CR}). It is enough to
show\cite{Matteo}%
\begin{equation}
\sum_{\mu \nu }a_{\mu }\mathcal{F}_{\text{CFisher}}^{\mu \nu }a_{\nu
}<\sum_{\mu \nu }a_{\mu }\mathcal{F}_{\text{QFisher}}^{\mu \nu }a_{\nu }
\end{equation}%
for arbitrary set of $a_{\mu }$. By introducing%
\begin{equation}
\mathcal{\tilde{L}}=\sum_{\mu }a_{\mu }\mathcal{L}^{\mu },
\end{equation}%
we define $\mathcal{F}_{\text{CFisher}}$ as%
\begin{align}
\mathcal{F}_{\text{CFisher}} \equiv &\sum_{n=1}^{N}\frac{1}{\text{Tr}\rho
\Pi _{n}}\left( \text{Tr}\left( \frac{1}{2}\left( \mathcal{\tilde{L}}\rho
+\rho \mathcal{\tilde{L}}\right) \Pi _{n}\right) \right) ^{2}  \notag \\
=&\sum_{n=1}^{N}\frac{1}{\text{Tr}\rho \Pi _{n}}\sum_{\mu \nu }\text{Tr}%
\left( \frac{a_{\mu }}{2}\left( \mathcal{L}^{\mu }\rho +\rho \mathcal{L}%
^{\mu }\right) \Pi _{n}\right)  \notag \\
&\times \text{Tr}\left( \frac{a_{\nu }}{2}\left( \mathcal{L}^{\nu }\rho
+\rho \mathcal{L}^{\nu }\right) \Pi _{n}\right)  \notag \\
=&\sum_{\mu \nu }a_{\mu }\mathcal{F}_{\text{CFisher}}^{\mu \nu }a_{\nu }.
\end{align}

First, we show the inequality%
\begin{align}
&\sum_{\mu \nu }a_{\mu }\mathcal{F}_{\text{CFisher}}^{\mu \nu }a_{\nu } 
\notag \\
=&\sum_{n=1}^{N}\frac{1}{\text{Tr}\rho \Pi _{n}}\left( \text{Tr}\left( \frac{%
1}{2}\left( \mathcal{\tilde{L}}\rho +\rho \mathcal{\tilde{L}}\right) \Pi
_{n}\right) \right) ^{2}  \notag \\
=&\sum_{n=1}^{N}\frac{1}{\text{Tr}\rho \Pi _{n}}\left( \text{Tr}\left( \text{%
Re}\mathcal{\tilde{L}}\rho \Pi _{n}\right) \right) ^{2}  \notag \\
\leq &\sum_{n=1}^{N}\frac{1}{\text{Tr}\rho \Pi _{n}}\left\vert \text{Tr}%
\left( \mathcal{\tilde{L}}\rho \Pi _{n}\right) \right\vert ^{2}.
\end{align}

Next, setting%
\begin{equation}
A^{\dagger }=\frac{\sqrt{\rho }\sqrt{\Pi _{n}}}{\sqrt{\text{Tr}\rho \Pi _{n}}%
},\qquad B=\sqrt{\Pi _{n}}\mathcal{\tilde{L}}\sqrt{\rho },
\end{equation}
we use the Schwartz inequality%
\begin{equation}
\left\vert \text{Tr}A^{\dagger }B\right\vert ^{2}\leq \text{Tr}A^{\dagger }A%
\text{Tr}B^{\dagger }B.
\end{equation}%
The left-hand side is%
\begin{align}
\left\vert \text{Tr}A^{\dagger }B\right\vert ^{2} =&\left\vert \text{Tr}%
\frac{\sqrt{\rho }\sqrt{\Pi _{n}}}{\sqrt{\text{Tr}\rho \Pi _{n}}}\sqrt{\Pi
_{n}}\mathcal{\tilde{L}}\sqrt{\rho }\right\vert ^{2}  \notag \\
=&\left\vert \text{Tr}\frac{\sqrt{\rho }\Pi _{n}\mathcal{\tilde{L}}\sqrt{%
\rho }}{\sqrt{\text{Tr}\rho \Pi _{n}}}\right\vert ^{2}  \notag \\
=&\left\vert \text{Tr}\frac{\mathcal{\tilde{L}}\rho \Pi _{n}}{\sqrt{\text{Tr}%
\rho \Pi _{n}}}\right\vert ^{2},
\end{align}%
while the right-hand side is%
\begin{align}
&\sum_{n=1}^{N}\text{Tr}A^{\dagger }A\text{Tr}B^{\dagger }B  \notag \\
=&\sum_{n=1}^{N}\text{Tr}\frac{\sqrt{\rho }\sqrt{\Pi _{n}}}{\sqrt{\text{Tr}%
\rho \Pi _{n}}}\frac{\sqrt{\Pi _{n}}\sqrt{\rho }}{\sqrt{\text{Tr}\rho \Pi
_{n}}}\text{Tr}\sqrt{\Pi _{n}}\mathcal{\tilde{L}}\sqrt{\rho }\sqrt{\rho }%
\mathcal{\tilde{L}}\sqrt{\Pi _{n}}  \notag \\
=&\sum_{n=1}^{N}\text{Tr}\frac{\sqrt{\rho }\Pi _{n}\sqrt{\rho }}{\text{Tr}%
\rho \Pi _{n}}\text{Tr}\sqrt{\Pi _{n}}\mathcal{\tilde{L}}\rho \mathcal{%
\tilde{L}}\sqrt{\Pi _{n}}  \notag \\
=&\sum_{n=1}^{N}\text{Tr}\frac{\rho \Pi _{n}}{\text{Tr}\rho \Pi _{n}}\text{Tr%
}\Pi _{n}\mathcal{\tilde{L}}\rho \mathcal{\tilde{L}}=\sum_{n=1}^{N}\text{Tr}%
\Pi _{n}\mathcal{\tilde{L}}\rho \mathcal{\tilde{L}}  \notag \\
=&\text{Tr}\mathcal{\tilde{L}}\rho \mathcal{\tilde{L}}=\text{Tr}\rho 
\mathcal{\tilde{L}}^{2}=\frac{1}{2}\mathcal{F}_{\text{QFisher}},
\end{align}%
where we have defined%
\begin{align}
&\mathcal{F}_{\text{QFisher}}^{\mu \nu }\equiv \frac{1}{2}\text{Tr}\rho 
\mathcal{\tilde{L}}^{2}=\frac{1}{2}\text{Tr}\rho \left( \sum_{\mu }a_{\mu }%
\mathcal{L}^{\mu }\right) ^{2}  \notag \\
=&\frac{1}{2}\sum_{\mu \nu }a_{\mu }\text{Tr}\left[ \rho \left\{ \mathcal{L}%
^{\mu },\mathcal{L}^{\nu }\right\} \right] a_{\nu }  \notag \\
=&\sum_{\mu \nu }a_{\mu }\mathcal{F}_{\text{QFisher}}^{\mu \nu }a_{\nu }.
\end{align}%
Hence, the quantum Cram\'{e}r-Rao inequality (\ref{CR}) is proved.

\subsection{Quantum Fisher information for a pure state}

\label{ApQFpure}

By inserting 
\begin{equation}
\rho =\left\vert \psi _{n}\right\rangle \left\langle \psi _{n}\right\vert ,
\end{equation}%
to Eq.(\ref{FisherRho}), we have%
\begin{align}
& \mathcal{F}_{\text{QFisher}}^{\mu \nu }  \notag \\
=& 2\text{Tr}\left\vert \psi _{n}\right\rangle \left\langle \psi
_{n}\right\vert (\frac{\partial \left\vert \psi _{n}\right\rangle
\left\langle \psi _{n}\right\vert }{\partial k_{\mu }}\frac{\partial
\left\vert \psi _{n}\right\rangle \left\langle \psi _{n}\right\vert }{%
\partial k_{\nu }}  \notag \\
& +\frac{\partial \left\vert \psi _{n}\right\rangle \left\langle \psi
_{n}\right\vert }{\partial k_{\nu }}\frac{\partial \left\vert \psi
_{n}\right\rangle \left\langle \psi _{n}\right\vert }{\partial k_{\mu }}) 
\notag \\
=& 2\text{Tr}\left\langle \psi _{n}\right\vert \frac{\partial \left\vert
\psi _{n}\right\rangle }{\partial k_{\mu }}\left\langle \psi _{n}\right\vert 
\frac{\partial \left\vert \psi _{n}\right\rangle }{\partial k_{\nu }}  \notag
\\
& +\left\langle \psi _{n}\right\vert \frac{\partial \left\vert \psi
_{n}\right\rangle }{\partial k_{\mu }}\frac{\partial \left\langle \psi
_{n}\right\vert }{\partial k_{\nu }}\left\vert \psi _{n}\right\rangle  \notag
\\
& +\frac{\partial \left\langle \psi _{n}\right\vert }{\partial k_{\mu }}%
\frac{\partial \left\vert \psi _{n}\right\rangle }{\partial k_{\nu }}+\frac{%
\partial \left\langle \psi _{n}\right\vert }{\partial k_{\mu }}\left\vert
\psi _{n}\right\rangle \frac{\partial \left\langle \psi _{n}\right\vert }{%
\partial k_{\nu }}\left\vert \psi _{n}\right\rangle  \notag \\
& +\left\langle \psi _{n}\right\vert \frac{\partial \left\vert \psi
_{n}\right\rangle }{\partial k_{\nu }}\left\langle \psi _{n}\right\vert 
\frac{\partial \left\vert \psi _{n}\right\rangle }{\partial k_{\mu }}%
+\left\langle \psi _{n}\right\vert \frac{\partial \left\vert \psi
_{n}\right\rangle }{\partial k_{\nu }}\frac{\partial \left\langle \psi
_{n}\right\vert }{\partial k_{\mu }}\left\vert \psi _{n}\right\rangle  \notag
\\
& +\frac{\partial \left\langle \psi _{n}\right\vert }{\partial k_{\nu }}%
\frac{\partial \left\vert \psi _{n}\right\rangle }{\partial k_{\mu }}+\frac{%
\partial \left\langle \psi _{n}\right\vert }{\partial k_{\nu }}\left\vert
\psi _{n}\right\rangle \frac{\partial \left\langle \psi _{n}\right\vert }{%
\partial k_{\mu }}\left\vert \psi _{n}\right\rangle .
\end{align}%
By using the relations%
\begin{align}
& \left\langle \psi _{n}\right\vert \frac{\partial \left\vert \psi
_{n}\right\rangle }{\partial k_{\mu }}\left\langle \psi _{n}\right\vert 
\frac{\partial \left\vert \psi _{n}\right\rangle }{\partial k_{\nu }}%
+\left\langle \psi _{n}\right\vert \frac{\partial \left\vert \psi
_{n}\right\rangle }{\partial k_{\mu }}\frac{\partial \left\langle \psi
_{n}\right\vert }{\partial k_{\nu }}\left\vert \psi _{n}\right\rangle  \notag
\\
=& \left\langle \psi _{n}\right\vert \frac{\partial \left\vert \psi
_{n}\right\rangle }{\partial k_{\mu }}\frac{\partial \left. \left\langle
\psi _{n}\right\vert \psi _{n}\right\rangle }{\partial k_{\nu }}=0,
\label{FA1} \\
& \left\langle \psi _{n}\right\vert \frac{\partial \left\vert \psi
_{n}\right\rangle }{\partial k_{\nu }}\left\langle \psi _{n}\right\vert 
\frac{\partial \left\vert \psi _{n}\right\rangle }{\partial k_{\mu }}%
+\left\langle \psi _{n}\right\vert \frac{\partial \left\vert \psi
_{n}\right\rangle }{\partial k_{\nu }}\frac{\partial \left\langle \psi
_{n}\right\vert }{\partial k_{\mu }}\left\vert \psi _{n}\right\rangle  \notag
\\
=& \left\langle \psi _{n}\right\vert \frac{\partial \left\vert \psi
_{n}\right\rangle }{\partial k_{\nu }}\frac{\partial \left. \left\langle
\psi _{n}\right\vert \psi _{n}\right\rangle }{\partial k_{\mu }}=0,
\label{FA2}
\end{align}%
we have%
\begin{align}
\mathcal{F}_{\text{QFisher}}^{\mu \nu }=& 2\text{Tr}\frac{\partial
\left\langle \psi _{n}\right\vert }{\partial k_{\mu }}\frac{\partial
\left\vert \psi _{n}\right\rangle }{\partial k_{\nu }}-\frac{\partial
\left\langle \psi _{n}\right\vert }{\partial k_{\mu }}\left\vert \psi
_{n}\right\rangle \left\langle \psi _{n}\right\vert \frac{\partial
\left\vert \psi _{n}\right\rangle }{\partial k_{\nu }}  \notag \\
& +\frac{\partial \left\langle \psi _{n}\right\vert }{\partial k_{\nu }}%
\frac{\partial \left\vert \psi _{n}\right\rangle }{\partial k_{\mu }}-\frac{%
\partial \left\langle \psi _{n}\right\vert }{\partial k_{\nu }}\left\vert
\psi _{n}\right\rangle \left\langle \psi _{n}\right\vert \frac{\partial
\left\vert \psi _{n}\right\rangle }{\partial k_{\mu }}  \notag \\
=& 2\left( \mathcal{F}^{\mu \nu }+\mathcal{F}^{\nu \mu }\right) =4g^{\mu \nu
}.
\end{align}%
This is Eq.(\ref{FIpure}) in the main text.%
\begin{align}
& \mathcal{U}^{\mu \nu }  \notag \\
=& -2i\text{Tr}\left\vert \psi _{n}\right\rangle \left\langle \psi
_{n}\right\vert (\frac{\partial \left\vert \psi _{n}\right\rangle
\left\langle \psi _{n}\right\vert }{\partial k_{\mu }}\frac{\partial
\left\vert \psi _{n}\right\rangle \left\langle \psi _{n}\right\vert }{%
\partial k_{\nu }}  \notag \\
& -\frac{\partial \left\vert \psi _{n}\right\rangle \left\langle \psi
_{n}\right\vert }{\partial k_{\nu }}\frac{\partial \left\vert \psi
_{n}\right\rangle \left\langle \psi _{n}\right\vert }{\partial k_{\mu }}) 
\notag \\
=& -2i\text{Tr}\left\langle \psi _{n}\right\vert \frac{\partial \left\vert
\psi _{n}\right\rangle }{\partial k_{\mu }}\left\langle \psi _{n}\right\vert 
\frac{\partial \left\vert \psi _{n}\right\rangle }{\partial k_{\nu }}  \notag
\\
& +\left\langle \psi _{n}\right\vert \frac{\partial \left\vert \psi
_{n}\right\rangle }{\partial k_{\mu }}\frac{\partial \left\langle \psi
_{n}\right\vert }{\partial k_{\nu }}\left\vert \psi _{n}\right\rangle  \notag
\\
& +\frac{\partial \left\langle \psi _{n}\right\vert }{\partial k_{\mu }}%
\frac{\partial \left\vert \psi _{n}\right\rangle }{\partial k_{\nu }}+\frac{%
\partial \left\langle \psi _{n}\right\vert }{\partial k_{\mu }}\left\vert
\psi _{n}\right\rangle \frac{\partial \left\langle \psi _{n}\right\vert }{%
\partial k_{\nu }}\left\vert \psi _{n}\right\rangle  \notag \\
& -\left\langle \psi _{n}\right\vert \frac{\partial \left\vert \psi
_{n}\right\rangle }{\partial k_{\nu }}\left\langle \psi _{n}\right\vert 
\frac{\partial \left\vert \psi _{n}\right\rangle }{\partial k_{\mu }}%
-\left\langle \psi _{n}\right\vert \frac{\partial \left\vert \psi
_{n}\right\rangle }{\partial k_{\nu }}\frac{\partial \left\langle \psi
_{n}\right\vert }{\partial k_{\mu }}\left\vert \psi _{n}\right\rangle  \notag
\\
& -\frac{\partial \left\langle \psi _{n}\right\vert }{\partial k_{\nu }}%
\frac{\partial \left\vert \psi _{n}\right\rangle }{\partial k_{\mu }}-\frac{%
\partial \left\langle \psi _{n}\right\vert }{\partial k_{\nu }}\left\vert
\psi _{n}\right\rangle \frac{\partial \left\langle \psi _{n}\right\vert }{%
\partial k_{\mu }}\left\vert \psi _{n}\right\rangle .
\end{align}%
Using the relations Eqs.(\ref{FA1}) and (\ref{FA2}), we have%
\begin{align}
\mathcal{U}^{\mu \nu }=& -2i\text{Tr}\frac{\partial \left\langle \psi
_{n}\right\vert }{\partial k_{\mu }}\frac{\partial \left\vert \psi
_{n}\right\rangle }{\partial k_{\nu }}-\frac{\partial \left\langle \psi
_{n}\right\vert }{\partial k_{\mu }}\left\vert \psi _{n}\right\rangle
\left\langle \psi _{n}\right\vert \frac{\partial \left\vert \psi
_{n}\right\rangle }{\partial k_{\nu }}  \notag \\
& -\frac{\partial \left\langle \psi _{n}\right\vert }{\partial k_{\nu }}%
\frac{\partial \left\vert \psi _{n}\right\rangle }{\partial k_{\mu }}+\frac{%
\partial \left\langle \psi _{n}\right\vert }{\partial k_{\nu }}\left\vert
\psi _{n}\right\rangle \left\langle \psi _{n}\right\vert \frac{\partial
\left\vert \psi _{n}\right\rangle }{\partial k_{\mu }}  \notag \\
=& -2i\left( \mathcal{F}^{\mu \nu }-\mathcal{F}^{\nu \mu }\right) =-4\Omega
^{\mu \nu }.
\end{align}%
This is Eq.(\ref{UhlPure}) in the main text.


\end{document}